\documentclass[10pt]{iopart}
\pdfoutput=1
\usepackage[english]{babel}

\usepackage{braket}
\usepackage{breqn}
\usepackage{iopams}
\usepackage{graphicx}
\usepackage{color}
\usepackage{cite}

\newcommand*{\refequa}[1]{(\ref{#1})}

\newcommand{\beq}{\begin{equation}}
\newcommand{\eeq}{\end{equation}}
\newcommand{\tobs}{t_{\rm obs}}

\newcommand{\PP}{\mathcal{P}}
\newcommand{\QQ}{\mathcal{Q}}

\newcommand{\LL}{ L_{\rm r} }  %% this is what used to be L'

\begin{document}
	\title[Large deviations and optimal control forces for hard particles in one dimension]
	{Large deviations and optimal control forces for diffusing hard particles in one dimension}
	\author{Jakub Dolezal$^1$, Robert L. Jack$^{1,2}$}
	\address{$^1$ DAMTP, Centre for Mathematical Sciences, Wilberforce Road, Cambridge CB3 0WA, United Kingdom}
	\address{$^2$ Department of Chemistry, Lensfield Road, Cambridge CB2 1EW, United Kingdom}
	%\date{2018}

	\begin{abstract}
		We analyse large deviations of the dynamical activity in one-dimensional systems of diffusing hard particles.
		Using an optimal-control representation of the large-deviation problem, we analyse effective interaction forces which can be added to the system, to aid sampling of biased ensembles of trajectories.  We find several distinct regimes, as a function of the activity and the system size: we present approximate analytical calculations that characterise the effective interactions in several of these regimes.   For high activity the system is hyperuniform and the interactions are long-ranged and repulsive.  For low activity, there is a near-equilibrium regime described by macroscopic fluctuation theory, characterised by long-ranged attractive forces.  There is also a far-from-equilibrium regime in which one of the interparticle gaps becomes macroscopic and the interactions depend strongly on the size of this gap.  We discuss the extent to which transition path sampling of these ensembles is improved by adding suitable control forces.
	\end{abstract}
	
	%\maketitle
	%\tableofcontents

	\section{Introduction}
	
	Large deviations of time-averaged quantities are becoming increasingly useful for understanding dynamical fluctuations in 
	physical systems~\cite{GallavottiFluctuatonTheorem,LebowitzFluctuationTheorem,Bodineau2004,Mehl2008,Derrida2007,Lecomte2007,Tailleur2007,Garrahan2007,Hedges2009,nemoto_optimizing_2018,Weber2014protein,Hurtado2014,YongJooDynamicalSymmetryBreaking}.  
	For example, consider an ergodic system in which time-averaged quantities converge almost surely to ensemble-averaged values.  Given some large time scale, the probability of a significant deviation between the time-average and the ensemble average is small but finite -- these rare events 
	%These events correspond to deviations from ergodic behaviour, and 
	are described by large-deviation theory~\cite{denH-book,Touchette2009}.  Despite their scarcity, analysis of these events has led to new insight into the behaviour of physical systems, and their dominant fluctuation mechanisms\cite{Ragone201712645,nemoto_optimizing_2018,bodineau_distribution_2005,Hedges2009,Weber2014protein,Jack2019-grow}.
	
	Early studies of these events focused on the entropy production in non-equilibrium systems, which is intrinsically linked to fluctuation theorems \cite{GallavottiFluctuatonTheorem,CrooksFluctuationTheorem,LebowitzFluctuationTheorem,SaitoFluctuationTheorem}.  Another direction has been the analysis of time-averaged currents, aiming towards a general theory of transport in non-equilibrium systems~\cite{Bodineau2004,Derrida2007,bertini_macroscopic,Gingrich2016}.
	Yet another line of enquiry has focused on glassy systems \cite{Garrahan2007,Garrahan2009,Hedges2009,speck_first-order_2012}, which have long-lived metastable states that hinder equilibration.  
	
	In studies of large deviations, there are numerous examples of dynamical phase transitions~\cite{bodineau_distribution_2005,Garrahan2007,Garrahan2009,Hedges2009,speck_first-order_2012,lecomte_inactive_2012,YongJooDynamicalSymmetryBreaking,Hirschberg2015,Shpiel2017,Janas2016,espigares2018-crit,nemoto_optimizing_2018}.  In simple terms, these occur when deviations from ergodic behaviour occur by mechanisms that differ qualitatively from the typical behaviour of the model.  For example, the rare events may involve spontaneous symmetry breaking, as in \cite{YongJooDynamicalSymmetryBreaking,nemoto_optimizing_2018,espigares2018-crit}.  In other cases, one encounters the phenomenology of first-order phase transitions, including phase coexistence\cite{Garrahan2007,Garrahan2009}.  
	
	Here we focus on a simple Brownian hard particle model (BHPM), which has rich fluctuation behaviour, including dynamical phase transitions~\cite{jack_hyperuniformity_2015,thompson_dynamical_2015}.  It consists of many hard particles diffusing in one dimension, and therefore has some similarities with the simple symmetric exclusion process (SSEP), generalised to continuous space.  Dynamical phase transitions in the SSEP have been analysed in detail~\cite{Derrida2007,appert-rolland_universal_2008,espigares2018-crit,YongJooDynamicalSymmetryBreaking,brewer_efficient_2018}.  In particular, its behaviour on very large (hydrodynamic) length scales is described by macroscopic fluctuation theory (MFT)~\cite{bertini_macroscopic}.  The general applicability of this theory means that its predictions apply also in the BHPM~\cite{jack_hyperuniformity_2015,thompson_dynamical_2015}.  A key prediction is that for large deviations with low values of the dynamical activity, the system becomes macroscopically inhomogeneous, which was identified in~\cite{jack_hyperuniformity_2015} as a form of phase separation.  
	
	This article extends previous work~\cite{jack_hyperuniformity_2015,thompson_dynamical_2015} on the BHPM in two main directions.  First, we
	analyse the macroscopically inhomogeneous regime, both numerically and analytically.  We generalise existing MFT predictions~\cite{bodineau_distribution_2005,lecomte_inactive_2012} to the BHPM and we show quantitative agreement with numerical results.  This theory is accurate for fluctuations that involve smooth large-scale modulations of the density. We also analyse fluctuations that are too large to be captured by MFT, where one observes the formation of an extensive region without any particles at all: a \emph{macroscopic gap}.  Based on previous results for kinetically constrained spin models~\cite{Bodineau2012cmp,Bodineau2012jsp,nemoto_finite-size_2017}, we propose a simple interfacial model that captures qualitative features of this regime.
	
	%combine numerical results with analytical theory, to show that the macroscopically inhomogeneous behaviour can be separated into two regimes.  The first regime is associated with smooth large-scale modulations of the density.  The second is associated with the formation of a macroscopic region without any particles at all: a \emph{macroscopic gap}.   These two types of rare events have probabilities that have different scalings as the system size tends to infinity.
	
	The second direction of this work is to show how the addition of control forces~\cite{Fleming85,OptimalControlRepresentationChetrite} to the equations of motion of the system can be used to improve numerical convergence. It is known \cite{OptimalControlRepresentationChetrite,jack_ptps} that for any given biased ensemble there is an optimal set of control forces for which numerical sampling of the rare events becomes trivial.  While these optimal forces cannot usually be computed in complex physical systems, it is expected that adding non-optimal control forces can also improve the convergence of numerical calculations, via a form of importance sampling~\cite{nemoto_population-dynamics_2016,nemoto_finite-size_2017,ray_exact_2018,Jacobson-arxiv}.  We use theoretical arguments to derive approximations to the optimal control force, in two regimes that we have identified.  We show that these control forces do indeed improve numerical performance, and this improvement is increasingly strong when we consider large systems.  (This is because our approximations to the optimal control forces are increasingly accurate for large systems.)
	
	The form of this paper is as follows.  Sec.~\ref{model} introduces the models that we consider, and some of the quantities that we will measure.  Sec.~\ref{biased} collects properties of biased ensembles of trajectories.  Sec.~\ref{sec:main-results} gives an overview of the main theoretical  results, before Sec.~\ref{sec:results-mft} and Sec.~\ref{sec:macro-gap} describe the detailed calculations for the two macroscopically inhomogeneous regimes that we identify.  We summarise our conclusions in Sec.~\ref{sec:conc}.

	\section{Models}\label{model}
	
	Consider $N$ hard particles moving in one dimension with periodic boundaries.   Each particle has size $l_0$ and the position of particle $i$ at time $t$ is $x_i(t)$.  We write $X=(x_1,x_2,\dots,x_N)$ for a configuration of the system and $\textbf{x}$ for a trajectory of the the system, over a time interval $[0,\tobs]$.
	The particle motion is stochastic and obeys detailed balance with respect to an equilibrium distribution
	\beq
	p(X) = \frac{1}{Z} \exp[ -\beta U(X) ]
	\label{equ:eqm}
	\eeq
	where $\beta$ is the inverse temperature, $Z$ is a normalisation constant, and $U$ is a pairwise additive potential energy
	$ U=\frac{1}{2}\sum_{i\neq j} v(x_i-x_j) $.
	As in~\cite{jack_hyperuniformity_2015, thompson_dynamical_2015},
	we consider two variants of the system, which have very similar behaviour.
	
	\subsection{Monte Carlo dynamics}
	\label{sec:mc}
	
	The MC variant of the model is a discrete-time Markov process.  On each step, a particle (say $i$) is chosen at random and we propose to move it to a new position $x_i+\Delta x$ where $\Delta x$ is uniformly distributed in $[-M,M]$.  Here, $M$ is a parameter of the model.  We use Glauber dynamics  so the move is accepted with probability 
	\beq
	p_{\rm acc}  = \frac{1}{1+\exp(\beta\Delta U)} % \min( 1, \exp(-\beta\Delta U) )
	\label{equ:glauber}
	\eeq
	where $\Delta U$ is the difference in energy between the current configuration and the proposed configuration.  If the move is rejected then the configuration remains the same.  After each attempted move, the time is incremented by $\delta t = M^2/(12ND_0)$ where $D_0$ is the single-particle diffusion constant, which is also a parameter of the model.  This ensures that in the dilute limit, particles diffuse independently between collisions, with diffusion constant $D_0$.  We use Glauber dynamics because this facilitates later analysis, when we add {control forces} to the system, see Sec.~\ref{sec:doob}.
	
	This variant of the model considers hard particles, so the interaction potential is
	\begin{equation}
	\label{equ:pair-inf}
	v(x)= \cases{ 0,  & $x>l_0$ \\ \infty,  &  $x<l_0$ }
	\end{equation}
	Suppose that the $j$th particle is selected to be moved in a given MC step, and suppose that the neighbouring particles have indices $p,q$.  The probability that the proposed move does not result in two particles overlapping is
	\beq
	\label{equ:def-ri}
	r_j^M = \frac{1}{2M} \left[ \min( M,|x_j - x_p| ) + \min( M,|x_j - x_q| )  \right]
	\eeq
	The superscript $M$ indicates that this quantity depends on the MC step size, it is a label (and not any kind of exponent).
	
	\subsection{Langevin dynamics}
	All numerical results in this work use MC dynamics.  However, it is convenient for theoretical analysis
	%For theoretical analysis it is convenient 
	to consider a Langevin equation
	\beq \label{langevin}
	\dot{x}_i = -\beta D_0 \nabla_i U + \sqrt{2D_0} \eta_i
	\eeq
	where $\eta_i$ is a standard Brownian noise.  In this case the pair potential should be differentiable: we assume a regularised version of (\ref{equ:pair-inf}) such that
	$v(x)=\infty$ for $|x|<l_0$.  Also there is some $l_1$ such that $v(x)=0$ for $|x|>l_0+l_1$, with $v(x)$ a continuous function for $l_0<x\leq l_1$, diverging as $x\to l_0$. This choice ensures that the separation between any pair of particles is always larger than $l_0$.
	
	The similarity between the MC and Langevin models can be justified in the following way.  In the Langevin model, take $l_1=l_0+M$ where $M$ is the step size in the MC model. 
	The two models behave equivalently in the limit $M\to0$: this can be verified by constructing the Fokker-Planck equation for the Langevin model and the corresponding master equation for the MC model, then taking the relevant limits. See \cite{gardiner2004handbook}.
	
	\subsection{Rescaled representation}
	\label{sec:rescaled}
	
	Since we consider hard particles in $d=1$, the ordering of particles in the system is preserved.\footnote{
		The particles exclude a volume $l_0$ which we assume throughout is larger than the MC step $M$.}
	One may always map such a model to a system of point particles that move in a spatial domain of size 
	%\rlj{RLJ: don't like $L'$, I am trying with $\LL$}
	\beq 
	\LL = L-Nl_0 \; .
	\eeq 
	%(again with periodic boundaries).  
	We insist that the particle positions are ordered with $x_1<x_2<\dots$ (modulo periodic boundaries) in which case the position of the $j$th point particle is
	%\beq
	$ 
	\tilde{x}_j = x_j - jl_0 .
	$
	%\label{equ:x-res}
	%\eeq
	For the MC variant of the model, the equilibrium distribution (\ref{equ:eqm}) reduces to an ideal-gas distribution for the positions $\tilde x$.  In some cases, this means that the rescaled system is simpler to analyse.  However, we emphasise that the rescaled system and the original system contain exactly the same information.
	
	In this rescaled representation, it is easy to see that varying $l_0$ in the original model simply shifts particles' positions by constants that are independent of time. It follows that many properties of the system (including trajectories of individual particles) are independent of $l_0$.

	\subsection{Dynamical activity}
	\label{sec:activity}
	
	In the following, we focus on ensembles of trajectories that are biased to low (or high) values of time-averaged measurements of dynamical activity.
	The definition of activity used in this work differs from~\cite{jack_hyperuniformity_2015,thompson_dynamical_2015} -- the choice used here does not change the qualitative behaviour but it makes it easier to analyse, both numerically and computationally.
	Large deviations for a different kind of dynamical activity have recently been analysed in a similar model~\cite{DasLimmer-arxiv}.
	
	%Here we explain how we measure the activity.
	The activity measures motion on a characteristic length scale $a$.  We introduce a dimensionless parameter 
	\beq
	\Phi_a = {\frac{Na}{\LL}}
	\label{equ:def-Phia}
	\eeq
	which is the ratio between $a$ and the average interparticle gap.
	For a trajectory $\textbf{x}$, we define
	\begin{equation}
	K[\mathbf{x}]= 
	\sum_{i=0}^{N}  \int_0^{\tobs} r_i^a(t) \,\mathrm{d}t
	\label{equ:def-K}
	\end{equation}
	where $r_i^a$ is the acceptance probability for an MC move of size $a$, as defined in (\ref{equ:def-ri}).   We allow the parameter $a$ that appears in the definition of $K$ to be different from the parameter $M$ that determines the size of MC moves, although our numerical results take $a=M$.
	Note also that while $K$ is defined in terms of the MC acceptance rate, it can be evaluated directly from particle trajectories, using (\ref{equ:def-K}).  Thus, $K$ is a well-defined quantity in the Langevin variant of the model, as well as in the MC variant.  Also, the value of $K$ only depends on gaps between adjacent particles and is therefore the same in the rescaled representation, or the original representation.

	It is useful to define an intensive (and dimensionless) version of $K$ by dividing by the number of particles and by $\tobs$:
	\begin{equation}
	k[\mathbf{x}]=\frac{K[\mathbf{x}]}{{N} t_{\mathrm{obs}}},
	\end{equation}

	In large systems, the gaps between adjacent particles are exponentially distributed with mean $\LL/N$.
	Hence the mean of $r_i^a$ is the probability that a randomly chosen gap ($y$) is larger than the proposed step ($z$):
	\beq
	\langle r_i^a \rangle_0 =  \int_0^a (1/a) \int_z^\infty (N/\LL) {\rm e}^{-yN/\LL} \,\mathrm{d}y\,  \mathrm{d}z
	\label{equ:mean-ri}
	\eeq
	(Here and throughout, $\langle \cdot \rangle_0$ indicates an average in the equilibrium state of the system.)
	The integral gives $\Phi_a^{-1}( 1 - {\rm e}^{-\Phi_a})$, so one has (for large systems, $N\to\infty$)
	\begin{equation}
	\langle k[\textbf{x}] \rangle_0 = \frac{1}{\Phi_a} (1-e^{-\Phi_a})
	\label{equ:ave-K}
	\end{equation}
	At low concentrations (small $\Phi_a$), particles diffuse almost independently and the activity $k$ is equal to unity.  For high concentrations the mean activity (per particle) is reduced; it approaches zero as $\Phi_a\to\infty$ (in which case particles do not move at all).

	\section{Biased Ensembles of Trajectories}\label{biased}
	
	This work focusses on the distribution of the intensive activity $k[\mathbf{x}]$ as $\tobs\to\infty$.  In a system with $N$ particles, large deviation theory for this time-averaged quantity means that its probability density scales as
	\beq
	p(k|\tobs,N) \sim {\rm e}^{-\tobs I(k) }
	\label{equ:ldp}
	\eeq
	where $I$ is the rate function.
	This is a large deviation principle, which holds for $\tobs\to\infty$ at fixed $N$.  

	Evaluation of $I(k)$ gives the probability of rare events where the time-averaged activity takes a non-typical value.  
	%To investigate these rare events, it is useful to define biased ensembles of trajectories.  
	This section outlines several results from large deviation theory as it applies to ensembles of trajectories~\cite{Derrida2007,Lecomte2007,jack_ptps,Chetrite2015}, including connections to optimal-control 
		theory~\cite{OptimalControlRepresentationChetrite,Jack2015b}, and its application for numerical sampling~\cite{Nemoto2014,nemoto_population-dynamics_2016}.
		Readers familiar with this material may prefer to skip directly to the summary of main results in Sec.~\ref{sec:main-results}.

	\subsection{Biased ensembles}
	
	We define biased 
		ensembles of trajectories according to standard methods~\cite{Derrida2007,Lecomte2007,Chetrite2015},
	by modifying the probabilities of trajectories of the system.  We use $\mathrm{d}P_0[\mathbf{x}]$ to indicate the (infinitesimal) probability that the system follows trajectory $\textbf{x}$.   The meaning of this notation is that the expectation value of some observable $O$ can be expressed as
	\beq
	\label{equ:ave-eq}
	\langle O \rangle_0 = \int O[\mathbf{x}] \, \mathrm{d}P_0[\mathbf{x}]
	\eeq
	where the integral runs over all possible trajectories, weighted by their probabilities.
	Now consider an ensemble in which
	the probability of trajectory $\textbf{x}$ is biased according to its activity:
	\begin{equation}\label{sensemble}
	\mathrm{d}P_{s}[\mathbf{x}]=\frac{e^{-sK[\mathbf{x}]}}{Z_{s}} \, \mathrm{d} P_0[\mathbf{x}] \;,
	\end{equation}	
	with $Z_s = \langle {\rm e}^{-sK[\mathbf{x}]} \rangle_0$ for normalisation.  By analogy with (\ref{equ:ave-eq}), averages in the biased ensemble are given by
	\beq
	\label{equ:ave-s}
	\langle O \rangle_s = \int O[\mathbf{x}] \,\mathrm{d}P_s[\mathbf{x}]  \; .
	\eeq
	
	Since $K$ is extensive in time, it is useful to invoke an analogy between these biased ensembles and canonical ensembles in statistical mechanics, see \cite{Chetrite2013,Touchette2009} for a discussion.  
	This motivates us to define the dynamical free energy,
	\beq\label{DFE}
	\psi(s)=\lim\limits_{\tobs \rightarrow \infty} \frac{1}{\tobs} \log \langle {\rm e}^{-sK[\mathbf{x}]}\rangle_0  \; .
	\eeq
	The average of the intensive activity in the biased ensemble is denoted by % $k(s)$, that is
	\beq
	{k}(s) = \langle k[\mathbf{x}] \rangle_s %  = \frac{-1}{N} \psi'(s) \;.
	\label{equ:k-of-s}
	\eeq
	(There should be no confusion between the mean activity $k(s)$ and the activity of an individual trajectory $k[\mathbf{x}]$.)
	The average is over trajectories of fixed length $\tobs$ so $k(s)$ depends implicitly on $\tobs$, as well as the parameters of the model. 
		Note also that $\lim_{\tobs\to\infty}k(s)=-\psi'(s)/N$, where the prime indicates a derivative.
	One reason that these biased ensembles are useful is that for large $\tobs$, typical trajectories taken from (\ref{sensemble}) are representative of the rare events 
	associated with (\ref{equ:ldp}), evaluated at $k=k(s)$~\cite{Lecomte2007,Garrahan2009,Chetrite2013,Chetrite2015}.
	
	In analogy with thermodynamics, this derivative $\psi'$ corresponds to the average value of an order parameter.  
	The free energy is related to the rate function $I$ by Legendre transform $I(k) = \sup [-sk-\psi(s)] $.
	If results for ${k}(s)$ and $\psi(s)$ are available from numerical data then the rate function may be estimated parametrically as
	\beq
	I\!\left({k}(s) \right) =   - s {k}(s) - \psi(s) \; .
	\label{equ:Ik-param}
	\eeq
	
	\subsection{Dependence of averages on time}
	\label{sec:Pave}
	From the definition of the biased ensemble in \refequa{sensemble}, it follows that this ensemble has transient regimes when the time $t$ is close to $t=0$ or $t=\tobs$.  These transients can be characterised theoretically following~\cite{Garrahan2009,jack_ptps,Chetrite2015}.  To this end, consider a general observable quantity $\hat z$ that can be measured at some single time $t$ (for example, $\hat z$ might be the distance between two particles).  The probability density for this quantity when evaluated at time $t=\tobs$ is
	\begin{equation}
	P_{s,\mathrm{end}}(z)= \left\langle \delta[z - \hat z(\tobs)] \right\rangle_s \; .
	\label{equ:pend}
	\end{equation}
	The probability density for $\hat z$ can also be averaged along the whole trajectory, which gives
	\begin{equation}
	P_{s,\mathrm{ave}}(z)= \frac{1}{\tobs} \int_0^{\tobs} \left \langle \delta [z - \hat z(t)] \right\rangle_s \mathrm{d}t \; .
	\label{equ:pave}
	\end{equation}
	This is the probability that $\hat z$ has value $z$, if we measure at a time $t$ chosen uniformly at random from $[0,\tobs]$.  These distributions depend implicitly on $\tobs$; their limits are well-defined as $\tobs\to\infty$.  The two distributions $P_{s,\mathrm{ave}}$, $P_{s,\mathrm{end}}$ are different in general; in particular, they have different limits as $\tobs\to\infty$ because $P_{s,\mathrm{end}}(z)$ characterises the transient regime while $P_{s,\mathrm{ave}}(z)$ characterises typical times, away from the transient regimes.
	
	\subsection{Conditioning of Doob, guiding forces, and optimal control theory}
	\label{sec:doob}
	
	It has been shown in recent years \cite{jack_ptps,Chetrite2015} that properties of biased ensembles of the form (\ref{sensemble}) can be reproduced by considering the typical (unbiased) dynamics of an ``auxiliary process'' that has been modified to include additional ``control forces''.  
	For the Langevin process (\ref{langevin}), the auxiliary process is 
	\beq \label{langevin-control}
	\dot{x}_i = -D_0 \nabla_i (\beta U+V_{\rm opt}) + \sqrt{2D_0} \eta_i
	\eeq
	where $V_{\rm opt}$ is an optimal control potential whose determination is discussed below.
	%It is called an optimal potential because the resulting model dynamics minimises the relative entropy between (\ref{sensemble}).
	
	In the following, we make extensive use of (non-optimal) control forces, to improve convergence of our numerical algorithms, following~\cite{Nemoto2014,nemoto_population-dynamics_2016,nemoto_finite-size_2017,ray_exact_2018}.  We now present the associated theory.
	%\subsubsection{Path measures with guiding forces}  
	%We give a brief summary of the use of guiding forces (or control forces) to transform between biased ensembles of trajectories.  
	Some details of derivations are given in~\ref{app:bias}.  
	
	The object of primary interest in this study is the biased probability distribution $P_s$ of (\ref{sensemble}).
		It is convenient to define a new biased distribution that is very close to $P_s$, but differs in the transient regimes close to $t=0$ and $t=\tobs$.
		Let $V$ be a control potential, similar to $V_{\rm opt}$ in (\ref{langevin-control}), but not necessarily optimal.
		Then define
		\begin{equation}
		\mathrm{d} \tilde{P}^V_s[\mathbf{x}] \propto \exp\left( \frac{1}{2} \Big[ V(X(0)) - V(X(\tobs))\Big] \right) \, \mathrm{d} {P}_s[\mathbf{x}] \;.
		\label{equ:PtVs-def}
		\end{equation}	
		The
		constant of proportionality in this equation is fixed by normalisation, we do not write it explicitly in order to have a compact notation.
		%The ensemble $\tilde{P}^V_s$ differs from the biased ensemble $P_s$ only in the transient regimes that were discussed in Sec.~\ref{sec:Pave}.  
		Since they differ only in transient regime, the distributions $P_s$ and $\tilde{P}^V_s$ are equivalent for long trajectories, in the sense that they yield the same results for $k(s)$ and $\psi(s)$.  Marginal distributions $P_{\rm ave}$ are also identical for $P_s$ and $\tilde{P}^V_s$ (as $\tobs\to\infty$), but $P_{\rm end}$ is different in general.\footnote{%
				One way to characterise the optimal controlled model is that it leads to $P_{\rm ave}=P_{\rm end}$~\cite{nemoto_population-dynamics_2016}, so there is no transient regime in that case.}%
	%This means that $V$ can be chosen at will, in order to facilitate numerical or analytical computations.
	
	The equivalence of  $P_s$ and $\tilde{P}^V_s$  means that we are free to choose $V$ in such a way that $\tilde{P}^V_s$ is easy to analyse, either numerically or theoretically.
		To this end, let $\tilde{P}^V[\mathbf{x}]$ be the probability of trajectory $\textbf{x}$ under the Langevin dynamics (\ref{langevin-control}), with $V_{\rm opt}$ replaced by $V$.
	Then, we show in \ref{app:bias} that $\tilde{P}^V_s$ can be interpreted as a biased ensemble for this controlled process, that is
	\begin{equation}
	\mathrm{d} \tilde{P}^V_s[\mathbf{x}] \propto \exp\left(  {\cal A}^{\rm{sym}}[\mathbf{x}] - s K[\mathbf{x}]  \right) \, \mathrm{d} \tilde{P}^V[\mathbf{x}] \;,
	\label{equ:PtVs}
	\end{equation}
	with 
	\beq
	{\cal A}^{\rm{sym}}[\textbf{x}]  = 
	\frac14 \sum_i \int_0^\tobs   \nabla_i V \cdot D_0 ( \nabla_i V + 2\beta \nabla_i U) - 2 D_0 \nabla_i^2 V  \, {\rm d} t \; .
	\label{equ:def-Asym}
	\eeq
	Similar results have been derived in~\cite{OptimalControlRepresentationChetrite,nemoto_population-dynamics_2016,nemoto_finite-size_2017,MPSGarrahan,ray_exact_2018}.  
	Note that (\ref{equ:PtVs}) applies for the Langevin model (\ref{langevin}), the analogous result for MC dynamics is given in~\ref{app:bias}.

	Comparing (\ref{equ:PtVs}) with (\ref{sensemble}), one sees that ${\rm d}P_0$ has been replaced by ${\rm d}\tilde{P}^V$, which indicates that the model has been modified by including the control potential $V$ in the equation of motion.  Also the exponential biasing factor in  (\ref{sensemble}) has been modified to include the action ${\cal A}^{\rm sym}$.  In numerical work, these modifications are simple to implement, so algorithms for analysing $P_s$ can also be used to analyse $\tilde{P}^V_s$.   This holds for any $V$ which allows an enormous flexibility~\cite{nemoto_population-dynamics_2016,nemoto_finite-size_2017,ray_exact_2018}.  
		%The meaning of (\ref{equ:PtVs}) is that ensemble $\tilde{P}^V_s$ (which we recall is equivalent to $P_s$) can be analysed by biasing $\tilde{P}^V$ with a factor ${\rm e}^{{\cal A}^{\rm sym}-sK}$.  This holds for any $V$ which allows an enormous flexibility~\cite{nemoto_population-dynamics_2016,nemoto_finite-size_2017,MPSGarrahan,ray_exact_2018}, particularly in numerical studies (see below).
		In particular,
	the optimal control ($V=V_{\rm opt}$) is the potential $V$ for which the factor $ {\cal A}^{\rm{sym}}[\mathbf{x}] - s K[\mathbf{x}]$ in (\ref{equ:PtVs}) evaluates to a constant value $\psi(s)\tobs$, independent of $\textbf{x}$.  
	This allows $V_{\rm opt}$ to be obtained by solving an eigenvalue problem, see~\ref{app:bias}.  If this optimal control potential is known then sampling from $\tilde{P}^V_s$ is trivial~\cite{Nemoto2014,Hartmann2012}.  More commonly, the optimal potential is not available, but even non-optimal controls can greatly improve the performance of numerical schemes~\cite{nemoto_population-dynamics_2016,nemoto_finite-size_2017,ray_exact_2018}.
	The nature of optimal control forces in some physical model systems is discussed in~\cite{Jack2015b}.
	%

	%\subsubsection{Effects of guiding forces}
	%\label{sec:control-effects}
	%
	%Equ.(\ref{equ:PtVs}) 
	%%(\ref{equ:MCPtV}) are 
	%is useful for numerical studies: instead of sampling  from $P_s$, one may alternatively choose some $V$ and sample from $\tilde P^V_s$.   This is a form of importance sampling.  For suitable choices of $V$, the numerical sampling may be easier.  In particular, if $V$ is chosen to be the optimal control then the exponential factor in (\ref{equ:PtVs}) is constant, so that sampling is trivial and simply involves generating representative trajectories of (\ref{langevin-control}).
	%
	%We emphasise that the distribution $\tilde{P}^V_s$ leads to the same results as $P_s$ when considering distributions of the form (\ref{equ:pave}), and also for statistics of time-averaged quantities including the activity $k[\mathbf{x}]$. However, $\tilde{P}^V_s$ differs from $P_s$ in transient regimes.  This means that the distribution $P_{\rm end}$ of (\ref{equ:pend}) depends on the control potential $V$.  An interesting case is when $V=V_{\rm opt}$, in which case $P_{s,\rm end} = P_{s,\rm ave}$.  This observation was used in \cite{nemoto_population-dynamics_2016} to infer suitable choices for $V$.
	
	\subsection{Sampling of biased path ensembles}
	
	%\emph{RLJ needs some rewriting thru here}
	
	%Older works have investigated different ways of improving the sampling of biased ensembles . 
	
	We use transition path sampling (TPS)~\cite{Bolhuis} to generate representative trajectories from $P_s$ and ${\tilde P}^V_s$.  
	%(An alternative method based on a cloning algorithm has also been widely used~\cite{Giardina2011,Lecomte_2007,DirectEval,ray_exact_2018}.  We use TPS in this work since its convergence properties and the associated numerical errors are easier to estimate.)
	Our TPS methodology is the same as \cite{thompson_dynamical_2015,Hedges2009}.
	To summarise, TPS is an MC method for sampling trajectories of a fixed length $\tobs$.  In each step, one starts with a trajectory and proposes to change it in some way.  This is called a TPS move.  The proposal is generated by direct simulation of the model of interest (see below).  The proposed trajectory is accepted or rejected according to a Metropolis criterion based on the relevant weighting factor [for example, ${\rm e}^{-sK}$ in the case of (\ref{sensemble})].  This MC method is designed to obey detailed balance, which ensures that it samples (\ref{sensemble}) or (\ref{equ:PtVs}), as required.   This is a key strength of the method; another advantage is that standard MC tests for numerical convergence can be applied, see for example Sec.~\ref{sec:mft-improvement}.

		There is some flexibility as to the specific choice of TPS moves.
		In this work,
	we use shifting moves~\cite{Bolhuis,thompson_dynamical_2015,Hedges2009} and the size of each shift is chosen uniformly from the range
	%The size of the shifting moves we use usually average around 
	$\tau_{\mathrm{B}}\pm0.5\tau_{\mathrm{B}}$, except where stated otherwise.  As usual in TPS,  proposing larger shifts is desirable for rapid exploration of trajectory space, but tends to lead to more TPS moves being rejected.  The best choice of shift size is a compromise between these two effects.
	%We have experimented with other sizes as well but found that if we shift too much we end up rejecting too many moves and wasting computational effort with generating long trajectories with low acceptance rates. On the other hand going too low means that the middle of the trajectory will change only very slowly ($\sqrt{m}\sim\tobs/\tau_{\rm{shift}}$).
	
We note that population dynamics (cloning) methods~\cite{Giardina2011,Lecomte_2007,DirectEval} have also been widely used for numerical studies of large deviations, and guiding (control) forces have also been used in that case~\cite{nemoto_population-dynamics_2016,nemoto_finite-size_2017,ray_exact_2018}.
		We comment on the strengths and weaknesses of the two approaches at the end of this work, in Sec.~\ref{sec:conc}.
	
	\begin{figure}
		\centering
		\includegraphics[width=1\linewidth]{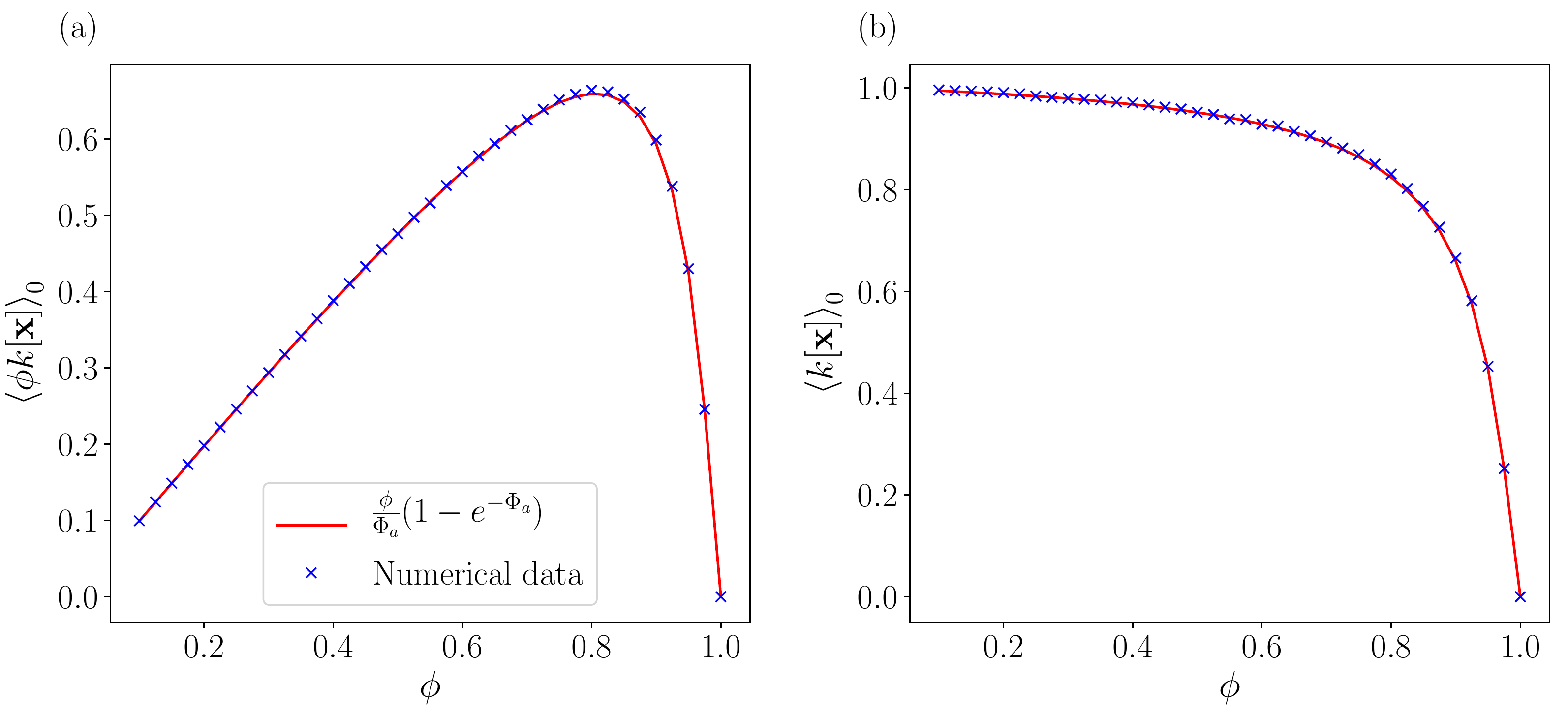}
		\caption{Average dynamical activity (per unit time), evaluated at equilibrium for $a=0.1l_0$. (a) Mean activity per unit length $\langle \phi k \rangle_0$ as a function of volume fraction $\phi$ for a system of size $L=40l_0$, compared with the theoretical prediction of (\ref{equ:ave-K}).
			(b) Corresponding activity, per particle. 
			%The dependence of of the biased observable $K$ per length (a) and per particle (b) respectively on $\phi=\frac{Nl_0}{L}$ for systems size $L=40l_0$. We compare them with the theoretical prediction for the $K$ of a homogeneous system with packing fraction $\phi$. The activity per length peaks at $\phi\approx0.8$, meaning that to lower activity the system splits into low and high density regions. 
		}
		\label{fig:ActDistro}
	\end{figure}

	%\section{Results and analysis}
	%
	%\rlj{Probably the following subsections should become sections, as they were before}
	
	\section{Overview of main results}
	\label{sec:main-results}

	We consider fluctuations of the dynamical activity $K$ in the BHPM, as defined in Sec.~\ref{sec:activity}.  
		All numerical results are obtained using the MC variant of the model.
		The behaviour of the mean activity
		is illustrated in Fig.~\ref{fig:ActDistro}.  
		We represent the data in two different ways.
		Fig.~\ref{fig:ActDistro}(a) shows
		$\langle K / (L\tobs) \rangle_0=\langle k\phi\rangle_0$, which is
		the average activity per unit length, as a function of the volume fraction $\phi=Nl_0/L$.  
		%This is the natural measure of the activity in the original (unrescaled) units.  
		%For small $\phi$, it increases proportional to the volume fraction (at fixed system size, adding more particles increases the activity).  For large $\phi$, the fraction of accepted moves goes down and the activity is reduced.
		Fig.~\ref{fig:ActDistro}(b) shows the activity per particle $\langle k \rangle$, and its dependence on $\phi$. 
		Note that the activity of a typical particle $\langle k\rangle$ decreases with volume fraction, but the activity per unit length is non-monotonic.
		(At small volume fractions, the activity is proportional to the number of particles and hence to $\phi$; on the other hand, it decreases for large volume fractions, because particles start to obstruct each other.)%
	
	\begin{figure}
		\centering
		\includegraphics[width=1\linewidth]{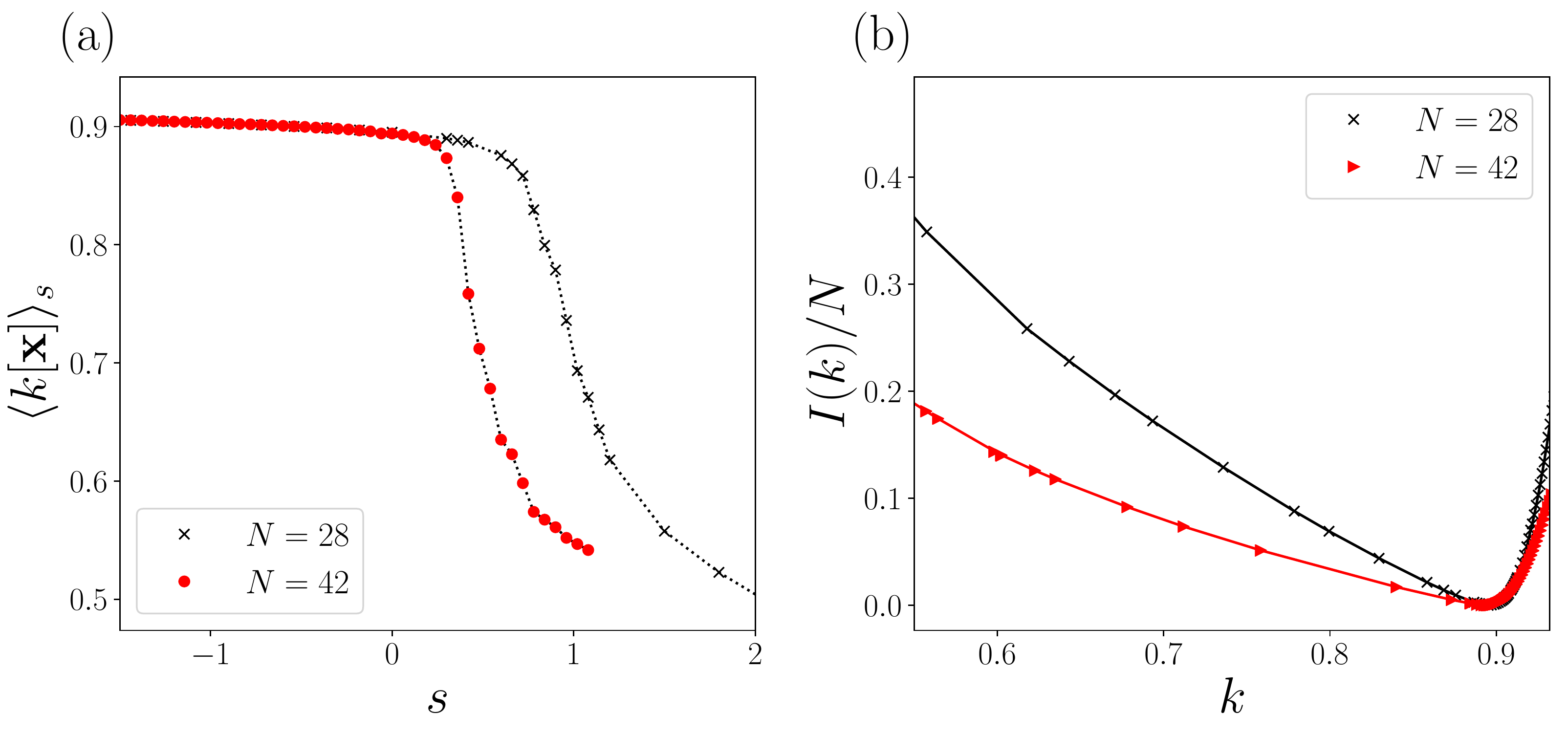}
		\caption{(a)~Average activity $\langle k[\textbf{x}]\rangle_{s}$ in the biased ensemble, obtained by TPS in a system with $N=(28,42)$ and $t_{\mathrm{obs}}=(10\tau_{\mathrm{B}},15.6\tau_{\mathrm{B}})$.  (b)~Corresponding estimate of the rate function, using (\ref{equ:Ik-param}).
		These results are analysed in more detail below, including a discussion of numerical uncertainties (error bars), see for example Fig.~\ref{fig:ScalingWrtN2tS1}.}
		\label{fig:ExtensionK}
	\end{figure}

	We now consider large deviations of $K$.
	Fig.~\ref{fig:ExtensionK} shows the behaviour of $k(s)$ and the corresponding estimate of the rate function [obtained by (\ref{equ:Ik-param})], for a representative state point $\phi=0.7$ in systems of $N=28$ and $N=42$ particles.  
	These results were obtained by TPS, we note that they depend on the trajectory length $\tobs$ which is quoted in units of the Brownian time,
		\beq
		\tau_{\rm B} =  l_0^2 /  (2D_0) \; .
		\eeq
		This is a natural unit of time in the model, and is comparable to the time required for a particle to diffuse its own size, $l_0$.
	
\begin{figure*}%x [b]
		\includegraphics[width=1\linewidth]{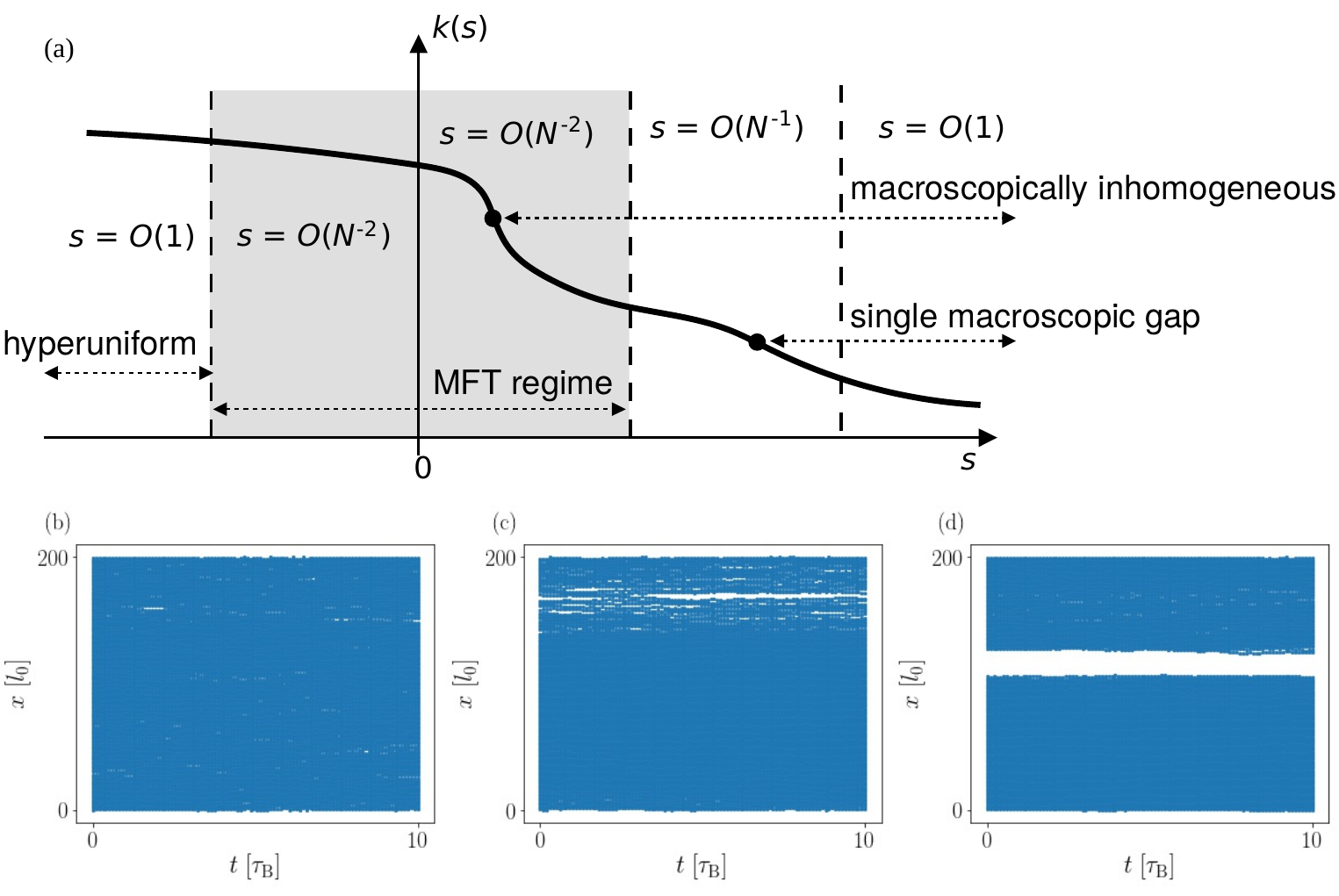}
		\\
		\caption{ (a) Sketch of the activity $k(s)$ as a function of the bias.  We concentrate in this work on three physical regimes:  (i)~homogeneous; (ii)~macroscopically inhomogeneous; and (iii) a system with a single macroscopic gap.  See the text for a discussion. 
			(b,c,d)~Representative trajectories of the system at $s=0, 0.18, 0.36$ respectively.  These trajectories illustrate the characteristics of the three regimes.  We take $N=160$ and $\tobs=100\tau_{\rm B}$, we show the behaviour for $0<t<10\tau_{\rm B}$ which is representative of the whole trajectory in these cases.
			%	(Bottom)The visual difference between a typical trajectory of the non-phase separated system ($s=0$) (a), the system in the regime with highly varying densities ($s=0.18\approx\mathcal{O}(N^{-2})$)(b) and in the regime with a macroscopic gap ($s=0.36\approx\mathcal{O}(N^{-1})$) (c). In all cases the packing fraction of the non-reduced system is $\phi=\frac{Nl_0}{L}=0.8$ and the number of particles $N=160$. 
		}
		\label{fig:PSState}
	\end{figure*}

	As noted in Section~\ref{biased},  $k(s)$ is analogous to an order parameter in thermodynamics.  This quantity decreases sharply for positive $s$.  
	As explained in~\cite{jack_hyperuniformity_2015,thompson_dynamical_2015}, this is a signature of a dynamical phase transition in the BHPM, which occurs in the limit $N,\tobs\to\infty$.
	Before embarking on a detailed analysis, we give a brief summary of the associated phenomena.
	The qualitative behaviour of $k(s)$ in a system with finite $N$ is shown in Fig.~\ref{fig:PSState}, which also shows typical trajectories of the system, as one passes through the phase transition.  At the phase transition, the system becomes inhomogeneous~\cite{jack_hyperuniformity_2015,thompson_dynamical_2015}.  In this work, we emphasise that (for this model) there are two distinct classes of inhomogeneous state.  There are states where the density is modulated in space, but particle spacings remain of order unity as $N\to\infty$.   However, for larger $s$ (smaller $k[\textbf{x}]$), there are states where a significant fraction of the available space in system is taken up by a single interparticle gap.   The two classes of inhomogeneous state are discussed in Secs.~\ref{sec:results-mft} and~\ref{sec:macro-gap}.
	
	In Sec.~\ref{sec:results-mft} we review and extend some previous 
	work~\cite{Bertini2005,Bertini2006,bodineau_distribution_2005,appert-rolland_universal_2008,lecomte_inactive_2012,jack_hyperuniformity_2015}, which shows that inhomogeneous states with spacings of order unity appear on taking $N\to\infty$ with $s = {O}(N^{-2})$.  This is the regime described by macroscopic fluctuation theory (which can also describe the behaviour for small negative values of the bias).  
	In this regime, the optimal control forces are long-ranged; they are attractive for $s>0$ and repulsive for $s<0$.  It is the attractive forces that drive the phase separation transition.
	%We show that the optimal control forces in the macroscopically homogeneous part of this involve interactions between particles that are long-ranged.  These forces are   
	We show that using control forces in numerical sampling significantly improves their efficiency.
	For $s<0$ the system always remains homogeneous; as explained in~\cite{jack_hyperuniformity_2015} it is hyperuniform when $s$ is negative and of order unity, see also Sec.~\ref{sec:hyperU}.
	%The attractive forces drive the formation of macroscopic inhomogeneities which appear above a critical value of $sN^2$.  
	%(We have not derived the optimal control forces in the inhomogeneous regime.)
	
	In Sec.~\ref{sec:macro-gap}, we discuss the behaviour on taking $N\to\infty$ with $s={O}(N^{-1}$).  We explain that this is the regime in which we expect a macroscopic gap to take up a finite fraction of the system.   
	%We predict properties of this gap size as a function of the bias $s$ and we show that near-optimal control forces can be derived by assuming that this potential depends only on the size of the largest gap.  
	By applying such control forces in numerics, we show that computational efficiency is significantly improved.  In fact, this improvement is much larger than for the MFT regime.  We discuss how parameters of the control force can be optimised for efficient sampling.
	
	We note that all these results apply in limits where $s\to0$ as $N\to\infty$.  The tractability of these limits arises because the biases that are applied to these ensembles of trajectories are weak.  For example, a central assumption of MFT~\cite{bertini_macroscopic} is that the system is in ``local equilibrium'', which means that any finite region of the system can be characterised through its local density and current, and that the bias has a negligible effect on the short-ranged correlations between microscopic particles.  In the regime with a macroscopic gap, there are deviations from local equilibrium but our central assumption is that these are localised near the interfaces, at the edges of the gap.
	%Physically, this means that most degrees of freedom in the model are somehow ``close to equilibrium''.  This simplifying observation enables the theoretical analyses.
	
	%Compared to these behaviours that operate on large scales, the behaviour for $s=\mathcal{O}(1)$ is harder to analyse in detail -- all kinds of interparticle correlation can occur.  We briefly discuss (in SEC) the hyperuniform behaviour that occurs for $s<0$ with $|s|=\mathcal{O}(1)$, see \cite{jack_hyperuniformity_2015}.
	
	\newcommand{\gammobs}{\gamma_{\rm obs}}
	
	\section{Diffusion governed (MFT) regime}
	\label{sec:results-mft}
	
	This section discusses the regime where MFT applies~\cite{bertini_macroscopic}.   
	We work in the rescaled representation of Sec.~\ref{sec:rescaled}.
	MFT is valid on large (hydrodynamic) length and time scales, which are related by a diffusive scaling.  That is, we introduce the (average) density $\overline{\rho}$ and define
	%Specifically, we take $N\to\infty$ with a diffusive scaling
	\beq
	L_r=N/\overline{\rho}, \qquad \tobs = \gammobs  \frac{\LL^2}{2D_0} .
	\label{equ:mft-L-tobs}
	\eeq 
	The hydrodynamic limit is $N\to\infty$ at fixed $\overline{\rho},\gammobs$.  One then takes a second limit, $\gammobs\to\infty$, in order to access the relevant large deviations.
	%where $\gamma$ is a constant of order unity.
	%Within MFT, one performs a rescaling of space and time, to arrive at a theory that is valid on 
	%large (hydrodynamic) length and time scales.  As the number of particles $N$ is increased, we scale $L_r\propto N$ and $\tobs \propto N^2$.  
	To arrive at a consistent theory, we also rescale the biasing parameter~\cite{appert-rolland_universal_2008} as
	%We work in the rescaled representation of Sec.~\ref{sec:rescaled} in which the natural (dimensionless) rescaled bias is 
	\beq
	\lambda =  s \LL^2 / D_0
	\label{equ:def-lambda}
	\eeq
	which is held constant as $N\to\infty$.
	%As we increase the system size, we keep $\lambda$ constant, so $s$ is reduced.  The effect is that we zoom-in on the jump-like feature in Fig.~\ref{fig:ExtensionK}(a).  

	\subsection{Theoretical analysis of density fluctuations and optimal control potential, using MFT}
	\label{sec:mft-theory}
	
	We consider the statistics of  
	the local density and current, which are the relevant hydrodynamic fields within MFT~\cite{bertini_macroscopic}. We analyse large deviations of the activity using a physical argument based on fluctuating hydrodynamics -- the same conclusions can also be reached other methods~\cite{appert-rolland_universal_2008}, the specific case of the BHPM is discussed in~\cite{jack_hyperuniformity_2015}.
	%We work in the rescaled representation of Sec.~\ref{sec:rescaled}.
	The statistics of the density and current may be characterised by writing Langevin equations:
	\begin{eqnarray}
	\dot \rho & = & - \mathrm{div}\, j \nonumber \\
	j &=& -D(\rho) \nabla\rho + \sqrt{2\sigma(\rho)} \eta
	\label{equ:rho-j-lang}
	\end{eqnarray}
	where $D(\rho)$ and $\sigma(\rho)$ are a diffusivity and a mobility, and $\eta$ is a space-time white noise with mean zero and $\langle \eta(x,t) \eta(x',t')\rangle = \delta(x-x') \delta(t-t')$. The specific forms of $D$ and $\sigma$ for the BHPM are discussed below.  We use Ito calculus.  
	
	The usual approach in MFT is to rescale the spatial domain $[0,\LL]$ into the unit interval $[0,1]$ and also to rescale time.  
	That is, define dimensionless coordinates on the hydrodynamic scale as $\hat{x}=x/\LL$ and $\hat{t}=t/\LL^2$,  the corresponding current is $\hat{\jmath}=j\LL$ (there is no rescaling of the density).
		Then (\ref{equ:rho-j-lang}) becomes
		\begin{eqnarray}
		(\partial/\partial \hat{t}) \rho & = & - \hat{\nabla} \cdot \hat{\jmath} \nonumber \\
		\hat{\jmath} &=& -D(\rho) \hat\nabla\rho + \sqrt{2\sigma(\rho)/\LL} \,\hat\eta
		\label{equ:rho-j-lang-hat}
		\end{eqnarray}
		where $\hat\nabla$ is a gradient with respect to $\hat x$, and $\hat\eta$ is a noise with zero mean and $\langle \hat{\eta}(\hat{x},\hat{t}) \hat{\eta}(\hat{x}',\hat{t}')\rangle = \delta(\hat{x}-\hat{x}') \delta(\hat{t}-\hat{t}')$. (Hence one sees that $\hat\eta = \LL^{3/2}\eta$.) It is apparent from (\ref{equ:rho-j-lang-hat})	that MFT is a weak-noise theory that is valid on large length scales.  For later comparison with numerics, it is convenient to quote all results in the original co-ordinates (without hats), but we emphasise that they are valid only on the hydrodynamic scale (which in this case will mean $s={O}(N^{-2})$, as usual in diffusive systems~\cite{appert-rolland_universal_2008}).   
	Within MFT it is consistent (by the local equilibrium assumption~\cite{bertini_macroscopic}) to approximate the activity $K$ from~(\ref{equ:def-K}) as% can be obtained as
	\beq
	K[\mathbf{x}] = \int_0^{\tobs} \int_0^{\LL} \kappa(\rho(x,t)) \, \mathrm{d}x  \mathrm{d}t
	\label{equ:K-rho}
	\eeq
	where $\kappa(\rho)$ is the average activity (per unit volume) of an equilibrium system with density $\rho$.  That is, $\kappa(\rho) = \frac{1}{\LL\tobs} \langle K \rangle_{0,\rho}$; for the BHPM then (\ref{equ:ave-K}) and $\rho=N/\LL$ imply that 
	\beq
	\kappa(\rho) = \frac{1}{a} (1-{\rm e}^{-a\rho}) \; .
	\label{equ:kappa}
	\eeq
	For activity fluctuations in the SSEP, one takes instead $\kappa=2\rho(1-\rho)$ in which case the following theoretical analysis is similar to~\cite{appert-rolland_universal_2008,lecomte_inactive_2012}.
	%The behaviour of this function in the specific case of interest here was shown in Fig.~\ref{fig:ActDistro}.  
	For the purposes of this discussion, the important feature is that $\kappa''(\rho)<0$.
	For positive $s$, this means that the system undergoes a continuous phase transition accompanied by spontaneous symmetry 
		breaking~\cite{bodineau_distribution_2005,appert-rolland_universal_2008,jack_hyperuniformity_2015}, see also~\cite{YongJooDynamicalSymmetryBreaking,Imparato2009,Shpiel2018}.
		For $s=0$ the system is homogeneous but for positive $s$, it becomes inhomogeneous, see Fig.~\ref{fig:PSState}.
		We now adapt previous MFT results to this setting before comparing with numerical results in Sec.~\ref{sec:mft-numeric}.

	\subsubsection{Homogeneous phase}
	\label{sec:mft-homog}
	%Following [cite] we anticipate that the noise has a weak effect on hydrodynamic length and time scales, so we expand the density about the homogeneous state as 
	As in~\cite{bodineau_distribution_2005,appert-rolland_universal_2008,Bodineau2008,jack_hyperuniformity_2015}, we analyse the homogeneous phase by writing $\rho(x,t) = \overline{\rho} + \delta\rho(x,t)$, and
	assuming that $\delta\rho$ is small.
	%Specifically, we follow ~\cite{jack_hyperuniformity_2015}. 
	%The inhomogeneous phase is analysed below, see also~\cite{lecomte_inactive_2012}.}
	%This approximation is valid in the .
	%, so $\lambda$ is not too large, and the system remains homogeneous.  
	From (\ref{equ:rho-j-lang}) we have $(\partial/\partial t) {\delta\rho}  =  - \mathrm{div}\, j$ with (at leading order in $\delta\rho$):
	\beq
	j = -D(\overline\rho) \nabla(\delta\rho) + 
	\sqrt{2\sigma(\overline\rho)} \eta \; .
	\eeq
	From (\ref{equ:K-rho}) then
	\beq
	K[\mathbf{x}]  =  \LL\tobs \kappa(\overline{\rho}) + \frac12 \int_0^{\tobs} \! \int_0^\LL \kappa''(\overline\rho) \delta\rho(x,t)^2 \, \mathrm{d}x  \mathrm{d}t \; .
	\eeq
	Fourier transforming as
	\beq
	\tilde\rho_q(t) = {\LL^{-1/2}} \int_{0}^{\LL}  \rho(x,t) \exp\left( - {\rm i}qx\right)dx
	\label{equ:fourier}
	\eeq
	%as in (\ref{equ:fourier}) 
	one has (for $q>0$)
	\beq
	(\partial/\partial t) \tilde {\rho}_q  =  - D(\overline{\rho}) q^2 \tilde \rho_q  
	+ q \sqrt{2\sigma(\overline\rho)} \tilde\eta_q
	\label{equ:dot-rhoq}
	\eeq
	where $\tilde\eta_q$ is a complex-valued Brownian noise.  [There is one noise for each positive wavevector, each noise is independent of all the others, 
	and $\langle \tilde\eta_q(t) \tilde\eta_{q}^*(t)\rangle = \delta(t-t')$.]
	Also,
	\beq
	K[\textbf{x}]  =  \LL\tobs \kappa(\overline{\rho}) + 
	\sum_{q> 0} \int_0^{\tobs}  \kappa''(\overline\rho)  \tilde \rho_q(t) \tilde \rho_{-q}(t) \,  \mathrm{d}t \; .
	\label{equ:K-rhoq}
	\eeq
	In (\ref{equ:dot-rhoq},\ref{equ:K-rhoq}), the different wavevectors are completely decoupled from each other.  The result is that each Fourier component of the density evolves independently by a (complex-valued) Ornstein-Uhlenbeck (OU) process with a bias proportional to $|\rho_{q}|^2$.  
	Biased ensembles for these OU processes can be analysed exactly by standard methods, see \ref{sec:ou} for details.   
	Using (\ref{equ:psi-ou}) with $\alpha = \kappa''(\overline\rho)$ %\rlj{/2}$ 
	and $\omega = D(\overline\rho) q^2$ and $\gamma=\sigma(\overline{\rho})q^2$, the result is that 
	\beq
	\fl\qquad
	\psi(s) =  -s\LL\kappa(\overline{\rho})  + \sum_{q>0} \left(
	D(\overline\rho)q^2 - \sqrt{ D(\overline\rho)^2q^4 + {2} s \kappa''(\overline\rho)\sigma(\overline\rho)q^2 } \right)
	\label{equ:psi-univ}
	\eeq
	which is equivalent to the results obtained in \cite{appert-rolland_universal_2008,jack_hyperuniformity_2015}. 
	The corresponding result for the activity is 
		\beq
		\fl\qquad
		k(s) = -N^{-1} \psi'(s) = \frac{\kappa(\overline{\rho})}{\overline{\rho}}  + 
		\frac{1}{N} \sum_{q>0} \frac{ \kappa''(\overline\rho)\sigma(\overline\rho)q^2  }{ [D(\overline\rho)^2q^4 + {2} s \kappa''(\overline\rho)\sigma(\overline\rho)q^2 ]^{1/2} }
		\label{equ:act-univ}
		\eeq
		which will be compared in Sec~\ref{sec:mft-numeric} with numerical data.  
		
		Note however that (\ref{equ:psi-univ})
	is
	valid only if the argument of the square root is positive which requires $2s \kappa''(\overline\rho) \sigma(\overline\rho) q^2  > -(Dq^2)^2$; otherwise the OU process predicts a divergence in the density fluctuations which signals a breakdown of the quadratic expansion in $\delta\rho$.  Recalling that $\kappa''<0$, this criterion is most stringent for the smallest wavevector $q=q_1=2\pi/\LL$ so one sees that the range of validity is $s<s_c$ with
	\beq
	s_{\rm c}   \LL^2 = -\frac{2\pi^2 D(\overline\rho)^2}{ \kappa''(\overline\rho) \sigma(\overline\rho)}
	\label{equ:sc-mft}
	\eeq
	as in \cite{jack_hyperuniformity_2015,appert-rolland_universal_2008}.  
	%Hence we identify $\lambda_c = -\frac{2\pi^2 D(\overline\rho)}{ \kappa''(\overline\rho) \sigma(\overline\rho)}$.  For $\lambda<\lambda_c$,
	For $s<s_c$, the sum in (\ref{equ:act-univ}) converges to a finite value so  $k(s)\to\kappa(\overline{\rho})/\overline{\rho}$ as $N\to\infty$, while the sum gives a finite-size correction that has a universal form within MFT~\cite{appert-rolland_universal_2008}.
	In this regime, the optimal-control potential required to generate typical trajectories of the biased ensemble is obtained from Equ.~(\ref{equ:Vopt-ou}) as
	\beq\label{equ:MFTBIAS}
	V[\rho] = \sum_{q>0}  \tilde v_q\tilde \rho_q^* \tilde\rho_{q}  
	\eeq
	with 
	\beq
	\tilde v_q = \frac{1}{D(\overline\rho)q} \left( \sqrt{ D(\overline\rho)^2q^2 + 2 s \kappa''(\overline\rho)\sigma(\overline\rho) } - D(\overline\rho)q \right) \; .
	\label{equ:tilde-vq}
	\eeq
	This corresponds to a pairwise-additive interaction whose pair potential $v(x)$ is given by the inverse Fourier transform of $v_q$.  If $s\neq0$ then $v_q$ diverges as $q\to0$ indicating that this interaction is long-ranged.  As discussed in~\cite{Popkov2010,Lazarescu2015}, the pair potential decays as $v(x) \sim 1/(\log x)$ for separations $x$ that are large compared to the particle spacing (but small compared to $\LL$).  The potential is attractive if $s\kappa''<0$ and repulsive if $s\kappa''>0$.  We again emphasise that this analysis requires a weak bias $s<s_c$.
	
	%Since the first Fourier mode is the most altered we cut off the higher ones in the optimal control force we use. In figure \ref{fig::MFTImprovement} we plot $\sigma^2_{\rm{TPS}}$ for both the unaltered dynamics and those biased by \ref{equ:MFTBIAS}. At positive $s$ we observe an improvement in sampling but at negative $s$ the improvement is negligible. 
	
		\subsubsection{Inhomogeneous phase}
		\label{sec:mft-inhom}
		For $s>s_c$, a slightly different approach is required, which is related to the Landau-like theory of~\cite{YongJooDynamicalSymmetryBreaking,Baek2018}
		as well as earlier work~\cite{bodineau_distribution_2005,lecomte_inactive_2012}.  For simplicity, we assume in this calculation that $D=D_0$ is a constant (independent of $\rho$): this situation holds for both the BHPM and the SSEP.  The generalisation to density-dependent $D$ is straightforward.
		
		We write the probability distribution (\ref{sensemble}) for the biased ensemble as
		\beq
		\mathrm{d}P_s[\mathbf{x}] \propto \exp( -S[\mathbf{x}] ) 
		\eeq
		where the action $S$ can be obtained from (\ref{equ:rho-j-lang}) or directly from MFT~\cite{bertini_macroscopic} as
		\beq
		S = \int_0^{\tobs} \int_0^{\LL} \left[ \frac{(j+D_0\nabla\rho)^2}{4\sigma(\rho)} + s\kappa(\rho) \right] \, \mathrm{d}x \, \mathrm{d}t
		\eeq
		where the integrand depends on $(x,t)$ through $\rho=\rho(x,t)$.
		We consider the system close to the transition and we derive a result analogous to Sec~3.2 of~\cite{lecomte_inactive_2012}.  Our method is slightly different from that work; note also that~\cite{lecomte_inactive_2012} considers specifically the case of the SSEP, where $\kappa=\sigma$ is a quadratic function of $\rho$.  The following calculation can be interpreted as the derivation of a Landau theory for a suitable order parameter: similar calculations for systems with open boundaries are considered in~\cite{YongJooDynamicalSymmetryBreaking,Baek2018,Baek2019-arxiv}.  
		
		We find the path that minimises the action $S$.  The minimum has $j=0$ and the associated density $\rho$ is independent of time, as one might expect since the biased ensemble is time-reversal symmetric.  Hence 
		\beq
		S=\frac{\gammobs}{2}  \int_0^\LL \left[ \frac{D_0\LL^2|\nabla\rho|^2}{4\sigma(\rho)} + \lambda\kappa(\rho) \right] \mathrm{d}x
		\label{equ:S-stationary}
		\eeq 
		where now $\rho=\rho(x)$ and we used (\ref{equ:mft-L-tobs}).  
		This action functional can be minimised numerically over the profile $\rho$, in this work we make an expansion~\cite{bodineau_distribution_2005,lecomte_inactive_2012} that is valid close to the critical point and allows an analytic treatment.
		In this regime one can capture the behaviour of the inhomogeneous phase by considering a density profile 
		\beq
		\rho(x) = \overline{\rho} + A \cos q_1x + B \cos 2q_1x
		\label{equ:rhoAB}
		\eeq
		where $A,B$ are variational parameters. 
		As noted above, the instability of the system originates in the smallest wavevector $q_1$ but it is necessary~\cite{lecomte_inactive_2012} to consider also the second-smallest wavector $2q_1$ to obtain accurate results for the inhomogeneous phase, even at leading order.  We take $A>0$ without loss of generality; the system has translational symmetry so one may equivalently replace $x\to x+\delta$ to obtain an equivalent profile, shifted in space.  
		
		Close to the transition, we anticipate that $A$ is small, and $B=O(A^2)$ is even smaller~\cite{lecomte_inactive_2012}.
		Inserting (\ref{equ:rhoAB}) in (\ref{equ:S-stationary}) we expand up to terms of order $A^4,A^2B,B^2$.  We introduce the short-hand $\sigma_0=\sigma(\overline{\rho})$ and similarly for derivatives such as $\sigma'_0=\sigma'(\overline{\rho})$, and also for $\kappa$.  The result is
		\beq
		S = \frac{L\gammobs}{2} \left[ \lambda\kappa_0 + \frac{\kappa''_0}{4}(\lambda-\lambda_c) A^2 +\frac{\mu_2}{2} B^2 + \mu_3 A^2 B + \frac{\mu_4}{2} A^4 \right]
		\eeq
		with $\lambda_c = \frac{D_0(q_1L)^2}{ -2\kappa''_0 \sigma_0}$, consistent with (\ref{equ:sc-mft},\ref{equ:def-lambda}) [recall $\kappa''<0$ and $q_1L=2\pi$] and
		\begin{eqnarray}
		\mu_2 & = &-\kappa_0''( 4\lambda_c - \lambda)/2 %  \lambda\kappa''_0/4  - \lambda_c\kappa''_0 % + D_0(q_1L)^2/(2\sigma_0)
		\nonumber\\
		\mu_3 & = & 3\lambda_c\kappa''_0\sigma'_0/(8\sigma_0) + \lambda\kappa^{(3)}_0/8 
		\nonumber\\
		\mu_4 & = &  -\lambda_c \kappa''_0 [  2(\sigma'_0/\sigma_0)^2 - (\sigma''_0/\sigma_0) ] / 16 + \lambda\kappa^{(4)}_0/32
		\end{eqnarray}
		where $\kappa^{(n)}_0$ is the $n$th derivative of $\kappa(\rho)$, evaluated at $\rho=\overline\rho$.
		The action is straightforwardly minimised over $B$, followed by minimisation over $A$.  We assume (as usual) that $\kappa_0''<0$ and that $(\lambda-\lambda_c)\ll\lambda_c$ is small so that $\mu_2>0$; we also require that  $\mu_4>\mu_3^2/\mu_2$, which is true for the BHPM (see below).
		Then for $\lambda<\lambda_c$ the action is minimised at $A=B=0$ and the system is homogeneous; while for $\lambda>\lambda_c$ the minimum occurs for 
		\beq
		\label{eqn:lecomteRes}
		A^2  = \frac{-\kappa''_0(\lambda-\lambda_c) }{ 4(\mu_4 - \mu_3^2/\mu_2) }
		\eeq
		This predicts the degree of inhomogeneity for $\lambda>\lambda_c$.
		One sees that $A=O(\lambda-\lambda_c)^{1/2}$ as one should expect, since $A$ is the order parameter for a $\phi^4$-like theory and we are making a mean-field analysis of the critical point, similar to~\cite{YongJooDynamicalSymmetryBreaking,Baek2018,Baek2019-arxiv}.
		Recall, we assumed in this derivation that $D$ is a constant, independent of $\rho$.\footnote{%
			To make contact with the analogous calculation for the SSEP in~\cite[Sec 3.2]{lecomte_inactive_2012} we take $\kappa=2\rho(1-\rho)$, $\sigma=\rho(1-\rho)$, $D=1$, noting that the definition of $\sigma$ in that work differs from ours by a factor of $2$.  Hence $\lambda_c=\pi^2/[2\rho(1-\rho)]$.
				After some algebra one finds $\mu_4 - (\mu_3^2/\mu_2)=\lambda_c/[8\rho^2(1-\rho)^2]$ and hence $A^2=8\rho^2(1-\rho)^2(\lambda-\lambda_c)/\lambda_c + O(\lambda-\lambda_c)^2$, consistent with that work
		}%
		
		These results closely resemble those of~\cite{espigares2018-crit,YongJooDynamicalSymmetryBreaking,Baek2018,Baek2019-arxiv} which apply in systems with open boundaries (not periodic).  Those works concentrated on transitions where a $Z_2$ particle-hole symmetry is spontaneously broken.  Here we do not assume particle-hole symmetry, instead the (periodic) system spontaneously breaks translational symmetry, which corresponds to a $U(1)$ symmetry (see below).  
		
		The average activity may then be estimated from (\ref{equ:K-rho}) by plugging in the most likely density profile, which yields (for $\lambda>\lambda_c$):
		\beq
		\fl\qquad
		\langle K \rangle_s \approx \LL\tobs \left[ \kappa_0 + \frac14 \kappa''_0 A^2 + O(\lambda-\lambda_c)^2 + O(L^{-1}) \right]
		\label{equ:K-mft-inhom}
		\eeq
		%This prediction will be compared with numerical results, below.
		The two corrections appear because the computation of $A,B$ is only valid when $\lambda-\lambda_c$ is small, and the restriction to the most likely profile neglects fluctuations which enter as corrections at $O(1/L)$.%
	
	\subsection{Application to the BHPM}
	\label{sec:mft-numeric}
	
	\subsubsection{Connection to MFT:}
	
	The results of Section~\ref{sec:mft-theory} are general within MFT, in the sense that we did not specify the functional dependence of $\sigma,\kappa$ on $\rho$ (we did assume $\kappa''<0$ but the case $\kappa''>0$ is a simple generalisation.)  We now consider the specific case of the BHPM, in the rescaled representation of Sec~\ref{sec:rescaled}. 
	For the Langevin variant of the model in the limit of a hard-core potential, this means that the statistics of the density field are identical to an ideal gas. (Particles are indistinguishable so collisions between hard particles are equivalent to events where the particles pass through each other.)  
	In this case 
	\beq
	D(\rho)=D_0, \qquad \sigma(\rho) = \rho D_0 \;.
	\label{equ:D-sig-bhpm}
	\eeq 
As discussed in Sec.~\ref{model}, the MC variant of the model is equivalent to the Langevin one in the limit of small steps $M\to0$.  
		Since $M$ is non-zero for our numerical work, we expect corrections to (\ref{equ:D-sig-bhpm}), but we neglect these in the following.
		(We expect them to affect the quantitative predictions of the theory, but they are unimportant at the level of accuracy that we consider.)
	%These will affect the quantitative predictions of the theory but we restrict our analysis to (\ref{equ:D-sig-bhpm}.}
	The expression for $\kappa$ is given in (\ref{equ:kappa}) which is independent of $M$; this yields $\kappa''(\rho) = -a{\rm e}^{-\rho a}$.  
	
	An interesting feature of the BHPM is that while the functions $D$ and $\sigma$ have ideal-gas behaviour, the nonlinear behaviour of $\kappa$ is still sufficient to drive the transition to an inhomogeneous state.  However, we emphasise that none of our theoretical analysis relies on the fact that $\sigma$ is linear.
		(Recall from above that the finite step size $M$ in our numerical work should result in corrections to $\sigma$, but this is not expected to affect the qualitative behaviour.)
	
	The MFT analysis requires that $\lambda$ is held constant as $\LL\to\infty$.
	Noting from (\ref{equ:def-Phia}) that $\rho a = \Phi_a$ and using (\ref{equ:sc-mft}) one sees that the homogeneous state is stable if
	\beq
	\lambda < \lambda_{\rm c} = \frac{2\pi^2 }{\Phi_a} {\rm e}^{\Phi_a} \; .
	\label{equ:lambda-crit}
	\eeq
	Hence, the calculation of Sec.~\ref{sec:mft-homog} is valid in the range $0<\lambda<\lambda_{\rm c}$, for which the control potential is attractive (so density fluctuations are enhanced).  It is also valid for negative $\lambda$, where the control potential is repulsive and density fluctuations are suppressed.  
	%As $\lambda\to-\infty$ (or alternatively taking $L\to\infty$ with fixed $s<0$), the system becomes hyperuniform, as we discuss below.
	
	For the inhomogeneous phase $0<(\lambda-\lambda_c)\ll \LL^2$ the results of Sec.~\ref{sec:mft-inhom} are relevant.
	A suitable (complex-valued) order parameter for the phase transition is 
	\beq 
	{\cal M} = \tilde\rho_{q_{1}}\LL^{-1/2} \; .
	\label{equ:MM-def}
	\eeq 
	Recalling (\ref{equ:fourier},\ref{equ:rhoAB}), we identify $ |{\cal M}|^2 $ with $(A^2/4)$ in Sec.~\ref{sec:mft-inhom}.
	The normalisation of (\ref{equ:fourier}) means that
	typical values of $\tilde\rho_q$  are $\mathcal{O}(1)$ in the homogeneous phase, so ${\cal M}\to0$ as $\LL\to\infty$.
	For the inhomogeneous phase then ${\cal M}$ is of order unity -- it is a complex number and its phase indicates the location of high and low-density regions in the system.  The system is invariant under translation so there is a $U(1)$ symmetry for the phase of ${\cal M}$, which is spontaneously broken when the system becomes inhomogeneous.

	\begin{figure}
		\includegraphics[width=1\linewidth]{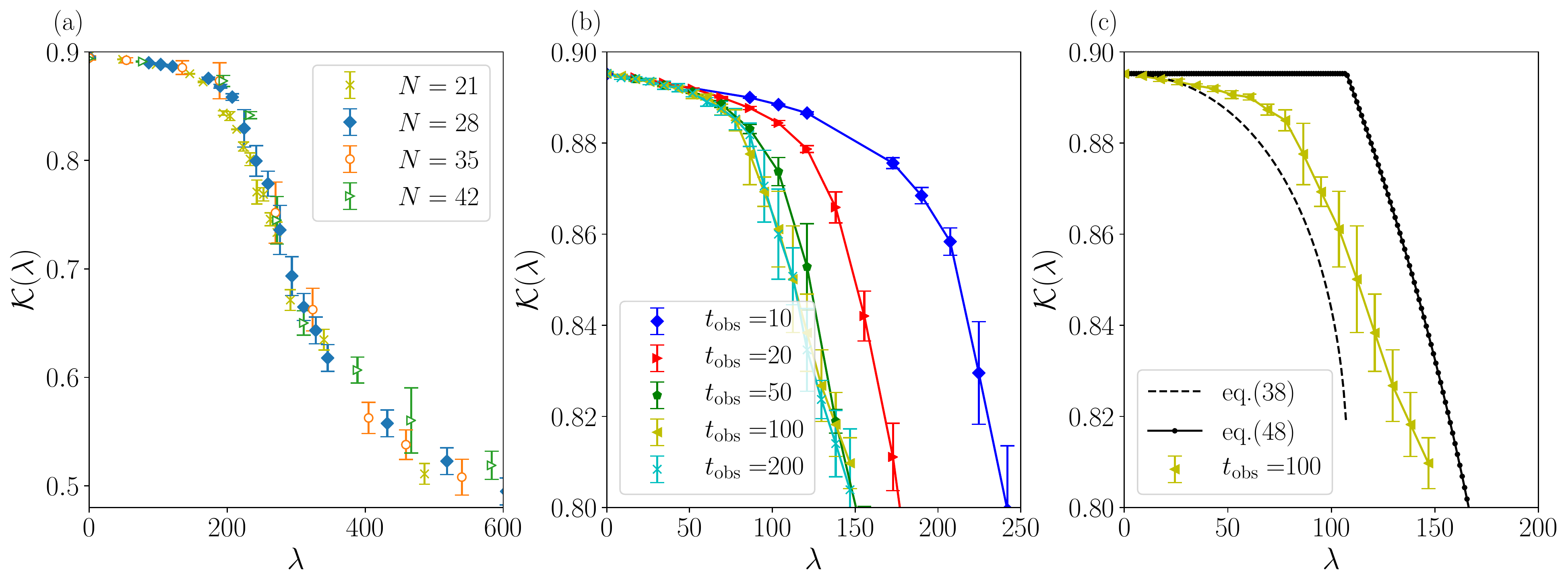}
		\caption{	(a)~Scaling plot of activity ${\cal K}(\lambda)$ showing data collapse when plotted as a function of the rescaled bias $\lambda$.   The dimensionless (rescaled) density is $\Phi_a\approx 0.233$ and $\gammobs \approx 0.070$ (which corresponds to $\tobs = 10\tau_{\rm B}$ for $N=28$).  The scaling function ${\cal K}(\lambda)$ depends weakly on $\lambda$ for $\lambda\lesssim200$, after which it decreases steeply (see  discussion in the main text).
			(b)~Dependence of the activity on $\tobs$ for $N=28$.  As this parameter increases, the decrease in ${\cal K}(\lambda)$ occurs at an increasingly small value of $\lambda$, which saturates (for large $\tobs$) at $\lambda\approx\lambda_{\rm c} \approx 90$. For $\tobs\gtrsim 100$ (in units of $\tau_{\rm B}$), the data are consistent with convergence to a limiting form; this corresponds to $\gammobs\gtrsim0.7$. (c) Comparison between the numerical results for the activity and the predictions of MFT. The prediction \refequa{equ:act-univ} applies for $\lambda<\lambda_c$; it includes a finite-size correction term that diverges at $\lambda_c$.  For $\lambda>\lambda_c$ we show a prediction based on \refequa{eqn:lecomteRes}, see the main text for a discussion.
		}
		\label{fig:ScalingWrtN2tS1}
	\end{figure}
	
	\subsubsection{Numerical results}
	To characterise the dependence of the mean activity on the bias $\lambda$ we define
	% Define also the average $\lambda$-dependent activity (per particle) 
	\beq
	{\cal K}(\lambda) = {k}(\lambda D_0/\LL^2) \; ,
	\eeq
	where the function $k(s)$ was introduced in (\ref{equ:k-of-s}). 
	We use TPS calculations to sample ensembles $P_s$ and $P_s^V$ as defined in Sec.~\ref{sec:doob}.  These computations are performed at fixed (finite) values of $N,\tobs$.  Numerical errors are discussed in Sec.~\ref{sec:mft-improvement}, below.  The methodology provides accurate results for the given values of $N,\tobs$; we compare these with the predictions of MFT that are valid in the hydrodynamic limit $N,\gammobs\to\infty$
	%
	% MFT predicts~\cite{lecomte_inactive_2012} that as $\gammobs,N\to\infty$ then ${\cal K}(\lambda)$ converges to a continuous function which has a discontinuity in its first derivative at a critical bias $\lambda=\lambda_{\rm c}$.  This limiting function is independent of $\lambda$ for $\lambda<\lambda_{\rm c}$ and then decreases for $\lambda>\lambda_{\rm c}$.  This prediction has been verified numerically for the SSEP in~\cite{espigares2018-crit,brewer_efficient_2018}.
	
	Numerical results for the BHPM are shown in  Fig.~\ref{fig:ScalingWrtN2tS1}. 
	In this case, results for $\lambda<250$ were obtained using TPS without any control forces, effects of control forces in this regime are discussed in Sec.~\ref{sec:mft-improvement}.
	For $\lambda>250$ we used the control force defined in \refequa{equ:VY} below, see section~\ref{sec:macro-gap} for a discussion.
	%\rlj{\emph{explain, using theory, and connect with Fig.~\ref{fig:PSState}}}
	%In particular, Fig.~\ref{fig:ScalingWrtN2tS1}(a) shows that ${\cal K}(\lambda)$ depends weakly on $\lambda$ when this parameter is small, before decreasing sharply for $\lambda\gtrsim 200$.    Also, Fig.~\ref{fig:ScalingWrtN2tS1}(b) shows how the results depend on $\gammobs$.  In particular, the critical value of $\lambda$ (that is, the value at which ${\cal K}(\lambda)$ starts to decrease) depends on $\gammobs$, but the results are consistent with convergence to a large-$\gammobs$ limit, for which the the critical value $\lambda_{\rm c} \approx 90$ in the simulations.  The numerical value of $\lambda_{\rm c}$ from the analytics is discussed below.
	%\subsubsection{Comparison of MFT with data for positive $s$:}
	%For the state point shown in Fig.~\ref{fig:ScalingWrtN2tS1} 

	The results of Fig.~\ref{fig:ScalingWrtN2tS1} are consistent with the asymptotic predictions of MFT.
	The state point is $\Phi_a=0.233$ so Equ.~(\ref{equ:lambda-crit}) predicts $\lambda_{\rm c}\approx107$, consistent with the data.  
	On general grounds one would expect $\lambda_{\rm c}$ to be of order unity; its large numerical value in this case arises partly from the factor of $2\pi^2$ in (\ref{equ:lambda-crit}) and partly from the fact that $\kappa''(\rho)$ in (\ref{equ:kappa}) is numerically small, for these parameters.
	
	Considering the results in more detail, Fig.~\ref{fig:ScalingWrtN2tS1}(a) shows that for fixed $\gammobs$, the function ${\cal K}(\lambda)$ shows a scaling collapse as $N$ is varied, consistent with the expected diffusive scaling.  (This scaling was less clear in~\cite{thompson_dynamical_2015}.  We suspect that this difference arises because the values of $\tobs$ used in~\cite{thompson_dynamical_2015} were not scaled with system size.)
		
		 Fig.~\ref{fig:ScalingWrtN2tS1}(b) shows data for a single system size, and increasing $\tobs$.  Taking $\tobs\to\infty$ at fixed $N$, one expects convergence of ${\cal K}$ to a limiting function: that is ${\cal K}(\lambda)\to \tilde{\cal K}_{N,\infty}(\lambda)$, where the subscripts indicate that $N$ is finite but $\tobs\to\infty$.  The system is finite so  $\tilde{\cal K}_{N,\infty}(\lambda)$ is smooth (analytic)~\cite{Chetrite2015}.   The data in Fig.~\ref{fig:ScalingWrtN2tS1}(b) are consistent with this theoretical prediction, as $\tobs\to\infty$.  In principle, convergence of this limit is expected for $\gammobs\gg1$; in practice, it is notable that convergence appears to be already achieved for $\gammobs\approx 0.7$.  We expect that this small numerical value occurs for similar reasons to the large numerical value of $\lambda_c$, particularly fact that the largest diffusional time scale is $\tau_L = 1/(D_0 q_1^2) = \LL^2/(4\pi^2 D_0)$ so that $D_0\tau_L/\LL^2=1/(4\pi^2)$ is numerically small.

		MFT makes predictions about the behaviour of $\cal K$ at large $N$.  In particular,
		taking $N\to\infty$ leads to $\tilde{\cal K}_{N,\infty}(\lambda)\to\tilde{\cal K}_{\infty,\infty}(\lambda)$, where
		the function $\tilde{\cal K}_{\infty,\infty}(\lambda)$ is predicted to be singular at $\lambda=\lambda_c$.   For $\lambda<\lambda_c$, Equ.~(\ref{equ:act-univ}) predicts that $\tilde{\cal K}_{\infty,\infty}$ is independent of $\lambda$; the same equation also predicts the first correction to the large-$N$ limit, as in~\cite{appert-rolland_universal_2008}. For $\lambda>\lambda_c$, a simple MFT prediction for $\tilde{\cal K}$ can be read from (\ref{equ:K-mft-inhom}), see also~\cite{lecomte_inactive_2012}.  The validity of this result is restricted to small $(\lambda-\lambda_c)$, because of the simple ansatz (\ref{equ:rhoAB}).  When comparing with numerics, we obtain a similar prediction by substituting (\ref{equ:rhoAB}) into (\ref{equ:K-rho}) and using (\ref{eqn:lecomteRes}) to fix $A$.  The integral in (\ref{equ:K-rho}) is performed numerically and yields a prediction for $\tilde{\cal K}_{\infty,\infty}$.  This is the prediction based on (\ref{eqn:lecomteRes}) that is shown in Fig.~\ref{fig:ScalingWrtN2tS1}(c); it matches (\ref{equ:K-mft-inhom}) when $(\lambda-\lambda_c)$ is small.
		
	\begin{figure}
		\includegraphics[width=1\linewidth]{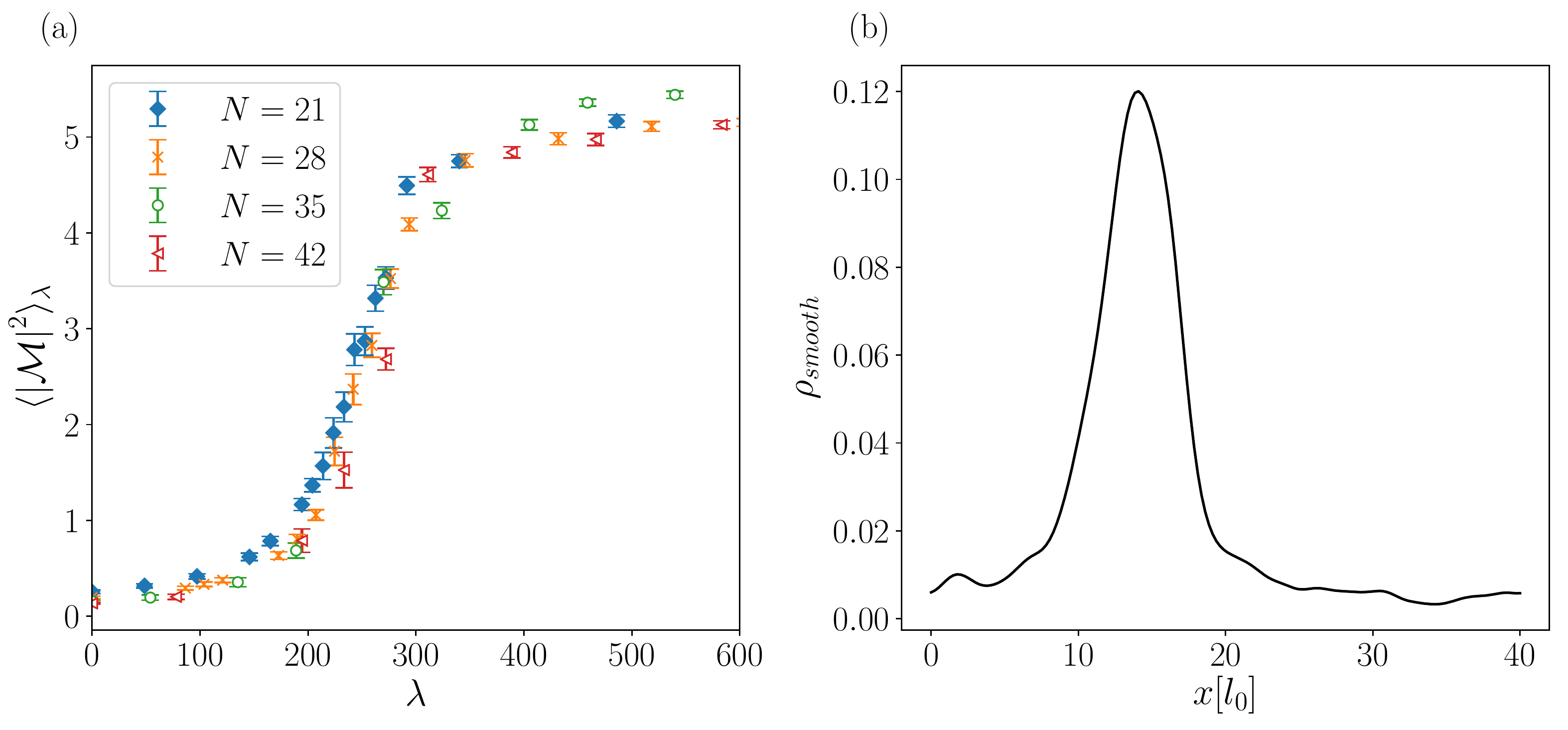}
		\caption{
			(a) The modulus of the complex order parameter $|{\cal M}|^2$, which is related to the first Fourier component of the density.  This increases from zero as the system becomes macroscopically inhomogeneous.  We take $\Phi_a=0.233$ and $\gammobs=0.070$ as in Fig.~\ref{fig:ScalingWrtN2tS1}.
			(b)~Smoothed density associated with the representative trajectory from Fig.~\ref{fig:PSState}(c), which has $N=160$ and $s=0.18$.  The system is macroscopically inhomogeneous but one interparticle gap does not yet dominate it.
		}
		\label{fig:ScalingWrtN2tS2}
	\end{figure}
		Fig.~\ref{fig:ScalingWrtN2tS1}(c)  compares a numerical estimate of $\tilde{\cal K}_{N,\infty}(\lambda)$ with these MFT predictions.  The finite-size correction term in (\ref{equ:act-univ}) is negative and diverges at $\lambda_c$, where the homogeneous theory is breaking down.  One sees that  (\ref{equ:act-univ}) gives the correct qualitative behaviour for small $\lambda$, but a quantitative agreement with numerical data would require consideration of higher-order corrections, see also~\cite{Shpiel2018}.  The theory behind (\ref{eqn:lecomteRes}) is valid as $N\to\infty$ and does not include any finite-size corrections; it captures the steep decrease in ${\cal K}(\lambda)$ but is not quantitative.  Following~\cite{YongJooDynamicalSymmetryBreaking,Baek2018,Baek2019-arxiv}, one expects a critical region $(\lambda-\lambda_c) = O(N^{-2/3})$ where neither of (\ref{equ:act-univ},\ref{eqn:lecomteRes}) is applicable; this is consistent with the data but a more detailed finite-size scaling analysis would be required to confirm it.

	Fig.~\ref{fig:ScalingWrtN2tS2}(a) shows the behaviour of the order parameter $\langle |{\cal M}|^2 \rangle_s$, which increases sharply at the transition, and takes a value of order unity in the inhomogeneous phase, consistent with the theory. 
	Fig.~\ref{fig:ScalingWrtN2tS2}(b) shows a smoothed representation of the density for the trajectory in Fig.~\ref{fig:PSState}(c), defined as
	\beq
	\rho_{\rm smooth}(x) = z^{-1} \sum_j \int_0^\tau \exp\left(-[x-\hat{x}_j(t)]^2/2 \right) \mathrm{d}t
	\eeq
	where the normalisation constant $z$ is chosen such that $\int \rho_{\rm smooth}(x) \mathrm{d}x=1$.  This shows that the density is macroscopically inhomogeneous, but we emphasise that the density is positive everywhere, which means that there is no macroscopic gap (see Section~\ref{sec:macro-gap}).
	We note that this system is quite far from the critical point, so the analysis of Sec.~\ref{sec:mft-inhom} is not sufficient to predict the density profile, consistent with the fact that it does not show a sinusoidal dependence on $x$.  Instead, the behaviour (in this rescaled representation) is that the density shows a rather narrow peak.  This might be analysed by minimisation of the action in (\ref{equ:S-stationary}) but we postpone such a calculation to future work.
	
	\ref{app:K-msd} shows similar results to those presented here, using the definition of the dynamical activity that was used in~\cite{jack_hyperuniformity_2015,thompson_dynamical_2015}.  The qualitative behaviour is the same.  
	%In particular, we see a good scaling collapse using the variables (\ref{equ:mft-L-tobs},\ref{equ:def-lambda}): this scaling was less clear in~\cite{thompson_dynamical_2015}.  We suspect that this difference arises because the values of $\tobs$ used in~\cite{thompson_dynamical_2015} were not scaled with system size.
	
	\begin{figure}
		\includegraphics[width=1\linewidth]{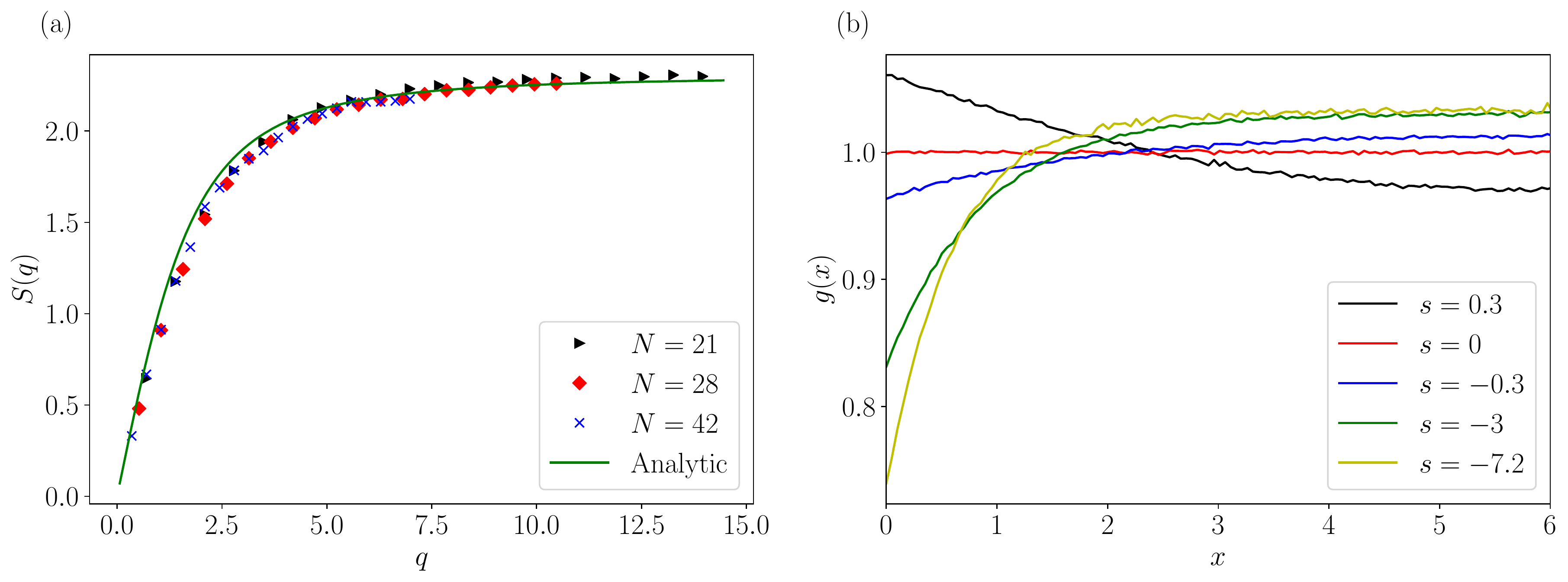}
		\caption{
			(a) The structure factor for negative $s$, compared with the theoretical prediction (\ref{equ:sq-analytic}).  We take $\gammobs=0.070$ and $\Phi_a=0.233$ as in Fig.~\ref{fig:ScalingWrtN2tS1}(a), and $s=-7.2$ (measured in units where $\tau_{\rm B}=1$).
			This $S(q)$ is suppressed at low $q$, consistent with hyperuniformity.
			(b) The corresponding pair correlation functions for the case $N=28$ and different values of $s$, as indicated.  For positive $s$ the particles tend to cluster and $g(x)$ is enhanced at contact; for negative $s$ the particles feel effective repulsion and $g(x)$ is suppressed.
		}
		\label{fig:hyperuniform}
	\end{figure}

	\subsubsection{Negative $s$ and hyperuniformity:}
	\label{sec:hyperU}
	
	Within MFT, biasing this system towards higher activity leads to hyperuniformity, as discussed in~\cite{jack_hyperuniformity_2015}.  This means that density fluctuations on large length scales are strongly suppressed~\cite{Torquato_2003}.  To measure this, define the structure factor
	%
	%When we bias the system towards higher activity with a negative $s$ it begins exhibiting hyperuniformity\cite{Garcia_Millan_2018,Torquato1,Torquato2}. Hyperuniformity is characterized by
	%the suppression of density fluctuations on large length
	%scales. One way of measuring hyperuniformity is by measuring the structure factor $S_q$ which is defined as
	\beq
	S(q)=\big\langle\tilde{\rho}_q\tilde{\rho}_{-q}\big\rangle_s.
	\eeq
	Consider the limit $L\to\infty$ so that $q$ can take arbitrarily small values, and write $S_\infty(q)=\lim_{L\to\infty}S(q)$.  A hyperuniform state is one where $\lim _{q\to0} S_\infty(q)=0$~\cite{Torquato_2003}.  Such states are not expected in finite-temperature equilibrium systems with short-ranged interactions (they require that the system should have a vanishing compressibility),  but there are many interesting examples that occur in systems away from equilibrium~\cite{Garcia_Millan_2018,Torquato1,Torquato2,Gabrielli-hyperU,Florescu2009}.   Within the MFT analysis of Sec.~\ref{sec:mft-homog}, hyperuniformity arises because the optimal control potential for $\lambda<0$ includes long-ranged repulsive forces, as may be deduced from~(\ref{equ:tilde-vq}).   Hyperuniformity is a well-known property of systems with such long-ranged forces~\cite{Lebowitz1983,Levesque2000}, where it is sometimes referred to as super-homogeneity~\cite{Gabrielli-hyperU}.
	Within the framework of Sec.~\ref{sec:mft-homog} and using (\ref{equ:pave-ou}), one sees that
	% $S_q$ is analytically found to be
	\beq
	S(q)=\frac{\sigma(\overline{\rho}) q}{\sqrt{D^2_0q^2 + 2s \sigma(\overline{\rho})\kappa''(\overline{\rho})}}.
	\label{equ:sq-analytic}
	\eeq
	see also \cite{Bodineau2008,jack_hyperuniformity_2015}. 
	This indicates that the system is hyperuniform for $s<0$.
	Fig.~\ref{fig:hyperuniform}(a) compares this prediction with the results from simulations, the suppression of $S(q)$ at small $q$ is clearly apparent.  The agreement is good -- we attribute the differences between theory and simulation to the fact that MFT requires $N,\gammobs\to\infty$ but these quantities are both finite in the numerical results.
	% for $S_q$ in figure \ref{fig:hyperuniform}(a). 
	%The low value of $S_q$ for the first few Fourier modes indicates a suppression of long range fluctuations and thus hyperuniformity.  
	Fig.\ref{fig:hyperuniform}(b) shows the pair correlation function 
	\beq
	g(x)=\frac{\big\langle\rho(x')\rho(x'+x)\big\rangle_s}{\overline{\rho}^2}
	\eeq 
	which is proportional to the probability that two particles have separation $x$ (in the rescaled representation of Sec.~\ref{sec:rescaled}).
	For the unbiased case ($s=0$) then $g(x)=1$ for all $x$.  On biasing to high activity $s>0$ one sees a reduction in $g(x)$ for small $x$, since particles feel an effective repulsion, which enhances the activity via (\ref{equ:def-K}).  Similarly, for a bias to low activity then $g(x)$ is enhanced for small $x$, consistent with an effective attraction.
	%$g(x)$ can also be viewed as the probability distribution of finding a particle distance $x$ away from another particle. 
	%The dip in $g(x)$ around 0 is caused by particles staying further away from each other in paths of high activity. This then translates into a hyperuniform state as the particles are less likely to clump up together and produce fluctuations in the density.
	
	\begin{figure}
		\includegraphics[width=0.7\linewidth]{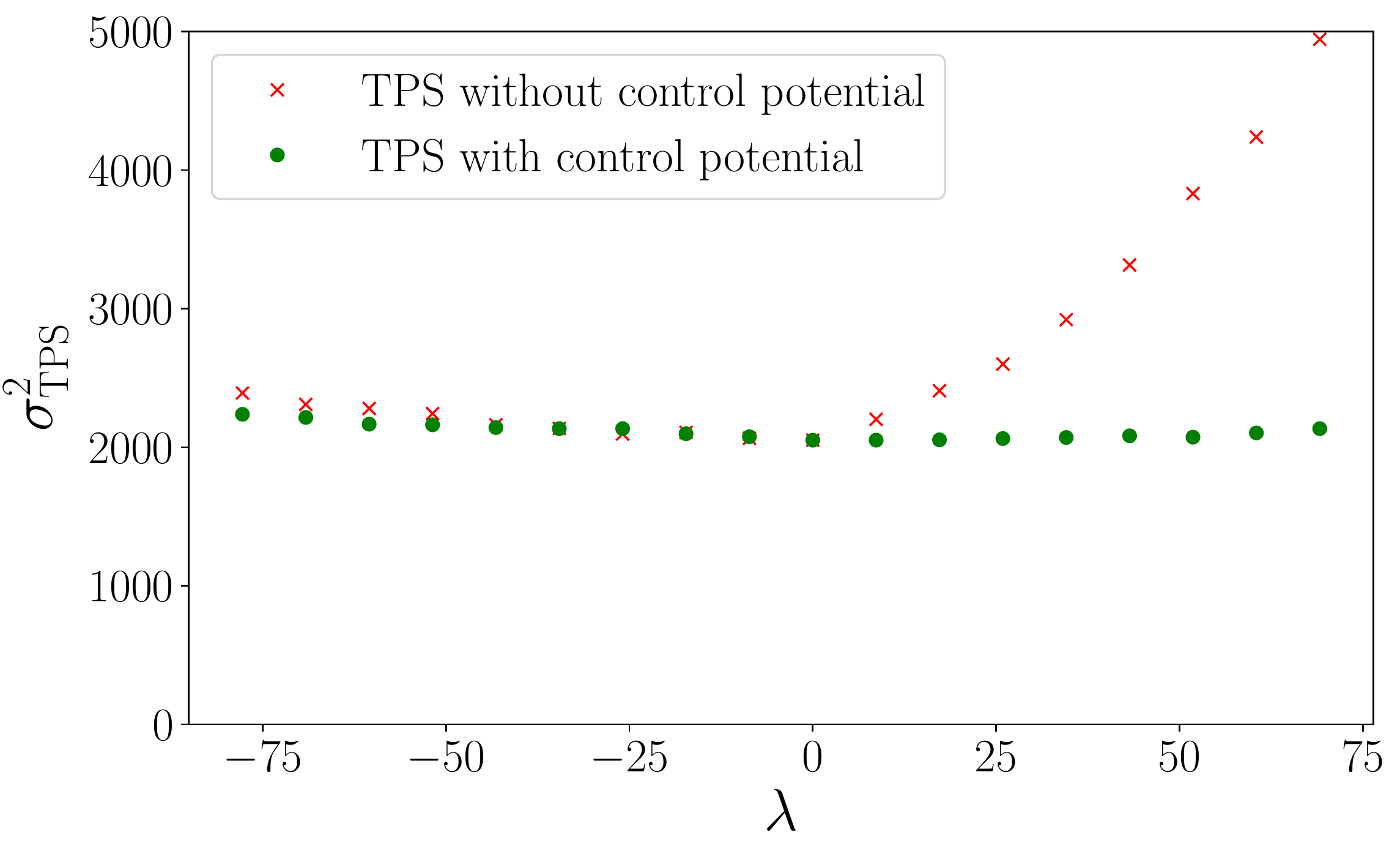}
		\caption{
			Improvement in TPS asymptotic variance by using guiding forces, in the MFT regime.  We take $N=28$, $\Phi_a=0.23$, and $\tobs=100\tau_{\rm B}$ (see also Fig.~\ref{fig:ScalingWrtN2tS1}(b)).
			In this Figure, the mean shift size used in the TPS was $\Delta t=5\tau_B$: in this parameter regime, this leads to near-optimal performance for TPS, both with and without guiding forces.  (For other parameter regimes, smaller shifts are necessary.  Hence our use of smaller shifts in other figures.)   For these parameters, the system becomes inhomogeneous for $\lambda\gtrsim\lambda_{\rm c}\approx 90$ -- these data are all within the homogeneous regime and the guiding force relies on this.
		}
		\label{fig::MFTImprovement}
	\end{figure}
	
	\subsection{Improved TPS by adding control forces}
	\label{sec:mft-improvement}
	
	%\subsubsection{Improved TPS sampling by adding control forces:}  
	
	\subsubsection{Convergence of TPS and error analysis}
	
	As discussed in Sec.~\ref{sec:doob}, we expect the addition of control forces to improve the efficiency of TPS sampling.  
	Since TPS is an MC method (which in mathematics would be called a Markov chain Monte Carlo (MCMC) method), analysis
		of convergence and numerical errors is straightforward~\cite{Asmussen-book}.
		To characterise the efficiency of the method,
	%measure the effectiveness of transition path sampling, 
	it is useful to compute how many TPS moves are required for trajectories to decorrelate from each other.
	%(note that this does not work if the trajectories are too short compared to the temporal correlations within the trajectories).  
	Let $\hat{K}_n$ be the value of the activity for the $n$th trajectory generated by TPS.  We define a block-averaged activity
	\beq
	\overline{K}_{n,m} = \frac{1}{m} \sum_{i=n}^{n+m} \hat{K}_i
	\label{equ:Kbar}
	\eeq
	As $m\to\infty$, this block average converges to $\langle K[\mathbf{x]} \rangle_s$.  Its variance  behaves as
	\beq
	\mathrm{Var}\left( \overline{K}_{n,m}  \right) = \frac{ \sigma_{\rm TPS}^2 }{m} + {\cal O}(1/m)^2
	\label{equ:sigTPS}
	\eeq
	where $\sigma^2_{\rm TPS}$ is the asymptotic variance~\cite{Asmussen-book}.  Smaller values of $\sigma_{\rm TPS}^2$ correspond to more efficient TPS sampling: in particular $\sigma_{\rm TPS}^2/\mathrm{Var}(\hat{K})$ can be used as a rough estimate of the number of TPS moves required to generate an independent sample.
	
	For small $m$ then all trajectories in the block are similar and one expects $\mathrm{Var}\left( \overline{K}_{n,m}  \right)$ to be close to $\mathrm{Var}(\hat{K})$, independent of $m$.  In our numerical analysis, we often plot 
	\beq\label{eqn::scalVar}
	\chi^{\rm TPS}_m = m\mathrm{Var}\left( \overline{K}_{n,m}  \right)
	\eeq 
	as a function of $m$, for which the expected behaviour is of the qualitative form
	\beq
	\chi^{\rm TPS}_m \approx \frac{ m \mathrm{Var}(\hat{K}) \sigma_{\rm TPS}^2 }{  m \mathrm{Var}(\hat{K}) + \sigma_{\rm TPS}^2  } \; .
	\label{equ:chi-theory}
	\eeq
	This quantity approaches $\sigma_{\rm TPS}^2$ as $m\to\infty$, as it should.  
	
	A suitable error bar for a numerical estimate of $K$ is then $\Delta K = \sigma_{\rm TPS}/N_{\rm TPS}^{1/2}$ where $N_{\rm TPS}$ is the total number of TPS moves over which the data is averaged.  This  error estimate accounts for the correlations between TPS moves.  All error bars for TPS measurements in this work are computed in this way, estimating $\sigma_{\rm TPS}$ by (\ref{equ:sigTPS}).
	
	\subsubsection{Numerical results for accelerated convergence}
	As explained above, the results for $\lambda<250$ in Fig.~\ref{fig:ScalingWrtN2tS1} were obtained using TPS without any control forces.
	We now show how the control forces that can be derived from MFT can lead to a more efficient estimate of the same result, in the homogeneous regime $\lambda<\lambda_c$.
	We have computed the asymptotic variance for the BHPM in the regime where the system is homogeneous. 
	As a simple control potential we take the first term in (\ref{equ:MFTBIAS}), so the control potential only depends on the first Fourier component of the density: 
	\beq
	V = \tilde v_{q_1} \rho_{q_1}^* \rho_{q_1}
	\label{equ:V-q1}
	\eeq
	where $q_1=2\pi/\LL$ is the smallest allowed wavevector.
	% and (? is $\tilde{v}$ given by (\ref{equ:tilde-vq}) or is it tuned?).  
	This choice for the control potential has no free parameters. 
	For these homogeneous states, it successfully captures the essential physical effect of the long-ranged control potential.   
	
	We emphasise once again that the TPS method is valid as a method for sampling from $P_s$, independent of whether control forces are used~\cite{Hedges2009,speck_first-order_2012}.  The question that we address  here is the rate of convergence, as characterised by the asymptotic variance.
	Fig.~\ref{fig::MFTImprovement} shows the improvement in TPS sampling obtained using the control potential (\ref{equ:V-q1}), which is significant for positive $\lambda$.
		All these results are in the homogeneous regime  $\lambda<\lambda_c$, where the simple control potential (\ref{equ:V-q1}) is applicable.  They provide clear evidence of a significant speedup, and demonstrate proof-of-principle for the method.  However, the small values of $s$ in this regime
		means that this is a regime where numerical sampling is relatively easy, and the speedup by the control forces is relatively modest. 
		%We will show in the next section that a much larger improvement is available when considering larger values of $s=O(1/N)$; the control forces that we use have a qualitatively different form in that case.
		Alternative methods for analysing convergence of TPS are discussed in~\ref{app:tps-conv}.
	
We note again that the results of Fig.~\ref{fig::MFTImprovement} are restricted to $\lambda<\lambda_c$.  For the inhomogeneous regime 
	($\lambda>\lambda_c$) the next section considers control forces that apply for $s=O(1/N)$, which corresponds to $\lambda=O(N)$.  We have not attempted to derive control forces for $\lambda=O(1)>\lambda_c$, this would be an interesting question within MFT.

	%Since the first Fourier mode is the most altered we cut off the higher ones in the optimal control force we use. In figure \ref{fig::MFTImprovement} we plot $\sigma^2_{\rm{TPS}}$ for both the unaltered dynamics and those biased by \ref{equ:MFTBIAS}. At positive $s$ we observe an improvement in sampling but at negative $s$ the improvement is negligible. 

	\section{The regime with a single macroscopic gap}\label{sec:macro-gap}
	%Also has a lot of fluff
	
	The results of Sec.~\ref{sec:results-mft} are based on MFT which is valid for $N\to\infty$ at fixed $\lambda$, as discussed above.
	%The theoretical results of Sec.~\ref{sec:mft-theory} are valid only for $\lambda<\lambda_{\rm c}$ but we emphasise that MFT is still applicable $\lambda>\lambda_c$, although obtaining theoretical results in that regime requires a deeper analysis~\cite{lecomte_inactive_2012}. 
	%The MFT theory is valid on taking $N\to\infty$ at fixed $\lambda$ (so $s=\mathcal{O}(N^{-2})$).  We emphasise that the general theory remains valid for $\lambda>\lambda^*$, even though the linear analysis of the previous section does not apply.  
	We note that minimisation of the MFT action $S$ in (\ref{equ:S-stationary}) always predicts that the density $\rho$ is finite everywhere, which means in turn that the gaps between particles almost surely have sizes of order $\overline{\rho}^{-1}$, as $\LL\to\infty$.  
	
	In contrast to this, Fig.~\ref{fig:PSState}(c) shows a trajectory in which a single gap takes up a finite fraction of the system. This section focusses on that regime. 
	As before, we work in the rescaled representation of Sec.~\ref{sec:rescaled}.
	At time $t$, suppose that the largest gap in the system has size ${\cal Y}(t)$.  We define $Y(t) = {\cal Y}(t)/\LL$, which is the fraction of the system occupied by this gap.
	If $Y(t)$ is order unity then the gap is macroscopic, in the sense that ${\cal Y}(t)=O(\LL)$. 
	
	To investigate this regime, define a new rescaled biasing parameter
	\beq
	h = {\frac{s\LL}{\bar\rho D_0} } %? s \LL a / D_0
	\label{equ:def-h}
	\eeq
	This rescaled bias $h$ is analogous to $\lambda$ of Section~\ref{sec:results-mft}.
	We consider the behaviour on taking $\LL\to\infty$ at fixed $h=O(1)$.  
	
	\begin{figure}
		\includegraphics[width=1\linewidth]{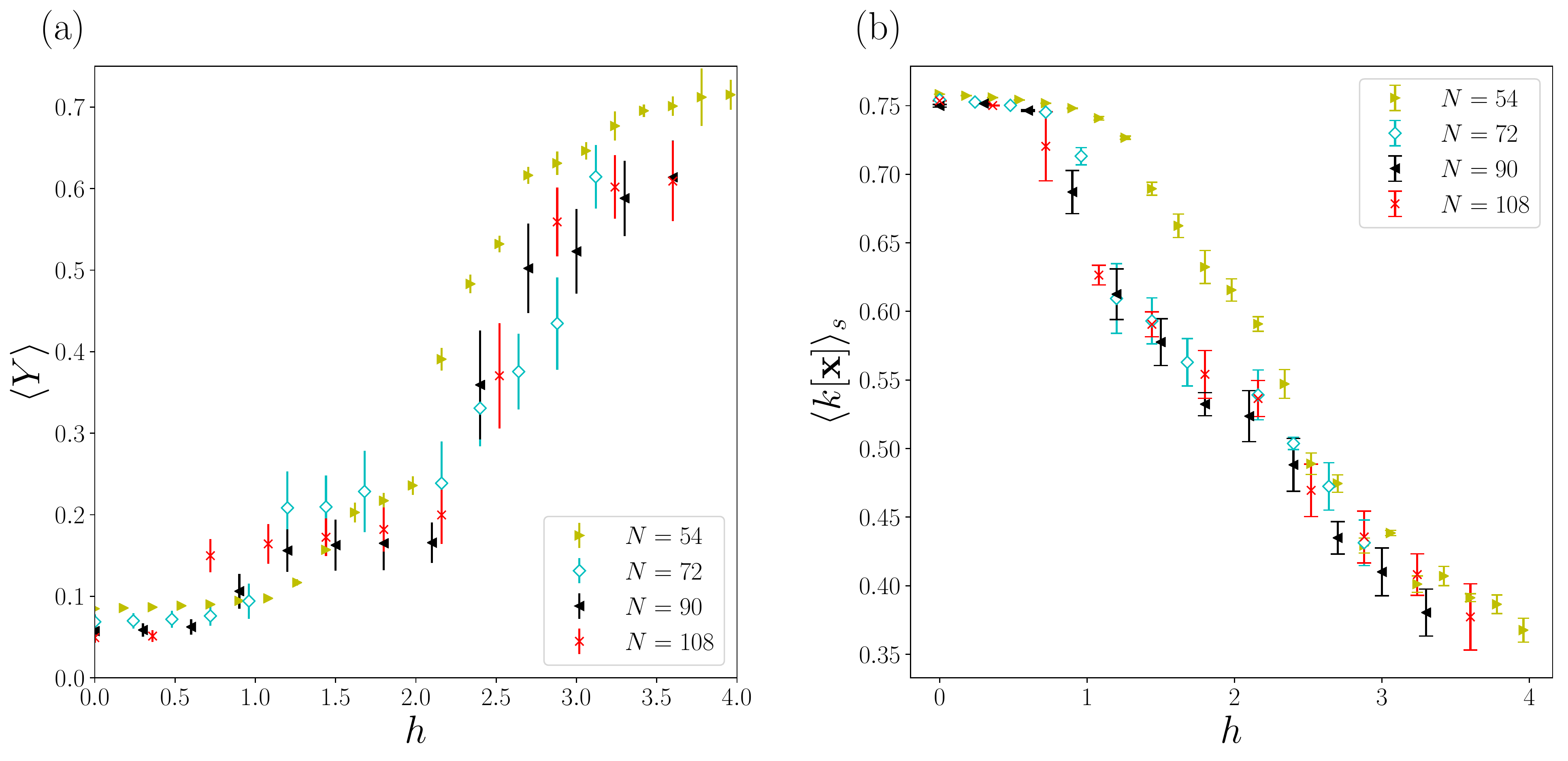}
		\caption{
			%\emph{is panel (a) doing any work here?}  (a) The typical density profile in a system $N=160$ and $L_r=40l_0$ in the largest gap regime. 
			(a) Largest gap size $\langle Y\rangle_s$ in biased ensembles.  The reduced packing fraction is $\Phi_a=0.6$. 
			(b) The activity per particle in the biased ensemble for the same systems. Around $h=2.5$ there is a change in the size of the largest gap and the derivative of $k(s)$.
			%Dependence of the size of the largest gap $Y$ on $h=\frac{sL_r}{\bar{\rho} D_0}$ for a system of $54$, $72$, $90$ and $108$ particles with reduced packing fraction $\Phi_a=\frac{Na}{\LL}=0.6$. 
			%From the graph it can be seen that the MFT phase transition causes the size of the gap to jump. As we increase system size at constant $\Phi$ the jump decreases in terms of system size so we postulate that it goes to $0$ in terms of system size as we take $L'$ to $\infty$. The same happens to the size of the largest gap in the homogeneous system in terms of system size. Thus in an infinite system only when $s\approx\mathcal{(O)}(1/L_r)$ does the largest gap become of order $L_r$. The theoretical approximation predicts the gap to open up at $h=5$ while simulations show that it happens at $h=2.5$. (c) The activity $k$ for 4 diffusively rescaled systems, these are the same systems as in (b). We would like to draw attention to the change in the derivative of $\langle k[\mathbf{x}]\rangle$ around the point where the size of the largest gap rapidly changes ($h\approx 2.5$). 
		}
		\label{fig:N160LargestHole}
	\end{figure}
	
	\subsection{Numerical results}
	
	Fig.~\ref{fig:N160LargestHole}(a) shows that for small $h$, the gap size $Y$ remains close to zero (in particular, for small fixed $h$, the average $\langle Y\rangle_s$ decreases with $\LL$).  However, for larger $h(\gtrsim 2)$ there is a sharp increase in $\langle Y \rangle_s$, which we interpret as opening of a single macroscopic gap.  [Recall again Fig.~\ref{fig:PSState}(c).]   
	
	To understand the behaviour for small $h$, we use extreme value theory to estimate the expected size of the largest gap.  
	The distribution of interparticle gaps is exponential with mean $\LL/N=\overline\rho^{-1}$.  Hence for large $N$ the largest gap $\cal Y$ has a Gumbel distribution with mean $(\log N + \gamma_{\rm E})/\overline\rho$ where $\gamma_{\rm E}\approx 0.577$ is the Euler-Mascheroni constant \cite{Haan_2010}.  Hence $\langle Y \rangle_0 = (\log N + \gamma_{\rm E})/N$ which for $N=90$ is $\approx0.06$, consistent with Fig.~\ref{fig:N160LargestHole}.
	
	Fig.~\ref{fig:N160LargestHole}(b) shows the behaviour of the activity.   As $h$ increases from zero, there is an initial sharp decrease in activity which corresponds to the MFT transition to an inhomogeneous state.   As $\LL\to\infty$, this transition would move towards $h=0$, because the critical point $\lambda=\lambda_{\rm c}$ discussed in Sec.~\ref{sec:results-mft} corresponds to $h=O(1/\LL)$.  However, the systems considered here are only moderately large, and the numerical value of $\lambda_{\rm c}$ is also quite large -- the result is that the MFT transition happens at $h\approx 1$ for the system sizes considered here.  In contrast, the largest gap opens at $h\approx 2.5$, where an additional feature in $k(s)$ is also observed (in the larger systems).  We now present a theoretical analysis of this regime, and we compare the resulting theory with these numerical results.
	
	\subsection{Theory -- interfacial model}
	
	We define a simple model that captures the qualitative behaviour of the system in the regime with a single large gap, building on recent work on kinetically constrained models~\cite{Bodineau2012cmp,Bodineau2012jsp,nemoto_finite-size_2017}.  We separate the system into a dense region and a large gap, and we focus on the behaviour at the edge of the gap, which is the interface between the two regions.  Hence we refer to this as an interfacial model.
	%The analysis follows previous work on kinetically constrained models, although the underlying physics has some important differences, as we discuss below.
	%To analyse this case, we make an ansatz, that typical trajectories of the system can be characterised by just one time-dependent random variable, which is $Y(t)\LL$, the size of the largest interparticle gap.  (The scaling is chosen so that $Y$ is of order unity when the gap is macroscopic, since this is the regime of interest.) 
	
	\subsubsection{Derivation of interfacial model:}
	To motivate the model, assume that configurations containing a large gap have 
	all the particles distributed in some (dense) region of size $\LL[1-Y(t)]$, and that they are distributed at random throughout this region.
	The mean distance between particles within the dense region is
	\beq
	\ell_Y = \frac{1-Y}{\overline{\rho}}
	\label{equ:def-Phit}
	\eeq
	with $\overline{\rho}=N/\LL$ as above.
	We model the dynamics of $Y$ by a Langevin equation where both the bias and the diffusion constant depend on $Y$:
	\beq
	\dot Y = \frac{b(Y) }{\LL} + \sqrt{\frac{2D_y(Y)}{\LL^2}} \eta
	\label{equ:Y-lang}
	\eeq
	Here $\eta$ is a standard Brownian noise.
	To fix the functions $b$ and $D_y$
	we use the MC variant of the BHPM to estimate the first and second moments of the change in the gap size $Y$, in a single MC move. 
	
	The gap size changes only when one of the particles on the edge of the gap has an accepted move.  Proposed MC moves that reduce $Y$  involve particles moving into the largest gap: these are accepted with probability $(1/2)$, by (\ref{equ:glauber}).  Proposed MC moves that increase $Y$ involve particles moving towards the dense region of the system: some of these moves will be rejected due to collisions between particles.  Since we assumed that particles are distributed at random in the dense region, the distance between neighbouring particles in this region is exponentially distributed with mean $\ell_Y$.
	Hence, for MC moves that act to increase $Y$, the fraction that is accepted is
	\beq
	\frac{1}{2A} \int_0^M {\rm e}^{-x/\ell_Y} {\rm d}x = \frac{\ell_Y}{2M} \left(1 - {\rm e}^{-M/\ell_Y} \right)
	\eeq
	where the factor of $2$ again comes from (\ref{equ:glauber}). 
	Hence, for MC moves in which the proposed particle is on the edge of the macroscopic gap, the mean change in the gap size is
	\beq
	\overline{\Delta x} = \frac{1}{4M} \int_{-M}^0 x {\rm d}x  + \frac{1}{4M} \int_0^M x {\rm e}^{-x/\ell_Y} {\rm d}x
	\eeq
	where we consider separately the situations where the gap size decreases (first term) or increases (second term).  
	The integrals can be computed exactly but we focus  on the limit where $M/\ell_Y$ is small (small MC moves).  This limit is sufficient to explain the main features of the model.  
	It yields
	\beq
	\overline{\Delta x} = - \frac{M^2}{12\ell_Y} + O(M^3) \; .
	\eeq
	Similarly the mean square displacement is
	\beq
	\overline{(\Delta x)^2} = \frac{M^2}{6} + O(M^3) \; .
	\eeq
	The relevant MC moves happen with rate $w_y = 2/\tau_0$ where the factor of $2$ arises because particles on either side of the macroscopic gap can both affect its size, and $\tau_0=M^2/(12D_0)$ is the time increment associated with one attempted MC move per particle (see also Sec.~\ref{sec:mc}).
	
	Using that the macroscopic gap is of size ${\cal Y}=Y\LL$ and taking $\LL\to\infty$ one 
	arrives at the Langevin equation (\ref{equ:Y-lang}) with
	$b(Y)=\overline{\Delta x}w_y$ and $D(Y)=\overline{(\Delta x)^2}w_y/2$. Hence (assuming as above that $M/\ell_Y\ll 1$):
	
	\begin{eqnarray}
	b(Y) & = - \frac{2\overline{\rho} D_0}{1-Y} 
	, \qquad
	D_y(Y) & = 2D_0 \; .
	\label{equ:by-Dy}
	\end{eqnarray}
	Since we assume that the particles are distributed at random in the dense region, the activity of a trajectory is [by analogy with (\ref{equ:K-rho})]
	\beq
	K[\mathbf{x}] = \LL \int_0^\tobs \kappa_y(Y(t))  \mathrm{d}t 
	\eeq
	with 
	$ 
	\kappa_y(Y) = (1-Y) \kappa( \overline\rho/(1-Y)  ) .
	$
	[To derive this, recall that $\kappa(\rho)$ is the activity per unit length for a system with density $\rho$.  Here, the dense region of the system has size $\LL(1-Y)$ and density $\overline\rho/(1-Y)$.]  Hence from (\ref{equ:kappa})
	\beq
	\kappa_y(Y) = \frac{1-Y}{a} \left( 1 - {\rm e}^{-\Phi_a/(1-Y)} \right) \; .
	\label{equ:kappa-y}
	\eeq
	The Fokker-Planck equation corresponding to (\ref{equ:Y-lang}) is 
	\beq
	\frac{\partial\PP}{\partial t} = -\frac{1}{\LL} \frac{\partial}{\partial Y} (  b\PP ) + \frac{1}{\LL^2} \frac{\partial^2}{\partial Y^2} ( D_y \PP )
	\label{equ:Y-fp}
	\eeq 
	where $\PP=\PP(Y)$ is the probability density for $Y$.  
	
	In kinetically constrained models~\cite{Bodineau2012cmp,Bodineau2012jsp,nemoto_finite-size_2017}, a similar interfacial model was derived, which gives semi-quantitative predictions for the system behaviour if $b$ and $D_y$ are taken as constants.  The system considered here is different in that $b(Y)$ has a diverging negative value as $Y\to 1$ -- this reflects the fact that as the largest gap approaches the size of the system, all the particles end up confined in a very small region.

	\subsubsection{Biased ensembles for the interfacial problem:}
	
	%The dynamical free energy
	We now analyse the effects of biasing to low dynamical activity in the interfacial model.
	The dynamical free energy $\psi(s)$ of the interfacial model is obtained by finding the largest $\psi$ that solves the following eigenproblem
	\beq
	\psi \PP = -\frac{1}{\LL} \frac{\partial}{\partial Y} (  b\PP ) + \frac{2D_0}{\LL^2} \frac{\partial^2 \PP}{\partial Y^2}  - s\LL \kappa_y \PP \; .
	\label{equ:fp-P-s}
	\eeq
	%where we used that $D_y=2D_0$, independent of $Y$.
	The diffusive term is suppressed by a factor of $1/\LL$ so we identify this as a small-noise problem that may be solved by saddle point methods.
	It is convenient to transform to a self-adjoint (Hermitian) form by ${\cal U}(Y) = \int_{Y_0}^Y b(Y')\LL/(2 D_0) {\rm d}Y'$ and defining $\QQ(Y) = \PP(Y) {\rm e}^{-{\cal U}(Y)}$.  (The reference point $Y_0$ can be chosen arbitrarily, so $\cal U$ is fixed only up to an additive constant.)  The eigenproblem (\ref{equ:fp-P-s}) becomes
	\beq
	\psi  \QQ =  \frac{2D_0}{\LL^2} \frac{\partial^2 \QQ}{\partial Y^2}  
	- \overline{\rho}^2 D_0 {\cal V} \QQ 
	\,  - \frac{1}{2\LL}  \frac{\partial  b}{\partial Y}  \QQ
	\label{equ:fp-Q}
	\eeq
	with a dimensionless potential 
	\beq
	{\cal V}(Y) = \frac{h}{\overline{\rho}} \kappa_y(Y) + \frac{1}{8\overline{\rho}^2D_0^2} b(Y)^2 \; .
	\eeq
	The final step of the derivation used (\ref{equ:def-h}).
	
	For large $\LL$, this eigenproblem can be solved by saddle-point methods.  The last term in (\ref{equ:fp-Q})  is negligible when $\LL$ is large.  Also, the dominant eigenfunction $\QQ$ is sharply-peaked at the minimum of ${\cal V}$, which we denote by 
	\beq
	Y^* = \mathrm{argmin}\; {\cal V}(Y) \; .
	\eeq  
	Also $\psi = -{\cal V}(Y^*) \overline{\rho}^2 D_0$.
	%The eigenfunction takes the form ${\cal Q}(Y) = {\rm e}^{-\LL f(Y)}$.
	Using (\ref{equ:by-Dy},\ref{equ:kappa-y}) we obtain
	\beq\label{equ::largestGap}
	{\cal V}(Y) = h \frac{1-Y}{\Phi_a} \left( 1 - {\rm e}^{-\Phi_a/(1-Y)} \right) + \frac{1}{2(1-Y)^2}
	\eeq
	% we take $f(Y^*)=0$ so that $f\geq0$ elsewhere.  
	%For the purposes of this work, it is sufficient to note that\dots
	%\dots we obtain the size of the largest gap (and the behaviour of $\langle k \rangle_s$) by minimising ${\cal V}$.
	Minimising this potential we find that $Y^*=0$ for small $h$, but there is a threshold $h_{\rm c}$ above which $Y^*$ becomes non-zero.  
	At the threshold, $Y^*$ increases continuously which means that ${\cal V}'(0)=0$ for $h=h_{\rm c}$.  
	The existence of a threshold is consistent with  Fig.~\ref{fig:N160LargestHole}, the accuracy of the detailed predictions will be discussed below.  
	Before that, we derive the effective potential that describes the state with $Y^*>0$.

	\subsubsection{Optimal control potential:}
	
	We present two possible methods for estimating the optimal control potential introduced in Sec.~\ref{sec:doob}.  The first is based on a physical argument: observe that the dense region of the system contains particles that are distributed as an ideal gas, so their pressure is
	\beq
	P_{\rm mech} = \frac{ \overline{\rho} }{ \beta (1-Y) } 
	\eeq
	Maintaining a gap of size $Y^*$ requires a control force that balances the pressure.  Since the eigenvector $\QQ$ is sharply-peaked at $Y^*$, the fluctuations of $Y$ are small in the biased ensemble, so the behaviour is relatively insensitive to the form of the control potential, as long as it produces the correct force in the typical states (which have $Y=Y^*$).  Hence, a control potential that reproduces the correct statistics for $Y$ is 
	\beq
	V(Y) =  - {\LL} Y c
	\label{equ:VY}
	\eeq
	where $c>0$ is a constant with units of inverse length -- its interpretation is that there is a constant force $c/\beta$ that acts to increase the gap size.  To determine $c$ we equate the force to the pressure required to stabilise a gap of size $Y^*$:
	\beq
	c = \frac{ \overline{\rho} }{ 1-Y^* }  \; .
	\label{equ:cy}
	\eeq
	In order to use (\ref{equ:VY}) with  (\ref{langevin-control}), the potential $V$ must be expressed as a function of the particle positions: this is 
	straightforward because $Y$ is the size of the largest interparticle gap, which is a simple function of the the particle positions.
	
	The second method for deriving a suitable control potential is the standard mathematical approach: consider the adjoint (Hermitian conjugate) of the eigenproblem (\ref{equ:fp-P-s}) which is
	\beq
	\psi {\cal F} = \frac{b}{\LL} \frac{\partial {\cal F}}{\partial Y  }  + \frac{D_y }{\LL^2} \frac{\partial^2 {\cal F}}{\partial Y^2}   - s \LL \kappa_y  {\cal F}
	\eeq
	Since the noise is weak, the expected solution is of the form ${\cal F}(Y) = {\rm e}^{-\LL g(Y)}$ and the optimal control potential may be identified from (\ref{eqn:DefVOpt}) as $V_{\rm opt}(Y)=2\LL g(Y)$.  Inserting the expected form for $\cal F$, retaining terms at leading order in $\LL^{-1}$, and using $\psi = -{\cal V}(Y^*) \overline{\rho}^2 D_0$ from above, one recovers $V_{\rm opt}'(Y^*) = -  \overline{\rho} \LL /(1-Y^*)$.  This is consistent with (\ref{equ:VY},\ref{equ:cy}) which together imply $V_{\rm opt}(Y) = -  \overline{\rho} \LL Y/(1-Y^*)$ for $Y\approx Y^*$.

	\begin{figure}
		\centering
		\includegraphics[width=1\linewidth]{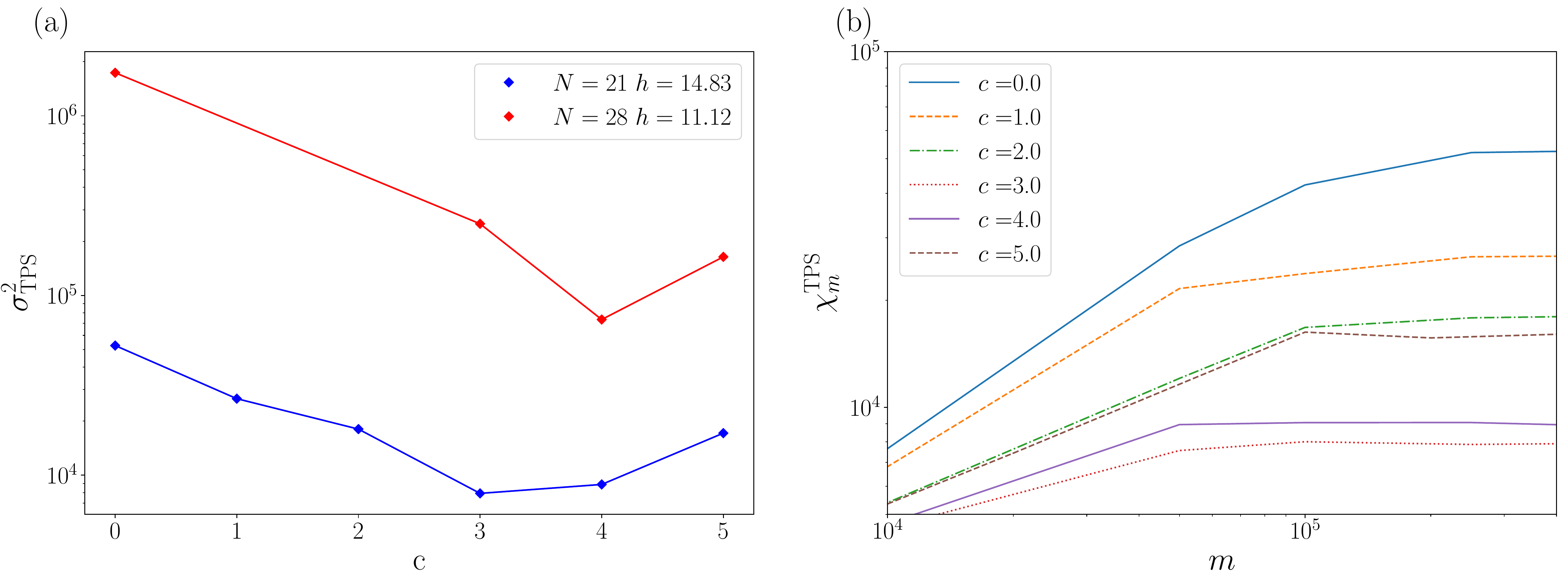}
		\caption{
			(a)~Asymptotic variance of the TPS method, as a function of the parameter $c$ used in the definition of the control force.  We take $\tobs=10\tau_{\rm B}$ and $\tobs=5.6\tau_{\rm B}$ for $N=28$ and $N=21$ respectively and in both $\Phi_a=0.233$, same as in \ref{fig:ScalingWrtN2tS1}. The control forces lead to a clear reduction in the variance, across a range of $c$.  We show results for two system sizes, at representative values of $h$ (always within the macroscopic-gap regime).
			(b)~For $N=21$ we show the scaled variance $\chi_m^{\mathrm{TPS}}$ as a function  of the block size $m$.  The behaviour is consistent with (\ref{equ:chi-theory}).
		}
		\label{fig:ImproveHole}
	\end{figure}

	\subsubsection{Comparison with numerical results}
	
	For the parameters shown in Fig.~\ref{fig:N160LargestHole}, Equ. \refequa{equ::largestGap} predicts $h_c=\Phi_a/(1-e^{-\Phi_a}-\Phi_a e^{-\Phi_a})\approx4.9$.  This overestimates the value of the bias at which a macroscopic gap appears.  The reason is clear if one considers the behaviour close to the threshold.  In the interfacial model, the state with $Y^*=0$ has the particles distributed homogeneously but the MFT analysis of Sec.~\ref{sec:results-mft} has already established that the system is not homogeneous for these values of the bias. 
	
	If the state with $Y^*=0$ is already inhomogeneous, one sees that the probability of opening up a macroscopic gap will be enhanced, because the gap will likely appear at a location where the density is already low.  Our conclusion is that the interfacial model predicts the existence of a threshold $h_{\rm c}$ at which a macroscopic gap appears, which is consistent with the numerical data.  However, the assumption within the model that the dense region of the system is homogeneous is not accurate enough for the model to deliver quantitative predictions.   In the following subsection, we show that despite these shortcomings, the optimal control potential predicted by the interfacial model is sufficiently accurate to significantly improve numerical sampling.  In this sense, the interfacial model does capture the essential physical features of the regime with a macroscopic gap.
	
	\subsection{Improvement in sampling by control forces}
	
	We have performed TPS sampling using the control potential (\ref{equ:VY}).  The relation (\ref{equ:cy}) is confirmed by our numerical results, in that a control potential with this value of $c$ leads to a typical largest gap of size $Y^*$.  Fig.~\ref{fig:ImproveHole} shows the improvement in TPS sampling that is obtained with this control potential, which is more than an order of magnitude, even for small systems.  The parameter $c$ in (\ref{equ:VY}) is varied, in order to obtain the maximal speedup.  
	For larger systems, the improvement increases rapidly -- we are not able to quantify the speedup because (for example) the results shown in Fig.~\ref{fig:N160LargestHole} would require a prohibitively large computational effort, if control forces were not used.
	The reason is that the macroscopic gaps that appear in those systems are extremely rare under the natural dynamics, so that
	TPS moves tend to be rejected if one uses a system without a control potential.  
	We also note from Fig.~\ref{fig:ImproveHole} that significant speedup is possible for control forces that are not optimal, as emphasised in~\cite{nemoto_population-dynamics_2016}.
	
	\begin{figure}
		%\centering
		\includegraphics[width=1\linewidth]{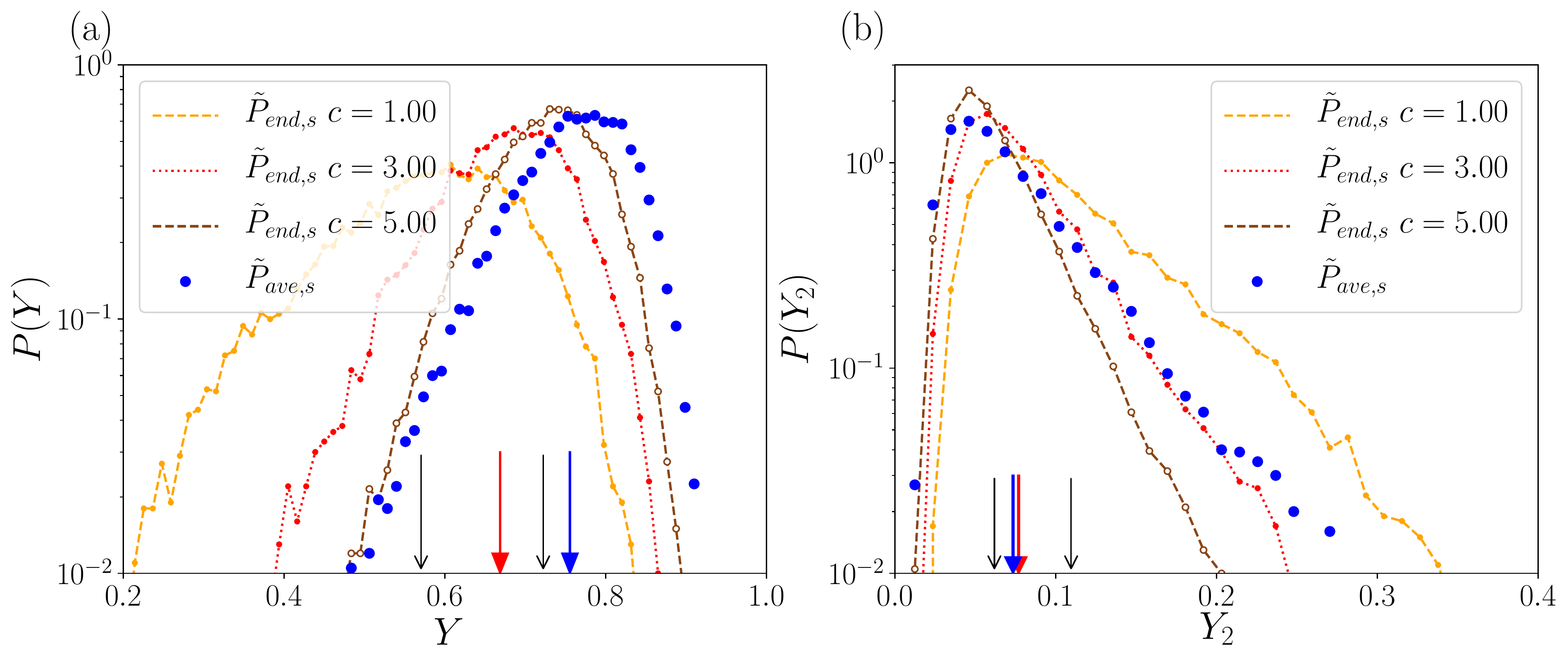}
		\caption{
			Distributions $P_{\rm ave}$ and $\tilde{P}_{\rm end}$ for two observable quantities.  Parameters are the same as Fig.~\ref{fig:ImproveHole}(b).  
			(a)~Distributions of the size $Y$ of the largest interparticle gap, for different strengths $c$ of the control force in (\ref{equ:VY}).    Vertical arrows indicate the means of the various distributions.  
			(b) Corresponding distributions for the size of the second-largest interparticle gap, $Y_2$.
		}
		\label{fig:ProbDistroscombine}
	\end{figure}
	
	This improvement that is available from control forces also enables us to investigate what value of $c$ is most effective for improved sampling.  As noted in Sec.~\ref{sec:doob}, if one uses the optimal guiding force, the distributions $P_{\rm ave}$ and ${\tilde P}_{\rm end}$ of (\ref{equ:pend},\ref{equ:pave}) coincide with each other, for all observable quantities.  Recall that $P_{\rm ave}$ is independent of the guiding force but ${\tilde P}_{\rm end}$ is evaluated in a system with control forces, which does depend on the choice of these forces.   It was suggested in~\cite{nemoto_population-dynamics_2016,nemoto_finite-size_2017} that a suitable method for choosing approximate (non-optimal) control forces is to adjust their parameters to make the distributions $P_{\rm ave}$ and ${\tilde P}_{\rm end}$ as similar as possible.  
	
	This hypothesis is tested in Fig.~\ref{fig:ProbDistroscombine}.   We first consider the distribution of $Y$, the largest interparticle gap.  In this case one sees that the control force that gives the best overlap of $P_{\rm ave}$ and ${\tilde P}_{\rm end}$ is $c=5$, which is larger than the force which gives the most efficient sampling (this is $c=3$, from Fig.~\ref{fig:ImproveHole}).   We also consider the distribution of $Y_2$, which is the second largest interparticle gap, measured relative to the system size $\LL$.  For this quantity, the distributions overlap best at $c=3$, where the sampling is most efficient. 

	The conclusion of this analysis is that maximising the overlap of $P_{\rm ave}$ and ${\tilde P}_{\rm end}$ for any single observable does not guarantee that the distributions for other observables should overlap.   This cautions against placing too much faith in the overlap of any single distribution, as an indicator of where sampling is most efficient. 
	
	As an alternative method for estimating which control force is optimal, one may consider the statistics of the action, as suggested in \cite{nemoto_optimizing_2018}.    Let $\langle B\rangle_{s,V}=\int B[\textbf{x}] d\tilde{P}^V_s[\textbf{x}]$ be the average of the observable $B$ with respect to the distribution $\tilde{P}^V_s$ of (\ref{equ:PtVs-def}).  If $V$ is the optimal control then
	%
	%An alternative measure with which to look at the improvement of sampling is the action in the modified ensemble as suggested in \cite{Nemoto2019}. If the optimal control force were to be applied to the system this action would equal the dynamical free energy. Since the force applied is not optimal the action is an approximation for the dynamical free energy
	\begin{equation}
	\lim_{\tobs\rightarrow\infty}\frac{1}{\tobs}\langle -sK+\mathcal{A}^{\mathrm{sym}}\rangle_{s,V} = \psi(s),
	\label{equ:sApsi}
	\end{equation}
	%where we define the average $\langle B\rangle_{s,V}=\int B[\textbf{x}] d\tilde{P}^V_s[\textbf{x}]$. The equality holds when $V$ is the optimal control force.
	%Minimising the difference is then indicative of the best control force as shown in figure \ref{fig:KaveAave} (a).
	The suggestion of~\cite{nemoto_optimizing_2018} is that optimising $V$ to achieve equality in (\ref{equ:sApsi}) can be used to obtain good sampling.  Note that there are many control forces that can achieve equality in (\ref{equ:sApsi}).  This situation is to be contrasted with the general inequality~\cite{OptimalControlRepresentationChetrite, Jacobson-arxiv}
	%
	%In previous theoretical works \cite{jack_ptps, OptimalControlRepresentationChetrite, kggw18} it has been shown that maximising 
	\begin{equation}
	\psi(s)
	%=\lim_{\tobs\rightarrow\infty}\frac{1}{\tobs}\log\langle \mathrm{e}^{-sK+\mathcal{A}^{\mathrm{sym}}}\rangle_{V}
	\geq\lim_{\tobs\rightarrow\infty}\frac{1}{\tobs}\langle -sK+\mathcal{A}^{\mathrm{sym}}\rangle_{V},
	\label{equ:sApsi2}
	\end{equation}
	where the average is with respect to the controlled dynamics, without any biasing: $\langle B\rangle_{V}=\int B[\textbf{x}] d\tilde{P}^V[\textbf{x}]$.
	In (\ref{equ:sApsi2}), equality can only be achieved if $V$ is the optimal control potential, this can be checked by noting that $\psi(s)=\lim_{\tobs\rightarrow\infty}\tobs^{-1} \log\langle \mathrm{e}^{-sK+\mathcal{A}^{\mathrm{sym}}}\rangle_{V}$ and using Jensen's inequality.  On this basis one might expect that maximising the right hand side of (\ref{equ:sApsi2}) would give the best sampling.
		\begin{figure}
			\centering
			\includegraphics[width=1\linewidth]{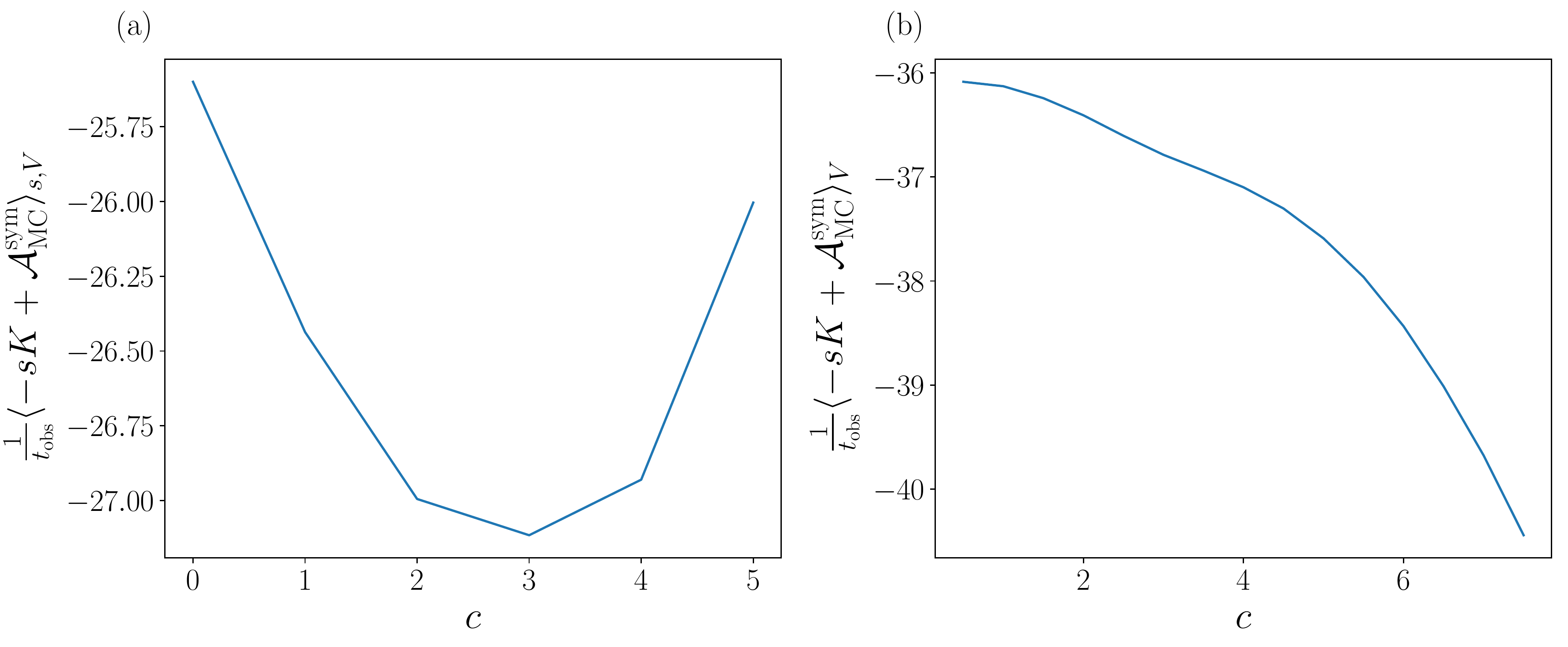}
			\caption{(a) The average of the action $\frac{1}{\tobs}\langle -sK+A^{\mathrm{sym}}\rangle_{s,V}$ that appears in (\ref{equ:sApsi}), for different biasing forces. All parameters are the same as in Fig.~\ref{fig:ImproveHole}, for which we estimate $\psi(s=1.9)\approx-32$.  (b) The (unbiased) average of the action in the controlled system that appears in (\ref{equ:sApsi2}).}
			\label{fig:KaveAave}
		\end{figure}
	Results for the averages in (\ref{equ:sApsi},\ref{equ:sApsi2}) are shown in Fig.~\ref{fig:KaveAave}.   Contrary to the situation in~\cite{nemoto_optimizing_2018}, there is no value of $c$ for which equality is achieved in (\ref{equ:sApsi}).  However, we note that the most efficient sampling takes place for $c=3$, which is the value where the average on the right hand side of (\ref{equ:sApsi}) is closest to $\psi(s)$,  consistent with the proposal~\cite{nemoto_optimizing_2018} that equality in (\ref{equ:sApsi}) is a desirable feature.  One also sees that the right hand side of (\ref{equ:sApsi2}) is decreasing in $\psi$ for all $c>1$.  Thus, $c=1$ gives the best bound on $\psi$ but it does not achieve the best sampling, contrary to the intuitive expectation stated above.   
	
	Based on Figs.~\ref{fig:ImproveHole},\ref{fig:ProbDistroscombine},\ref{fig:KaveAave}, our conclusion in this Section is that no single prescription seems satisfactory for determining the best choice of control force $V$ in practical situations such as this one, and some trial-and-error is still necessary in this process.
	
	%
	%, over the set of all control forces should yield the optimal control force. We do not explore all possible forces but in our sampling range this method does not yield the best possible replacement or even have a maximum as shown in figure \ref{fig:KaveAave} despite the guiding force improving the sampling.

	\section{Conclusions}
	\label{sec:conc}
	
	We have given a detailed analysis of the behaviour that was summarised in Fig.~\ref{fig:PSState}, including discussion of hyperuniform states that appear when states are biased to high activity, and inhomogeneous states with low activity.  We have discussed the existence of two inhomogeneous regimes, with $s=O(N^{-2})$ (MFT regime) and $s=O(N^{-1})$ (macroscopic interparticle gap).
	
	We have shown that control forces can be used to improve numerical sampling of these ensembles~\cite{nemoto_population-dynamics_2016,nemoto_finite-size_2017,ray_exact_2018,DasLimmer-arxiv}.  In the MFT regime where the system is homogeneous, these optimal control forces are long-ranged.  On biasing to low activity, these forces are attractive and drive the formation of inhomogeneous states.  Using these guiding forces (in the homogeneous state) leads to an improvement in sampling efficiency.   However, for these small values of $s$, the effect of the bias is weak, so sampling is already possible without these forces.   
	
	In the regime with a macroscopic interparticle gap, we have argued that a form of interfacial model can capture some features of the system, similar to~\cite{nemoto_finite-size_2017}.   Using this model to infer a suitable control force leads to an improvement in sampling efficiency that is more than a factor of 10 in small systems.  For large systems, the computations that we present would be prohibitively expensive without these control forces.  We have discussed how the parameters of the control force might be optimised.  In particular, we find that the simple criterion of \cite{nemoto_finite-size_2017,nemoto_population-dynamics_2016}, to match the distributions $P_{\rm ave}$ and $P_{\rm end}$ is not optimal for the cases considered here: since the control force is very simple we have instead optimised its free parameter by hand.  Further work would be valuable, to understand how to infer control forces that improve sampling efficiency.  Some generic aspects of optimally-controlled models are discussed in~\cite{Jack2015b}, where they are referred to as auxiliary models.
	
	Comparing the TPS method with cloning methods~\cite{Giardina2011,Lecomte_2007,DirectEval,nemoto_population-dynamics_2016,nemoto_finite-size_2017,ray_exact_2018}, we note that while TPS gives results for finite $\tobs$, cloning provides direct access to the limit $\tobs\to\infty$, which is the limit where large-deviation theory applies. On the other hand, the detailed-balance property of TPS~\cite{Bolhuis} means that it samples directly from $P_s$ or $\tilde{P}_s^V$ as defined in (\ref{sensemble},\ref{equ:PtVs-def}).  By contrast, cloning methods do not sample directly from a target distribution; they do allow estimation of averages with respect to these distributions but the associated statistical estimators have systematic errors (bias) which only disappear as the population size tends to infinity~\cite{Hidalgo2017-ii,brewer_efficient_2018}. Estimation of statistical uncertainties is also simpler for TPS, see Sec.~\ref{sec:mft-improvement}.
	There are also other methods for sampling large deviations, some of which require accurate representations of an optimally-controlled dynamics in order to achieve accurate results~\cite{Nemoto2014,Ferre2018,Jacobson-arxiv}, see also~\cite{DasLimmer-arxiv}.  Such methods are attractively simple, but accurate representations of optimally-controlled dynamics may be challenging in complex systems.  We emphasise once more that the role of control forces in TPS and cloning is to improve convergence, but accurate results are still available without obtaining the optimal control force.
	
	On physical grounds, it is notable that the rich physics of inhomogeneous and hyperuniform states in the BHPM occurs for 
	very small values of the bias parameter $s$, which are at either $O(N^{-2})$ or $O(N^{-1})$.  The strong response of the system to these biasing fields has its origin in hydrodynamic modes.  Many theories of biased ensembles assume the existence of a gap in the spectrum of the generator of the relevant stochastic process.  Here the gap size is vanishing as $N\to\infty$, because of slow (diffusive) hydrodynamic modes.  The MFT approach is to rescale (speed up) time so that one is restricted to hydrodynamic time scales, but the gap for the generator is restored.  
	
	The fact that the behaviour originates on the hydrodynamic scale also explains why MFT predictions are universal, in that they depend on diffusive scalings but not on microscopic details of the model.  The predictions for the behaviour for $s=\mathcal{O}(N^{-1})$ are not universal in the same sense, but the simplicity of the interfacial model indicates that they may arise generically in systems with sharp interfaces between coexisting phases (see also~\cite{nemoto_finite-size_2017}).
	
	\ack We thank Takahiro Nemoto and Vivien Lecomte for discussions about the use of control forces to aid numerical sampling.
    Jakub Dolezal was supported by a studentship from the EPRSC, reference EP/N509620/1.
	The figures in this article were made using Matplotlib \cite{Hunter:2007}.
	
	\begin{appendix}
		
		\section{Biased path ensembles}
		\label{app:bias}
		
		This appendix includes a derivation of (\ref{equ:PtVs}) and its analogue for MC dynamics.  
		
		\subsection{Langevin dynamics}
		
		As in the main text, 
		let $\mathrm{d} \tilde{P}^V[\mathbf{x}]$ be the probability for trajectory $\mathbf{x}$ under the controlled process (\ref{langevin-control}), but with $V_{\rm opt}$ replaced by some general (possibly non-optimal) potential $V$.  Then standard path-integral arguments (see eg \cite{nemoto_population-dynamics_2016}) show that
		\begin{equation}
		\mathrm{d}\tilde{P}^V[\mathbf{x}] \propto \exp\left( -{\cal A}[\mathbf{x}] \right) \, \mathrm{d} P_0[\mathbf{x}] \;,
		\label{equ:tilde-P}
		\end{equation}	
		where the normalisation constant has been omitted for ease of writing and 
		\beq
		{\cal A}[\textbf{x}]  = \frac14 \sum_i \int_0^\tobs 2 \dot{x}_i \cdot \nabla_i V + \nabla_i V \cdot  D_0 ( \nabla_i V + 2\beta \nabla_i U) {\rm d} t \; .
		\label{equ:def-A}
		\eeq  
		%which may be obtained by standard path-integral methods,  [cite].
		Combining (\ref{sensemble}) with (\ref{equ:tilde-P}) yields
		\begin{equation}
		\mathrm{d} {P}_s[\mathbf{x}] \propto \exp\left( {\cal A}[\mathbf{x}] - s K[\mathbf{x}] \right) \, \mathrm{d} \tilde{P}^V[\mathbf{x}] \;,
		\label{equ:Ps-AK}
		\end{equation}	
		which  means that the ensemble (\ref{sensemble}) which  was obtained from $P_0$ by biasing $K$ can also be obtained (exactly) by a suitable biasing of $\tilde{P}^V$. 
		Recalling
		that the product of $\dot{x}$ and $\nabla V$ in (\ref{equ:def-A}) is to be interpreted in the Ito sense and using Ito's formula for $\mathrm{d} V/\mathrm{d}t$, we obtain
		$ {\cal A}[\mathbf{x}]  =  \frac{1}{2}[ V(\tobs) - V(0)]  +{\cal A}^{\rm{sym}}[\textbf{x}]  $
		where ${\cal A}^{\rm{sym}}$ is given by (\ref{equ:def-Asym}).  Using this with (\ref{equ:Ps-AK}) and (\ref{equ:PtVs-def}) yields (\ref{equ:PtVs}).
		
		Finally, using (\ref{equ:def-Asym}) with the observation that $ {\cal A}^{\rm{sym}}[\mathbf{x}] - s K[\mathbf{x}]$ is constant and equal to $\psi(s)\tobs$ for $V=V_{\rm opt}$,    one sees that  
		\beq
		V(X)=-2\log u(X) 
		\label{eqn:DefVOpt}
		\eeq
		where $u$ is the solution with largest eigenvalue $\psi$ of the eigenproblem 
		\beq
		\sum_i \left[ D_0 \nabla_i^2 u - (\beta D_0 \nabla_i U) \cdot  \nabla_i u - s r_i^a  u \right]  = \psi  u \; .
		\eeq  
		This is a tilted Fokker-Planck equation in its adjoint form, see for example~\cite{Chetrite2015}.

		\subsection{MC dynamics}
		
		Analogous formulae hold for the (discrete-time) MC variant of the model. Let $p(X_k|X_{k-1},i_k)$ be the probability that the system is in state $X_k$ at step $k$, given that it was in state $X_{k-1}$ at step $k-1$, and that the particle proposed to be moved on step $k$ was $i_k$.  Note that this $p$ is a normalised probability for $X_k$ and $p(X_k|X_k,i_k)$ is generically finite.  Also let $p^V(X_k|X_{k-1},i_k)$ be the analogous quantity for the controlled model.  Then the analogue of $ {\cal A}^{\rm{sym}}$ is
		\beq\label{equ:MCPtV}
		{\cal A}^{\rm{sym}}_{\rm MC}[\mathbf{x}] = - \sum_k \log \frac{p^V(X_{k+1}|X_k,i_k)}{p(X_{k+1}|X_{k},i_k)} - \frac12 [ V(X_{k+1})- V(X_k) ]
		\eeq
		For the logarithm to be finite, it is important that $p(X_{k+1}|X_k,i_k)$ should not be zero (except if $p^V(X_{k+1}|X_k,i_k)=0$ also).  This is the reason to use the Glauber criterion in (\ref{equ:glauber}) instead of the Metropolis condition (because using Metropolis may result in $p(X|X,i)=0$ for some choices of $X,i$ but $p^V(X|X,i)\neq0$).
		
		Note also that ${\cal A}^{\rm{sym}}_{\rm MC}$ depends on which moves were proposed, as well as the actual sequence of states in the trajectory.  If the control force $V$ has a very simple form then it is possible to write an equation similar to (\ref{equ:MCPtV}), in which the action depends only on the actual sequence of states.  This gives some improvement in numerical sampling but is restricted to simple cases, for example where $V$ is a linear potential.
		
		% \emph{RLJ: the following is NOT clear} In our simulations we calculate the integral analytically for the rejection probability when possible (when the optimal force is simple enough). When the integral does not have a analytic solution we consider only the same move in the unbiased model and let the TPS calculate the integral by sampling. 
		
		For the numerical results for the MC variant of the BHPM, we use control forces that are introduced by replacing $\beta\Delta U$ in (\ref{equ:glauber}) by $\beta\Delta U + \Delta V$ where $\Delta V$ is the change in the control potential, for the proposed move.  We note that the optimal auxiliary model for such a system would require that we take instead
		\beq
		p_{\rm acc}  = \frac{\exp[-\psi\delta t-s(r^a(t_+)+r^a(t_-))\delta t/2-\Delta V_{\rm opt}/2]}{1+\exp(\beta\Delta U)} 
		\label{equ:p-acc-aux}
		\eeq
		where $r^a(t_\pm)$ are the values of $r^a$ just before and after the proposed move.  For small $\Delta V$ and small $\delta t$, this is equivalent to replacing  $\beta U\to(\beta U+V)$ in (\ref{equ:glauber}) and it is also equivalent to the Langevin case.\cite{gardiner2004handbook}.
		%chap4, gardiner2004handbookchap7}
		
		We emphasise that for any $V$, the TPS method targets the distribution $\tilde{P}^V_s$ and provides accurate results (as long as sufficiently many TPS moves are performed).  However, it is possible that we might have observed faster convergence if we had used (\ref{equ:p-acc-aux}) instead of simply including  the control potential in (\ref{equ:glauber}).  To check this, we tested an algorithm based on (\ref{equ:p-acc-aux}) for several representative cases; the differences in performance were within the statistical uncertainty of our estimates of the asymptotic variance.
		
		%We have checked that the results shown here are very little affected if we use instead an auxiliary dynamics as in (\ref{equ:p-acc-aux}), with $V_{\rm opt}$ replaced by $V$.
		
		\section{Complex Ornstein-Uhlenbeck processes}
		\label{sec:ou}
		
		We collect some results for biased ensembles constructed from Ornstein-Uhlenbeck (OU) processes, see for example~\cite[Sec.~6.2]{Chetrite2015}.
		Suppose that $z$ is a complex number which evolves by the complex OU process
		\beq
		\dot z = - \omega z + \eta \sqrt{2\gamma} 
		\label{equ:z-lang}
		\eeq
		where $\omega,\gamma$  are real positive constants
		and $\eta$ is a complex-valued white noise.  That is, $\eta=\eta_r + {\rm i}\eta_i$ with real-valued noises $\eta_r,\eta_i$ that satisfy $\langle \eta_r(t) \eta_r(t') \rangle = \frac12 \delta(t-t') = \langle \eta_i(t) \eta_i(t') \rangle$ and $\langle \eta_i(t') \eta_r(t) \rangle = 0$.  Then also $\langle \eta(t) \eta^*(t') \rangle = \delta(t-t')$.  Writing $z=x+{\rm i}y$  one has independent equations of motion for $x$ and $y$.  The corresponding Fokker-Planck equation for the probability density $P=P(x,y)$ is
		\beq
		\dot{P} = \partial_x ( \omega x P ) + \partial_y ( \omega y P ) + \frac{\gamma}{2} ( \partial_x^2 + \partial_y^2 ) P
		\eeq
		whose stationary distribution is $P_0 \propto \exp(-\omega(x^2+y^2)/\gamma)$.  
		Alternatively one may use the calculus of complex variables and consider a probability density defined as $Q=Q(z,z^*)$ which obeys
		\beq
		\dot{Q} = \partial_z ( \omega z Q ) + \partial_{z^*} ( \omega z^* Q ) + 2\gamma \partial_z \partial_{z^*} Q
		\eeq
		The stationary solution is $Q_0\propto \exp(-\omega z^* z /\gamma)$ which is (of course) equivalent to $P_0$ as given above.  The following results can be derived by considering separately the real and imaginary parts of $z$ but we use the complex variable representation, which simplifies the analysis.
		
		For biased ensembles of the form (\ref{sensemble}) with $K = \alpha \int_0^\tobs z^*(t) z(t) \mathrm{d}t$, the dynamical free energy can be obtained by solving the eigenproblem
		\beq
		\psi Q = \partial_z ( \omega z Q ) + \partial_{z^*} ( \omega z^* Q ) + 2\gamma \partial_z \partial_{z^*} Q - s \alpha  z^* z Q
		\eeq
		It is easily verified that the eigenfunction with maximal eigenvalue is
		\beq
		Q\propto \exp\left[-\frac{z^* z}{2\gamma} \left(  \sqrt{\omega^2 + 2s\alpha \gamma} + \omega \right) \right]
		\eeq
		which is valid for $2s\alpha \gamma > -\omega^2$ (otherwise the eigenvalues are not bounded above and the dynamical free energy does not exist).
		The corresponding eigenvalue is 
		\beq
		\psi(s) = \omega -  \sqrt{\omega^2+2\alpha s\gamma}
		\label{equ:psi-ou}
		\eeq  
		To obtain the optimal control force one should solve the adjoint eigenproblem
		\beq
		\psi {\cal F} = - \omega z \partial_z  {\cal F}  - \omega z^* \partial_{z^*} {\cal F} + \frac{\gamma}{2} \partial_z \partial_{z^*} {\cal F} - s \alpha  z^* z {\cal F}
		\eeq
		whose solution is ${\cal F} \propto \exp\left[-\frac{z^* z}{2\gamma} \left( \sqrt{\omega^2 + 2s\alpha \gamma} - \omega \right) \right]$.  
		Note that ${\cal F} \propto Q/Q_0$ which follows because the underlying equation is reversible (obeys detailed balance).
		The optimal control potential is $V_{\rm opt} = -2\log {\cal F}$ (up to an arbitrary additive constant) which yields 
		\beq
		V_{\rm opt} = \frac{z^* z}{\gamma} \left( \sqrt{\omega^2 + 2s\alpha \gamma} - \omega \right)  \; .
		\label{equ:Vopt-ou}
		\eeq
		Away from transient regions, the distribution of $z$ in the biased ensemble is $P_{\rm ave}(z,z^*) \propto {\cal F} Q $ so
		\beq
		P_{\rm ave}(z,z^*) \propto \exp\left(-\frac{z^* z}{2\gamma}  \sqrt{\omega^2 + 2s\alpha \gamma}  \right)
		\label{equ:pave-ou}
		\eeq
		For the discussion here, the case of primary interest is when $s\alpha<0$, in which case the control potential $V_{\rm opt}$ has negative curvature and guides the system towards increasingly large values of $z$.  
		As $s\alpha$ tends to $-\omega^2/(2\gamma)$ one sees that the variance of $P_{\rm ave}$ diverges.  If the original equation (\ref{equ:z-lang}) was derived by linearisation at small $z$, then this divergence indicates the breakdown of the linear approximation, within the biased ensemble.  
		This is the situation discussed in Sec.~\ref{sec:mft-theory}.

		\section{Convergence of TPS}
		\label{app:tps-conv}
		
		\begin{figure}
			\centering
			\includegraphics[width=0.7\linewidth]{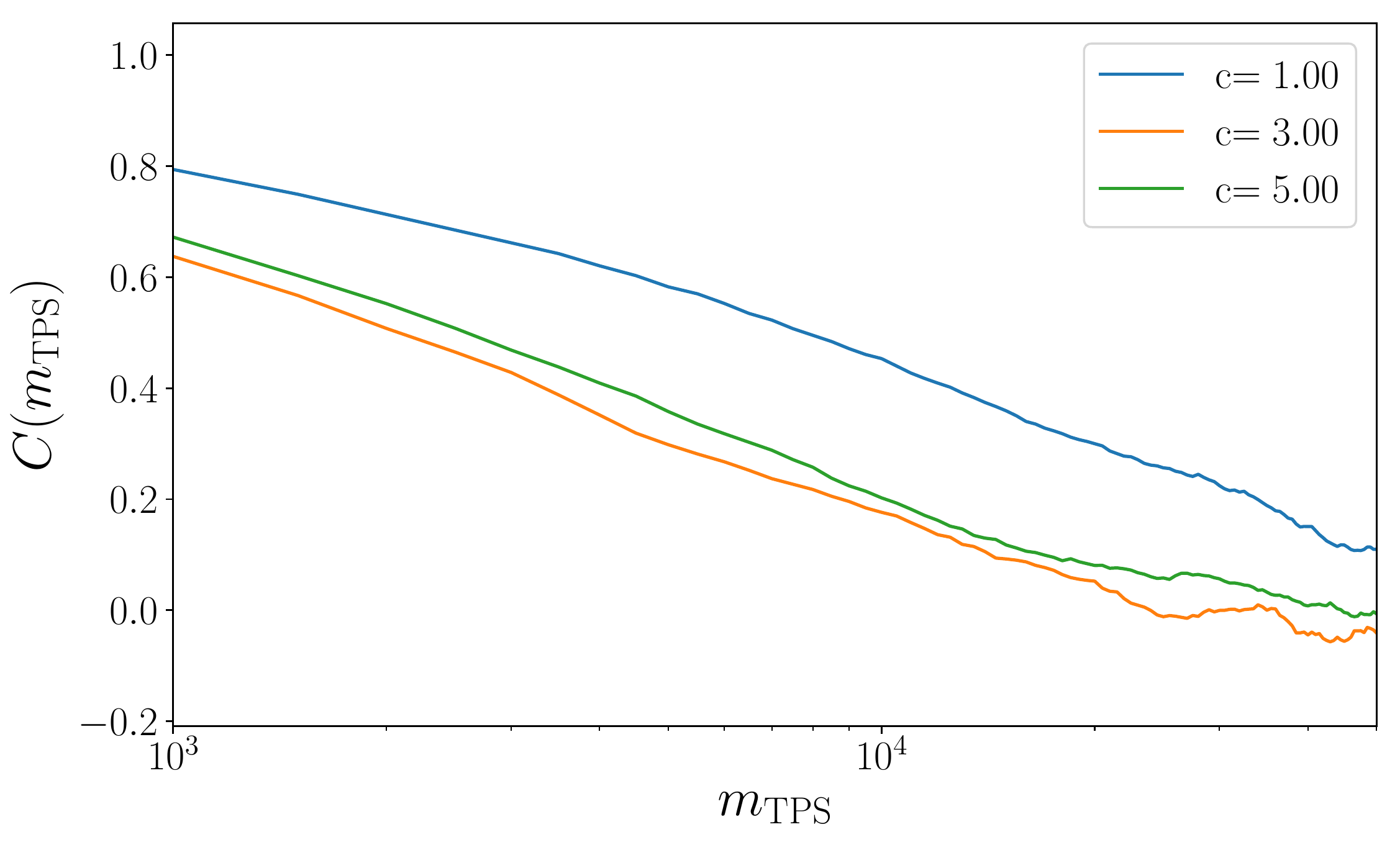}
			\caption{ Autocorrelation $C(m_{\mathrm{TPS}})$ of the TPS sampling method for different biasing forces.  The state points are those of Fig.~\ref{fig:ImproveHole}, for $N=21$.  The autocorrelation function decays faster for $c=3,5$, indicating that sampling is more effective when these control forces are included, consistent with Fig.~\ref{fig:ImproveHole}
			}
			\label{fig:AutoCorr}
		\end{figure}
		
		%With guiding forces we often have only an estimate for the force. Thus to determine which of the one's considered is the best to use there are a number of things that can be done.
		
		In order to measure the improvement in sampling that is achieved by guiding (control) forces, we discuss in the main text the asymptotic variance $\sigma^2_{\rm TPS}$, see (\ref{equ:sigTPS}).  This quantity requires a large amount of TPS data to evaluate it, but does give a reliable estimate of the effort required to obtain an independent sampling from a biased trajectory ensemble.
		As an alternative, we also consider the autocorrelation function.  In the notation of (\ref{equ:Kbar}) let
		\beq
		C(m) = \langle \hat{K}_i \hat{K}_{i+m} \rangle -  \langle \hat{K}_i \rangle \langle \hat{K}_{i+m} \rangle 
		\eeq
		where the average is over many realisations of the TPS algorithm.  One sees that 
		$\chi^{\rm TPS}_m =  \sum_{i,j=1}^m \langle \hat{K}_i \hat{K}_{j} \rangle - \langle \hat{K}_i \rangle \langle \hat{K}_{j} \rangle $ is related to a sum of $C(n)$ over the lag time $n$.
		Fig.~\ref{fig:AutoCorr} shows results for this correlation function.  As in figure \ref{fig:ImproveHole}, one concludes that the sampling is most effective for $c=3$, since the correlations decay most quickly when the control force has this strength.  Compared with the asymptotic variance $\sigma^2_{\rm TPS}$, results for the autocorrelation function are somewhat easier to obtain in practice.  The difficulty is that $\sigma^2_{\rm TPS}=\sum_{m=-\infty}^{\infty} C(m)$ has contributions from weak correlations at large $m$: accurate estimation of these (weak) correlations requires very long TPS runs.

		Another approach is to consider what fraction of TPS moves are accepted, and how this is affected by the guiding forces.  In general, TPS acceptance rates are not reliable as indicators of convergence.  For example short shifting moves lead to slow decorrelation of the trajectories, while longer trajectories may decorrelate the trajectory more quickly, even if the acceptance probability is somewhat lower.  Hence, if a control force leads to acceptance of longer shifting moves then this can still improve sampling, even at the cost of a lower overall  acceptance rate.
		%This can be seen in figure \ref{fig:ActvEnt} where out of the three considered forces $c=1$ has the highest acceptance rate but $c=3$ has the lowest asymptotic variance. 
		
		Despite these limitations, there is useful information available by monitoring TPS acceptance rates. 
		For TPS with control forces in place, it follows from (\ref{equ:PtVs}) that a proposed trajectory is accepted with probability
		\beq
		\rm{min}\left(1,{\rm e}^{\Delta {\cal A}^{\rm sym}_{\rm MC} - s\Delta K} \right)
		\eeq
		where $\Delta K$ is the change in activity between the original and proposed trajectory, and similarly $\Delta {\cal A}^{\rm sym}$ is the change in ${\cal A}^{\rm sym}$.  For TPS with the optimal control potential then ${\cal A}^{\rm sym}_{\rm MC} - s\Delta K=\psi(s)$ for every trajectory so the acceptance probability is unity.  That is
		\beq
		\Delta {\cal A}^{\rm sym} - s\Delta K = 0 \; .
		\label{equ:opt-A-sK}
		\eeq
		
		\begin{figure}
			\centering
			\includegraphics[width=1\linewidth]{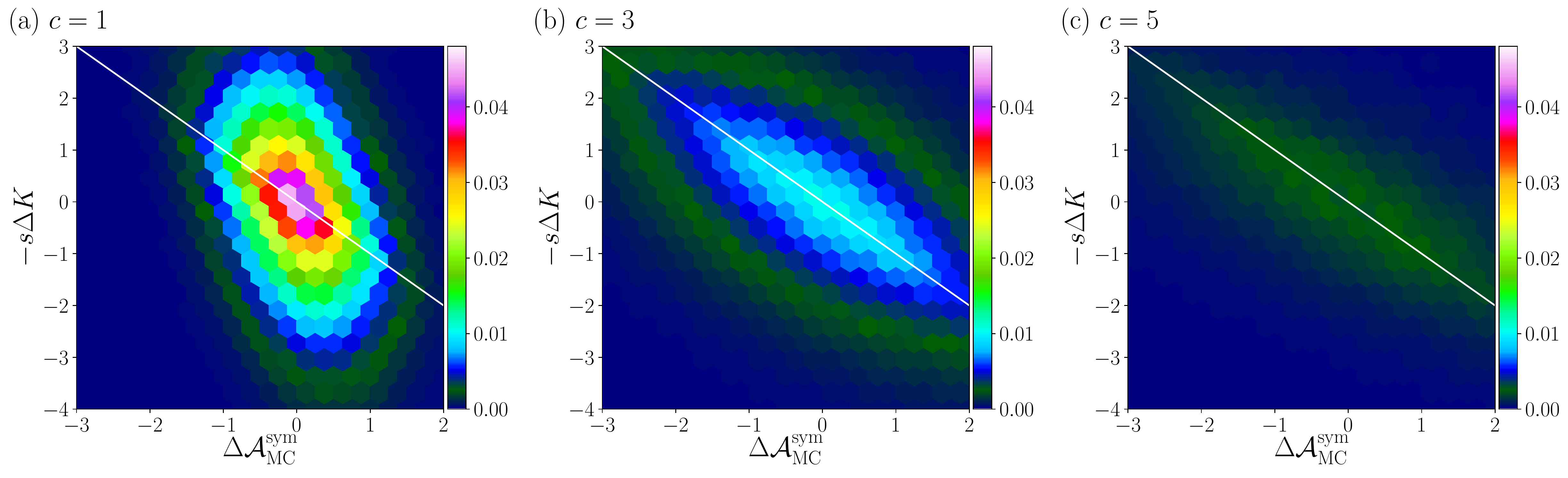}
			\caption{The probability density of $-s\Delta K$ and $\Delta\cal{A}^{\rm{sym}}_{\mathrm{MC}}$ for accepted TPS moves, with the control force (\ref{equ:VY}) in place.  (a) $c=1$; (b) $c=3$, which leads to the most efficient sampling; (c) $c=5$.
				Other parameters are $\Phi_a=0.233$, $N=21$, $\tobs=5.6\tau_{\rm B}$, and $h=11$ as in Fig.~\ref{fig:ImproveHole}(b). }
			\label{fig:ActvEnt}
		\end{figure}
		
		Joint probability density functions for accepted values of $\Delta{\cal A}^{\rm sym}_{\rm MC}$ and $s\Delta K$ are shown in Fig.~\ref{fig:ActvEnt}.
		%the proposed changes of these quantities weighted by their acceptance probability is shown in figure \ref{fig:ActvEnt}, and 
		The relationship (\ref{equ:opt-A-sK}) is indicated.  There are two effects at play here.  For control forces that are close to optimal, the distribution concentrates close to (\ref{equ:opt-A-sK}).  On the other hand, larger control forces tend to suppress the total acceptance, because the forces are not optimal.  The most efficient sampling occurs in an intermediate regime. In this case, we find that the the optimal regime is when the typical values of $s\Delta K$ and $\Delta{\cal A}^{\rm sym}$ are of similar sizes.
		
		%Firstly the more  it is with the linear relationship the closer it is to the optimal guiding force. However this is counteracted by the lowering acceptance rate due to these guiding forces only being approximate. Despite this the biasing force with $c=3$ still improves the sampling eightfold over the system with no guiding force.

		%As noted in chapter $\ref{sec:macro-gap}$ an overlap of the different probabiblity distributions of observables of the path and their probability distributions $P_{\mathrm{ave}}$ and $P_{\mathrm{end}}$ as in figure $\ref{fig:ImproveHole}$ often indicates the optimal control force. Considering just the biased observable is not optimal though as we may over-correct and ruin the agreement of other observables.  
		
		% All of these methods require to have run some simulations already. So in general it is a good idea to start off with a small computationally cheap system and compare some versions of the approximate guiding force with the unguided system for some initial information about their accuracy.

		\section{An alternative measure of dynamical activity}
		\label{app:K-msd}
		
		\begin{figure}
			\centering
			\includegraphics[width=1\linewidth]{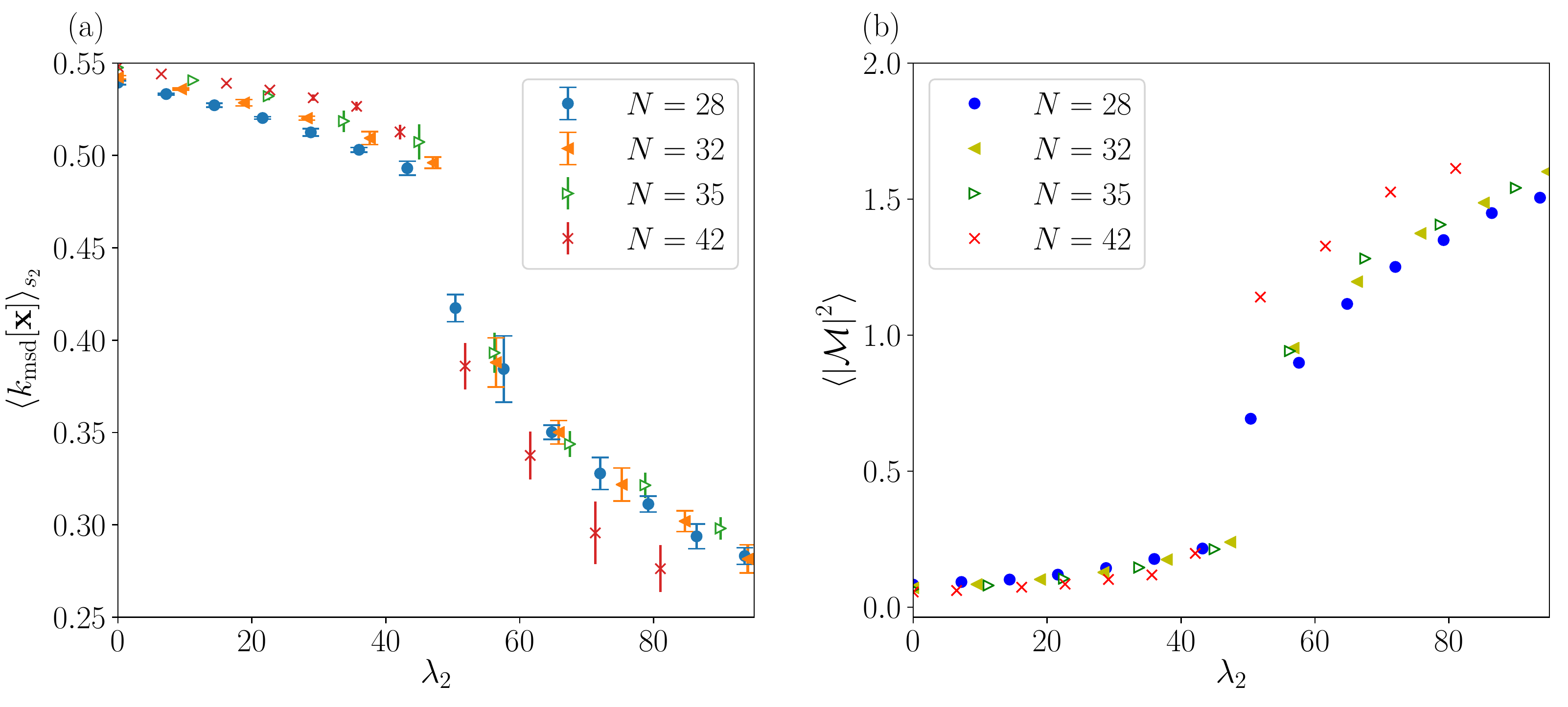}
			\caption{(a) Activity per particle per unit time $k_{\rm{msd}}$ for diffusively rescaled systems. (b) Modulus of the complex order parameter which is related to the first Fourier component of density. Both figures come from systems with $\gamma_{\mathrm{obs}}\approx0.0128$ and $\Phi_a=0.233$.  }
			\label{fig:KMSD}
		\end{figure} 
		
		Previous work has considered large deviations of the dynamical activity in this system~\cite{jack_hyperuniformity_2015,thompson_dynamical_2015}, but using a different measure of activity, which is defined in terms of squared particle displacements.  One separates the time interval $[0,\tobs]$ into $S$ segments, each of length $\Delta t = \tobs/S$.  Then define
		
		\begin{equation}\label{KMSD}
		K_{\mathrm{msd}}[\mathbf{x}]=\frac{1}{2D_0} \sum_{j=1}^{S}\sum_{i=1}^{N}\vert x_i(t_j)-x_i(t_{j-1})-\Delta \bar{x}_j\vert^2,
		\end{equation}
		where $t_j=j\Delta t$ and $\Delta \bar{x}_j$ is the displacement of the centre of mass of all particles, between times $t_{j-1}$ and $t_j$.  The activity $K$ of (\ref{equ:ave-K}) depends on the characteristic length $a$, while $ K_{\mathrm{msd}}$ depends on the parameter $\Delta t$.  To obtain a corresponding length one may define $a_{\rm msd} = \sqrt{2D_0\Delta t}$ where $D_0$ is the single-particle diffusion constant.  For a direct comparison between $K$ and $K_{\rm msd}$ it is natural to take $a_{\rm msd} \approx a$ since this means that both activity measures are sensitive to motion on the same length scales.  
		
		Analogous to (\ref{sensemble}) we define a biased ensemble with a bias parameter $s_2$, as
		\begin{equation}\label{s2ensemble}
		\mathrm{d}P_{s_2}[\mathbf{x}]=\frac{e^{-s_2K_{\rm msd}[\mathbf{x}]}}{Z(s_2)} \, \mathrm{d} P_0[\mathbf{x}] \; .
		\end{equation}	
		
		Also define $\lambda_2 = s_2 \LL^2 / D_0$, analogous to (\ref{equ:def-lambda}), and $k_{\rm msd} = K_{\rm msd}/(N\tobs)$.
		With these definitions, Fig.~\ref{fig:KMSD} shows that the ensemble of (\ref{s2ensemble}) has the same qualitative features as biasing by $K$.
		Specifically, Fig.~\ref{fig:KMSD}(a) is analogous to Fig.~\ref{fig:ScalingWrtN2tS1}(a) and Fig.~\ref{fig:KMSD}(b) is analogous to Fig.~\ref{fig:ScalingWrtN2tS2}(a).
		The data collapses when plotting these results as a function of $\lambda_2$, consistent with the MFT predictions.

	\end{appendix}
	
	\section*{Bibliography}
	\bibliographystyle{unsrt}
	\bibliography{dev,Bibliography}

\begin{thebibliography}{10}

\bibitem{GallavottiFluctuatonTheorem}
G.~Gallavotti and E.~G.~D. Cohen.
\newblock Dynamical ensembles in nonequilibrium statistical mechanics.
\newblock {\em Phys. Rev. Lett.}, 74:2694--2697, 1995.

\bibitem{LebowitzFluctuationTheorem}
Joel~L. Lebowitz and Herbert Spohn.
\newblock A gallavotti--cohen-type symmetry in the large deviation functional
  for stochastic dynamics.
\newblock {\em J. Stat. Phys.}, 95(1):333--365, 1999.

\bibitem{Bodineau2004}
T.~Bodineau and B.~Derrida.
\newblock Current fluctuations in nonequilibrium diffusive systems: An
  additivity principle.
\newblock {\em Phys. Rev. Lett.}, 92:180601, 2004.

\bibitem{Mehl2008}
Jakob Mehl, Thomas Speck, and Udo Seifert.
\newblock Large deviation function for entropy production in driven
  one-dimensional systems.
\newblock {\em Phys. Rev. E}, 78(1), 2008.

\bibitem{Derrida2007}
Bernard Derrida.
\newblock Non-equilibrium steady states: fluctuations and large deviations of
  the density and of the current.
\newblock {\em J. Stat. Mech.}, 2007(07):P07023, 2007.

\bibitem{Lecomte2007}
V.~Lecomte, C.~Appert-Rolland, and F.~van Wijland.
\newblock Thermodynamic formalism for systems with markov dynamics.
\newblock {\em J. Stat. Phys.}, 127(1):51, 2007.

\bibitem{Tailleur2007}
Julien Tailleur and Jorge Kurchan.
\newblock Probing rare physical trajectories with lyapunov weighted dynamics.
\newblock {\em Nature Phy.}, 3(3):203--207, 2007.

\bibitem{Garrahan2007}
J.~P. Garrahan, R.~L. Jack, V.~Lecomte, E.~Pitard, K.~van Duijvendijk, and
  F.~van Wijland.
\newblock Dynamical first-order phase transition in kinetically constrained
  models of glasses.
\newblock {\em Phys. Rev. Lett.}, 98:195702, 2007.

\bibitem{Hedges2009}
L.~O. Hedges, R.~L. Jack, J.~P. Garrahan, and D.~Chandler.
\newblock Dynamic order-disorder in atomistic models of structural glass
  formers.
\newblock {\em Science}, 323(5919):1309, 2009.

\bibitem{nemoto_optimizing_2018}
Takahiro Nemoto, \'Etienne Fodor, Michael~E. Cates, Robert~L. Jack, and Julien
  Tailleur.
\newblock Optimizing active work: Dynamical phase transitions, collective
  motion, and jamming.
\newblock {\em Phys. Rev. E}, 99:022605, 2019.

\bibitem{Weber2014protein}
Jeffrey~K. Weber, Robert~L. Jack, Christian~R. Schwantes, and Vijay~S. Pande.
\newblock Dynamical phase transitions reveal amyloid-like states on protein
  folding landscapes.
\newblock {\em Biophys. J.}, 107(4):974--982, 2014.

\bibitem{Hurtado2014}
Pablo~I. Hurtado, Carlos~P. Espigares, Jes{\'u}s~J. del Pozo, and Pedro~L.
  Garrido.
\newblock Thermodynamics of currents in nonequilibrium diffusive systems:
  Theory and simulation.
\newblock {\em J. Stat. Phys.}, 154(1):214--264, 2014.

\bibitem{YongJooDynamicalSymmetryBreaking}
Yongjoo Baek, Yariv Kafri, and Vivien Lecomte.
\newblock Dynamical symmetry breaking and phase transitions in driven diffusive
  systems.
\newblock {\em Phys. Rev. Lett.}, 118:030604, 2017.

\bibitem{denH-book}
Frank den Hollander.
\newblock {\em Large deviations}.
\newblock American Mathematical Society, Providence, RI, 2000.

\bibitem{Touchette2009}
H.~Touchette.
\newblock The large deviation approach to statistical mechanics.
\newblock {\em Phys. Rep.}, 478(1-3):1, 2009.

\bibitem{Ragone201712645}
Francesco Ragone, Jeroen Wouters, and Freddy Bouchet.
\newblock Computation of extreme heat waves in climate models using a large
  deviation algorithm.
\newblock {\em Proceedings of the National Academy of Sciences}, 115(1):24--29,
  2017.

\bibitem{bodineau_distribution_2005}
T.~Bodineau and B.~Derrida.
\newblock Distribution of current in nonequilibrium diffusive systems and phase
  transitions.
\newblock {\em Phys. Rev. E}, 72(6), 2005.

\bibitem{Jack2019-grow}
Robert~L. Jack.
\newblock Large deviations in models of growing clusters with symmetry-breaking
  transitions.
\newblock {\em Phys. Rev. E}, 100:012140, Jul 2019.

\bibitem{CrooksFluctuationTheorem}
Gavin~E. Crooks.
\newblock Entropy production fluctuation theorem and the nonequilibrium work
  relation for free energy differences.
\newblock {\em Phys. Rev. E}, 60:2721--2726, 1999.

\bibitem{SaitoFluctuationTheorem}
Keiji Saito and Abhishek Dhar.
\newblock Generating function formula of heat transfer in harmonic networks.
\newblock {\em Phys. Rev. E}, 83:041121, 2011.

\bibitem{bertini_macroscopic}
Lorenzo Bertini, Alberto De~Sole, Davide Gabrielli, Giovanni Jona-Lasinio, and
  Claudio Landim.
\newblock Macroscopic fluctuation theory.
\newblock {\em Reviews of Modern Physics}, 87(2):593--636, 2015.

\bibitem{Gingrich2016}
Todd~R Gingrich, Jordan~M Horowitz, Nikolay Perunov, and Jeremy~L. England.
\newblock Dissipation bounds all steady-state current fluctuations.
\newblock {\em Phys. Rev. Lett.}, 116:120601, 2016.

\bibitem{Garrahan2009}
Juan~P Garrahan, Robert~L Jack, Vivien Lecomte, Estelle Pitard, Kristina van
  Duijvendijk, and Fr\'{e}d\'{e}ric van Wijland.
\newblock First-order dynamical phase transition in models of glasses: an
  approach based on ensembles of histories.
\newblock {\em J. Phys. A}, 42(7):075007, 2009.

\bibitem{speck_first-order_2012}
Thomas Speck, Alex Malins, and C.~Patrick Royall.
\newblock First-{Order} {Phase} {Transition} in a {Model} {Glass} {Former}:
  {Coupling} of {Local} {Structure} and {Dynamics}.
\newblock {\em Phys. Rev. Lett.}, 109(19), 2012.

\bibitem{lecomte_inactive_2012}
Vivien Lecomte, Juan~P Garrahan, and Fr{\'e}d{\'e}ric van Wijland.
\newblock Inactive dynamical phase of a symmetric exclusion process on a ring.
\newblock {\em J. Phys. A}, 45(17):175001, 2012.

\bibitem{Hirschberg2015}
Ori Hirschberg, David Mukamel, and Gunter~M Sch{\"u}tz.
\newblock Density profiles, dynamics, and condensation in the {ZRP} conditioned
  on an atypical current.
\newblock {\em J. Stat. Mech.}, 2015(11):P11023, nov 2015.

\bibitem{Shpiel2017}
Ohad Shpielberg, Yaroslav Don, and Eric Akkermans.
\newblock Numerical study of continuous and discontinuous dynamical phase
  transitions for boundary-driven systems.
\newblock {\em Phys. Rev. E}, 95:032137, Mar 2017.

\bibitem{Janas2016}
Michael Janas, Alex Kamenev, and Baruch Meerson.
\newblock Dynamical phase transition in large-deviation statistics of the
  kardar-parisi-zhang equation.
\newblock {\em Phys. Rev. E}, 94:032133, Sep 2016.

\bibitem{espigares2018-crit}
Carlos P\'erez-Espigares, Federico Carollo, Juan~P. Garrahan, and Pablo~I.
  Hurtado.
\newblock Dynamical criticality in open systems: Nonperturbative physics,
  microscopic origin, and direct observation.
\newblock {\em Phys. Rev. E}, 98:060102, 2018.

\bibitem{jack_hyperuniformity_2015}
Robert~L. Jack, Ian~R. Thompson, and Peter Sollich.
\newblock Hyperuniformity and {Phase} {Separation} in {Biased} {Ensembles} of
  {Trajectories} for {Diffusive} {Systems}.
\newblock {\em Phys. Rev. Lett.}, 114(6), 2015.

\bibitem{thompson_dynamical_2015}
Ian~R. Thompson and Robert~L. Jack.
\newblock Dynamical phase transitions in one-dimensional hard-particle systems.
\newblock {\em Phys. Rev. E}, 92(5), 2015.

\bibitem{appert-rolland_universal_2008}
C.~Appert-Rolland, B.~Derrida, V.~Lecomte, and F.~van Wijland.
\newblock Universal cumulants of the current in diffusive systems on a ring.
\newblock {\em Phys. Rev. E}, 78(2), 2008.

\bibitem{brewer_efficient_2018}
Tobias Brewer, Stephen~R. Clark, Russell Bradford, and Robert~L. Jack.
\newblock Efficient characterisation of large deviations using population
  dynamics.
\newblock {\em J. Stat. Mech.}, 2018(5):053204, 2018.

\bibitem{Bodineau2012cmp}
T.~Bodineau and C.~Toninelli.
\newblock Activity phase transition for constrained dynamics.
\newblock {\em Comm. Math. Phys.}, 311(2):357--396, 2012.

\bibitem{Bodineau2012jsp}
Thierry Bodineau, Vivien Lecomte, and Cristina Toninelli.
\newblock Finite size scaling of the dynamical free-energy in a kinetically
  constrained model.
\newblock {\em J. Stat. Phys.}, 147(1):1--17, 2012.

\bibitem{nemoto_finite-size_2017}
Takahiro Nemoto, Robert~L. Jack, and Vivien Lecomte.
\newblock Finite-size scaling of a first-order dynamical phase transition:
  {Adaptive} population dynamics and an effective model.
\newblock {\em Phys. Rev. Lett.}, 118(11), 2017.

\bibitem{Fleming85}
Wendell~H. Fleming.
\newblock A stochastic control approach to some large deviations problems.
\newblock In Italo~Capuzzo Dolcetta, Wendell~H. Fleming, and Tullio Zolezzi,
  editors, {\em Recent Mathematical Methods in Dynamic Programming}, pages
  52--66, Berlin, Heidelberg, 1985. Springer Berlin Heidelberg.

\bibitem{OptimalControlRepresentationChetrite}
Raphael Chetrite and Hugo Touchette.
\newblock Variational and optimal control representations of conditioned and
  driven processes.
\newblock {\em J. Stat. Mech.}, 2015:P12001, 2015.

\bibitem{jack_ptps}
Robert~L. Jack and Peter Sollich.
\newblock Large deviations and ensembles of trajectories in stochastic models.
\newblock {\em Prog. Theor. Phys. Supp.}, 184:304--317, 2010.

\bibitem{nemoto_population-dynamics_2016}
Takahiro Nemoto, Freddy Bouchet, Robert~L. Jack, and Vivien Lecomte.
\newblock Population-dynamics method with a multicanonical feedback control.
\newblock {\em Phys. Rev. E}, 93(6):062--123, 2016.

\bibitem{ray_exact_2018}
Ushnish Ray, Garnet Kin-Lic Chan, and David~T. Limmer.
\newblock Exact fluctuations of nonequilibrium steady states from approximate
  auxiliary dynamics.
\newblock {\em Phys. Rev. Lett.}, 120(21), 2018.

\bibitem{Jacobson-arxiv}
Daniel Jacobson and Stephen Whitelam.
\newblock Direct evaluation of dynamical large-deviation rate functions using a
  variational ansatz.
\newblock arXiv:1903.06098.

\bibitem{gardiner2004handbook}
C.~W. Gardiner.
\newblock {\em Handbook of stochastic methods for physics, chemistry and the
  natural sciences}.
\newblock Springer-Verlag, Berlin, second edition, 2004.

\bibitem{DasLimmer-arxiv}
A.~Das and D.~Limmer.
\newblock arXiv:1909.03589.

\bibitem{Chetrite2015}
Rapha\"{e}l Ch\'{e}trite and Hugo Touchette.
\newblock Nonequilibrium markov processes conditioned on large deviations.
\newblock {\em Ann. Henri Poincar\'{e}}, 16:2005, 2015.

\bibitem{Jack2015b}
R.~L. Jack and P.~Sollich.
\newblock Effective interactions and large deviations in stochastic processes.
\newblock {\em Eur. Phys. J.: Spec. Topics}, 224(12):2351--2367, Sep 2015.

\bibitem{Nemoto2014}
Takahiro Nemoto and Shin-ichi Sasa.
\newblock Computation of large deviation statistics via iterative
  measurement-and-feedback procedure.
\newblock {\em Phys. Rev. Lett.}, 112:090602, 2014.

\bibitem{Chetrite2013}
Rapha\"el Chetrite and Hugo Touchette.
\newblock Nonequilibrium microcanonical and canonical ensembles and their
  equivalence.
\newblock {\em Phys. Rev. Lett.}, 111:120601, 2013.

\bibitem{MPSGarrahan}
Juan~P Garrahan.
\newblock Classical stochastic dynamics and continuous matrix product states:
  gauge transformations, conditioned and driven processes, and equivalence of
  trajectory ensembles.
\newblock {\em J. Stat. Mech.}, 2016(7):073208, 2016.

\bibitem{Hartmann2012}
Carsten Hartmann and Christof Sch{\"u}tte.
\newblock Efficient rare event simulation by optimal nonequilibrium forcing.
\newblock {\em J. Stat. Mech.}, 2012(11):P11004, 2012.

\bibitem{Bolhuis}
Peter~G. Bolhuis, David Chandler, Christoph Dellago, and Phillip~L. Geissler.
\newblock Transition path sampling: Throwing ropes over rough mountain passes,
  in the dark.
\newblock {\em Annual Review of Physical Chemistry}, 53(1):291--318, 2002.

\bibitem{Giardina2011}
Cristian Giardina, Jorge Kurchan, Vivien Lecomte, and Julien Tailleur.
\newblock Simulating rare events in dynamical processes.
\newblock {\em J. Stat. Phys.}, 145(4):787--811, 2011.

\bibitem{Lecomte_2007}
Vivien Lecomte and Julien Tailleur.
\newblock A numerical approach to large deviations in continuous time.
\newblock {\em J. Stat. Mech.}, 2007(03):P03004, 2007.

\bibitem{DirectEval}
Cristian Giardin\`a, Jorge Kurchan, and Luca Peliti.
\newblock Direct evaluation of large-deviation functions.
\newblock {\em Phys. Rev. Lett.}, 96:120603, 2006.

\bibitem{Bertini2005}
L~Bertini, A~De~Sole, D~Gabrielli, G~Jona-Lasinio, and C~Landim.
\newblock Current fluctuations in stochastic lattice gases.
\newblock {\em Phys. Rev. Lett.}, 94(3), 2005.

\bibitem{Bertini2006}
L.~Bertini, A.De Sole, D.~Gabrielli, G.~Jona-Lasinio, and C.~Landim.
\newblock Non equilibrium current fluctuations in stochastic lattice gases.
\newblock {\em J. Stat. Phys.}, 123(2):237--276, 2006.

\bibitem{Imparato2009}
A.~Imparato, V.~Lecomte, and F.~van Wijland.
\newblock Equilibriumlike fluctuations in some boundary-driven open diffusive
  systems.
\newblock {\em Phys. Rev. E}, 80:011131, Jul 2009.

\bibitem{Shpiel2018}
Ohad Shpielberg, Takahiro Nemoto, and Jo\~ao Caetano.
\newblock Universality in dynamical phase transitions of diffusive systems.
\newblock {\em Phys. Rev. E}, 98:052116, Nov 2018.

\bibitem{Bodineau2008}
T.~Bodineau, B.~Derrida, V.~Lecomte, and F.~van Wijland.
\newblock Long range correlations and phase transitions in non-equilibrium
  diffusive systems.
\newblock {\em J. Stat. Phys.}, 133(6):1013--1031, 2008.

\bibitem{Popkov2010}
Vladislav Popkov, Gunter~M. Sch\"{u}tz, and Damien Simon.
\newblock Asep on a ring conditioned on enhanced flux.
\newblock {\em J. Stat. Mech.}, 2010(10):P10007, 2010.

\bibitem{Lazarescu2015}
Alexandre Lazarescu.
\newblock The physicist's companion to current fluctuations: one-dimensional
  bulk-driven lattice gases.
\newblock {\em J. Phys. A}, 48(50):503001, 2015.

\bibitem{Baek2018}
Yongjoo Baek, Yariv Kafri, and Vivien Lecomte.
\newblock Dynamical phase transitions in the current distribution of driven
  diffusive channels.
\newblock {\em J. Phys. A}, 51(10):105001, 2018.

\bibitem{Baek2019-arxiv}
Yongjoo Baek, Yariv Kafri, and Vivien Lecomte.
\newblock Finite-size and finite-time effects in large deviation functions near
  dynamical symmetry breaking transitions.
\newblock arXiv:1905.05486.

\bibitem{Torquato_2003}
Salvatore Torquato and Frank~H. Stillinger.
\newblock Local density fluctuations, hyperuniformity, and order metrics.
\newblock {\em Phys. Rev. E}, 68:041113, 2003.

\bibitem{Garcia_Millan_2018}
R.~Garcia-Millan, G.~Pruessner, L.~Pickering, and K.~Christensen.
\newblock Correlations and hyperuniformity in the avalanche size of the oslo
  model.
\newblock {\em {EPL} (Europhysics Letters)}, 122(5):50003, 2018.

\bibitem{Torquato1}
Chase~E. Zachary, Yang Jiao, and Salvatore Torquato.
\newblock Hyperuniform long-range correlations are a signature of disordered
  jammed hard-particle packings.
\newblock {\em Phys. Rev. Lett.}, 106:178001, 2011.

\bibitem{Torquato2}
Yang Jiao, Timothy Lau, Haralampos Hatzikirou, Michael Meyer-Hermann, Joseph~C.
  Corbo, and Salvatore Torquato.
\newblock Avian photoreceptor patterns represent a disordered hyperuniform
  solution to a multiscale packing problem.
\newblock {\em Phys. Rev. E}, 89:022721, 2014.

\bibitem{Gabrielli-hyperU}
A.~Gabrielli, B.~Jancovici, M.~Joyce, J.~L. Lebowitz, L.~Pietronero, and
  F.~Sylos~Labini.
\newblock Generation of primordial cosmological perturbations from statistical
  mechanical models.
\newblock {\em Phys. Rev. D}, 67:043506, Feb 2003.

\bibitem{Florescu2009}
Marian Florescu, Salvatore Torquato, and Paul~J. Steinhardt.
\newblock Designer disordered materials with large, complete photonic band
  gaps.
\newblock {\em Proc. Natl. Acad. Sci. USA}, 106(49):20658--20663, 2009.

\bibitem{Lebowitz1983}
Joel~L. Lebowitz.
\newblock Charge fluctuations in coulomb systems.
\newblock {\em Phys. Rev. A}, 27:1491--1494, Mar 1983.

\bibitem{Levesque2000}
D.~Levesque, J.~J. Weis, and J.~L. Lebowitz.
\newblock Charge fluctuations in the two-dimensional one-component plasma.
\newblock {\em Journal of Statistical Physics}, 100(1):209--222, 2000.

\bibitem{Asmussen-book}
S.~Asmussen and Peter~W. Glynn.
\newblock {\em Stochastic Simulation: Algorithms and Analysis}.
\newblock Springer, New York, 2007.

\bibitem{Haan_2010}
Laurens de~Haan and Ana Ferreira.
\newblock {\em Extreme Value Theory: An Introduction (Springer Series in
  Operations Research and Financial Engineering)}.
\newblock Springer, 2010.

\bibitem{Hidalgo2017-ii}
Esteban Guevara~Hidalgo, Takahiro Nemoto, and Vivien Lecomte.
\newblock Finite-time and finite-size scalings in the evaluation of
  large-deviation functions: Numerical approach in continuous time.
\newblock {\em Phys. Rev. E}, 95:062134, Jun 2017.

\bibitem{Ferre2018}
Gregoire Ferr\'{e} and Hugo Touchette.
\newblock Adaptive sampling of large deviations.
\newblock {\em J. Stat. Phys.}, 172(6):1525--1544, 2018.

\bibitem{Hunter:2007}
J.~D. Hunter.
\newblock Matplotlib: A 2d graphics environment.
\newblock {\em Computing In Science \& Engineering}, 9(3):90--95, 2007.

\end{thebibliography}
\end{document}